\documentclass[fp,twocolumn]{jpsj3}
\usepackage{txfonts}
\usepackage{array}
\usepackage{bm}
\usepackage{url}

\makeatletter
\newcommand\newblock{\hskip .11em\@plus.33em\@minus.07em}
\makeatother

\title{Community Structure and Its Stability on a Face-to-Face Interaction Network in Kyoto City}

\author{Yu Ohki\thanks{ohki.yu.65a@st.kyoto-u.ac.jp}, Yuichi Ikeda\thanks{ikeda.yuichi.2w@kyoto-u.ac.jp}, and Hitomi Tanaka}
\inst{Graduate School of Advanced Integrated Studies in Human Survivability, Kyoto University, 1 Yoshida-nakaadachi-cho, Sakyo-ku, Kyoto 606-8306, Japan} 

\abst{As social behavior plays an essential role in people's lives, the features of face-to-face interaction networks must be examined to understand people's social behavior. In this study, we focused on the stable community structure of a face-to-face interaction network because it explains the persistent communities caused by the stationary communication patterns of citizens and visitors in a city. We regarded citizens and visitors as two kinds of particles and the community as a phase and theorized the stability of the community structure using the equilibrium conditions among communities. We formulated the chemical potentials of the communities and examined whether they were in equilibrium under the assumption of a canonical ensemble. We estimated the chemical potentials of persistent communities and found that these values matched within approximately 10\% error for each day. This result indicates that the cause of persistent communities is the stability of community structure.}

\begin{document}
\maketitle

\section{Introduction}

Social behavior plays an essential role in people's lives. People are socially connected through various means of communication. Social ties among people are represented by networks. Face-to-face interaction networks shed light on an important aspect of social life, because face-to-face interaction is one of the major means of forming and maintaining social ties. Hence, the features of face-to-face interaction networks must be examined to understand people's social behavior.

Recently, social networks via cell phones and social networking services (SNSs) have been broadly studied. However, large-scale face-to-face interaction networks have not yet been studied because of the cost of data collection. Networks with only a few hundred people have been constructed by conventional methods using a radio frequency identification (RFID) sensor and Bluetooth with a cell phone \cite{RN54,RN43}. However, substantial amounts of data collected by cell phones and SNSs have been used to construct large-scale social networks. The analysis of these networks has revealed the structural characteristics of social networks and their impact on people's lives. These studies revealed the relationships between people's social connections and their economic status\cite{RN46,RN47,RN53,RN64}, social segregation among people with different attributes\cite{RN62,Rn63}, and universal structural features of social networks\cite{Xu2019,Muchnik2013}. We must overcome the limitations of constructing large-scale face-to-face interaction networks and comprehensively understand face-to-face interaction networks \cite{RN71}. 

A communication pattern between citizens living in the city and visitors from outside forms a face-to-face interaction network. We question whether this face-to-face interaction network has persistent communities on weekdays because the commuting flow in the city is steady, and the communication pattern between citizens and visitors seems to be stationary. In this study, persistent communities were defined as communities that formed in a particular location in the city. We note that persistent communities do not necessarily mean that the nodes comprising communities represent the same population. The persistence of communities is an important characteristic of a face-to-face interaction network in the city. The existence of persistent communities indicates that people in a major part of the network have stable social connections among themselves, while people in other parts are exposed to unsteady social interaction. People tend to isolate themselves from social connections in non-persistent communities. Social isolation is a serious problem that needs to be measured \cite{Ohki2021}. Previous studies using activity pattern analysis have examined communication patterns among people with different attributes\cite{RN55,RN57,RN58}. However, these studies have not analyzed the stationarity of communication through face-to-face inter- action using the network science approach. We aim to find persistent communities from real data analysis to understand the characteristics of communication patterns in the city. 

Persistent communities have factors that enable their stable existence. We consider thermodynamic stability to be the cause of the persistence of the community. Therefore, we expect that persistent communities in face-to-face interaction networks are satisfied under thermodynamic stability conditions. Previous studies have analyzed the stability of communities to evaluate the quality of community analysis. Some methods have been proposed for measuring stability using network structures, such as clusters and reciprocity\cite{RN23,RN31}, node centrality within communities relative to the outside\cite{RN41}, groups of nodes that are invariant over time in the network\cite{RN33,RN35}, consensus of results from multiple methods of community analysis\cite{RN32}, and autocorrelation of dynamic Markov processes\cite{RN34}. However, these stability measures do not represent the cause of the community's stable existence. We developed a theory to examine stable community structures in terms of thermodynamics. This theory provides a thermodynamic explanation of the stability of persistent communities in face-to-face interaction networks. It is also possible to adapt the developed theory of stable community structure to any kind of complex system because a network structure universally exists in any complex system.

We aim to achieve three goals: (i) to construct a large-scale face-to-face interaction network using mobility data, (ii) to identify persistent communities consisting of citizens and visitors, and (iii) to explain the cause of persistent communities according to the theory of stable community structure using a statistical mechanics model. We used mobility data to construct a large-scale face-to-face interaction network in Kyoto City, Japan, and explored the persistent communities. We calculated the chemical potentials of the communities based on the statistical mechanical model. We then explain the stability of the community structure that satisfies the equilibrium condition among the communities.

This study contributes to our understanding of social networks in several ways. First, we developed a method to construct a large-scale face-to-face interaction network using mobility data, which makes it easy to collect a large amount of data. Second, we show a persistent community structure in Kyoto City, which indicates stationary communication patterns between citizens and visitors. Third, we adapt the theory to the results of real data analysis and find the community structure among persistent communities that satisfy the thermodynamic stability conditions. This theory of a stable community structure has been missing in previous network science studies. We consider this to explain the cause of the persistence of communities.

The remainder of this paper is organized as follows. In Sect. 2, we describe a method for constructing a large-scale face-to-face interaction network using mobility data, and identify persistent communities from data analysis. We then develop the theory of a stable community structure and formulate the chemical potential in each community in Sect. 3. In Sect. 4, we present the results of the network construction, community analysis, and estimation of chemical potential. Section 5 discusses the stable community structure of the face-to-face interaction networks. Finally, we conclude the paper in Sect. 6.

\section{Face-to-face interaction network}

We used mobility data acquired from the global positioning system (GPS) of the cell phones in Kyoto City, Japan. We describe methods to construct a face-to-face interaction network using the data, analyze the basic network features and community structure, and identify persistent communities over the data period.

\subsection{Data}

The data were collected by Agoop Corp.\cite{Agoop} over 39 days, including weekdays in February and April 2019. People's movements were recorded by GPS on a personal level in Kyoto City. Table I summarizes these data. The data included 33,238 people and 1,716,164 logs on an average for each day. Each log included the location of a person at a given time. The data items were person ID, time, latitude, longitude, 100 m mesh ID, GPS measurement accuracy, and the estimated residential area. All data were anonymized, and each participant had one unique ID. The ID was consistent for one day but varied daily, and the personal trajectory could only be tracked within one day. 

\begin{figure}[h!]
 (a) 
 \begin{center}
 \includegraphics[width=80mm]{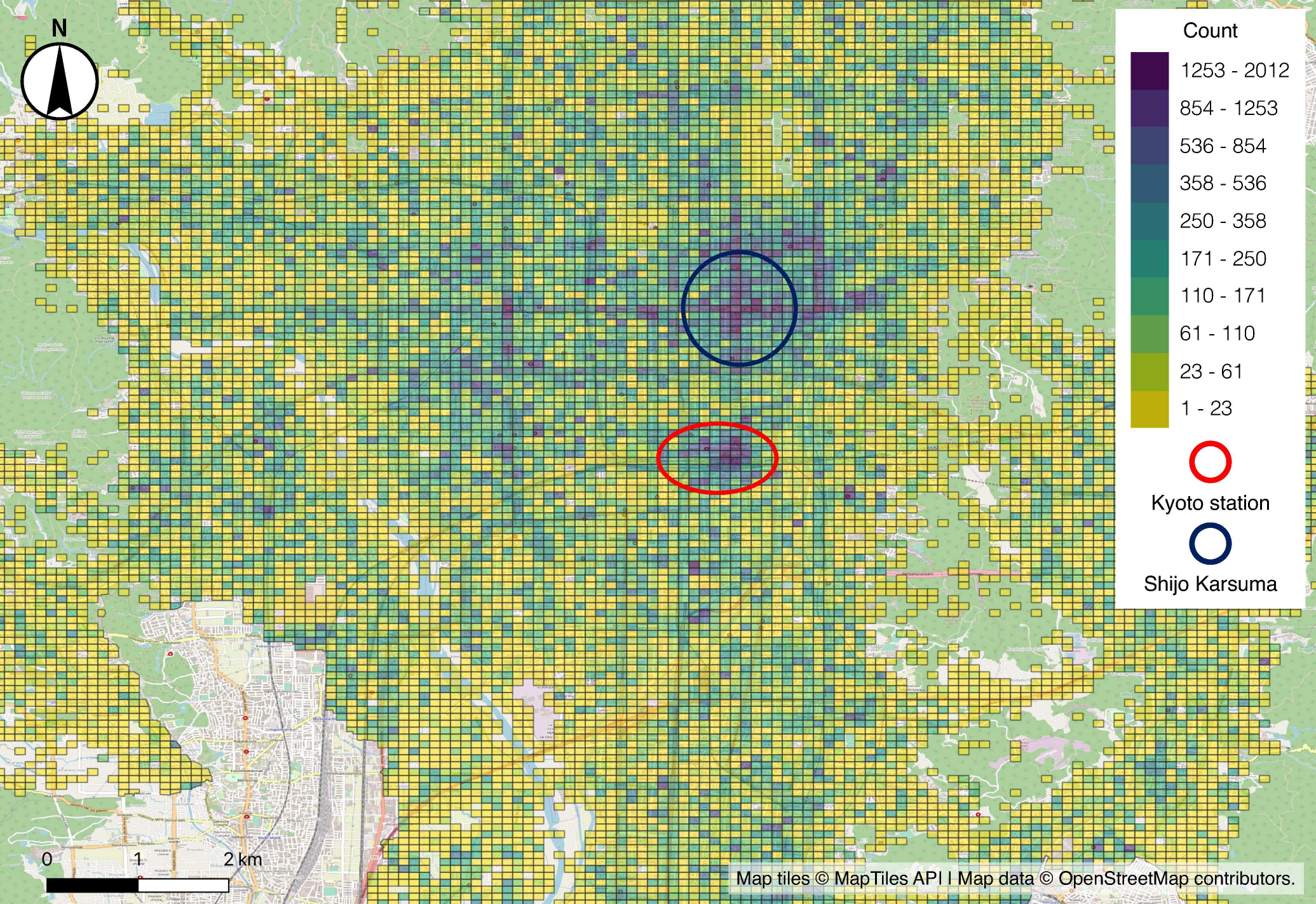} \\
 \end{center}
 (b) 
 \begin{center}
 \includegraphics[width=80mm]{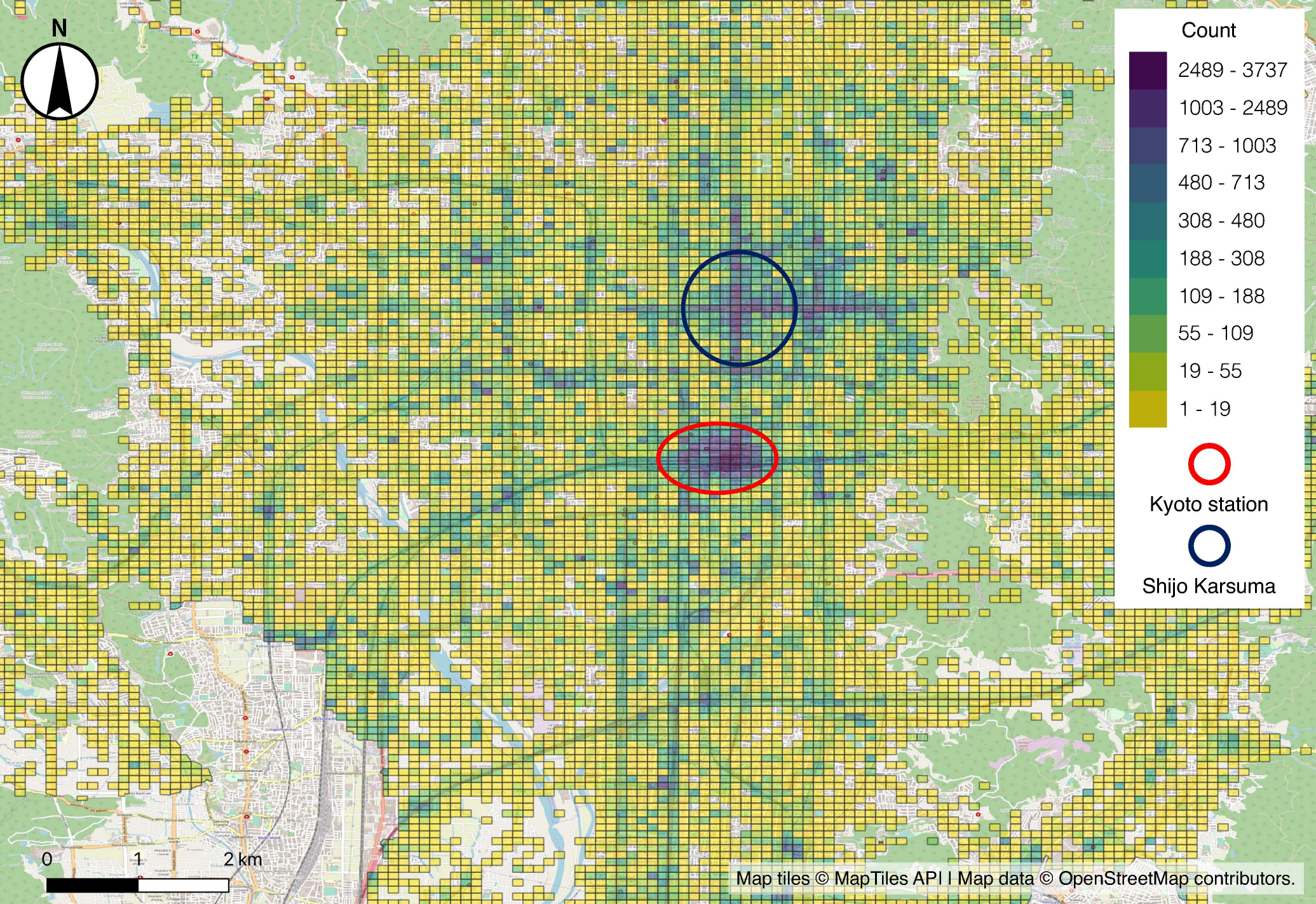}
 \caption{\label{figure1} Spatial Distribution of the number of logs of the mobility data of citizens and visitors during the activity period (09h00--16h59) on Tuesday, April 2, 2019. The red and blue circles indicate the locations of \textit{Kyoto station} and \textit{Shijo Karasuma}, respectively, which are city centers in Kyoto City: (a) Citizens, (b) Visitors.}
\end{center} 
\end{figure}

We labeled people who were estimated to live in Kyoto City as citizens and others as visitors. Figures \ref{figure1}(a) and \ref{figure1}(b) show the spatial distribution of the number of logs of people's mobility data during the activity period (09h00--16h59) on Tuesday, April 2, 2019. The activity area of the citizens tended to spread widely, whereas visitors were concentrated in the city centers, such as \textit{Kyoto station} and \textit{Shijo Karasuma}.

\begin{table}[h]
\caption{\label{table1}Summary of mobility data.}

\begin{tabular}{@{}lllll}
\hline
 & Number & Number & Number & Number \\
 & of people & of citizens & of visitors & of logs \\
\hline
Mean & 33238 & 18282 & 14956 & 1716164 \\
SD & 1934 & 782 & 1262 & 30112 \\
CV & 0.058 & 0.043 & 0.084 & 0.018 \\
\hline
\end{tabular} \\
SD: standard deviation, CV: coefficient of variation.
\end{table}

\subsection{Network construction}

\begin{figure}[h]
\begin{center}
 \includegraphics[width=55mm]{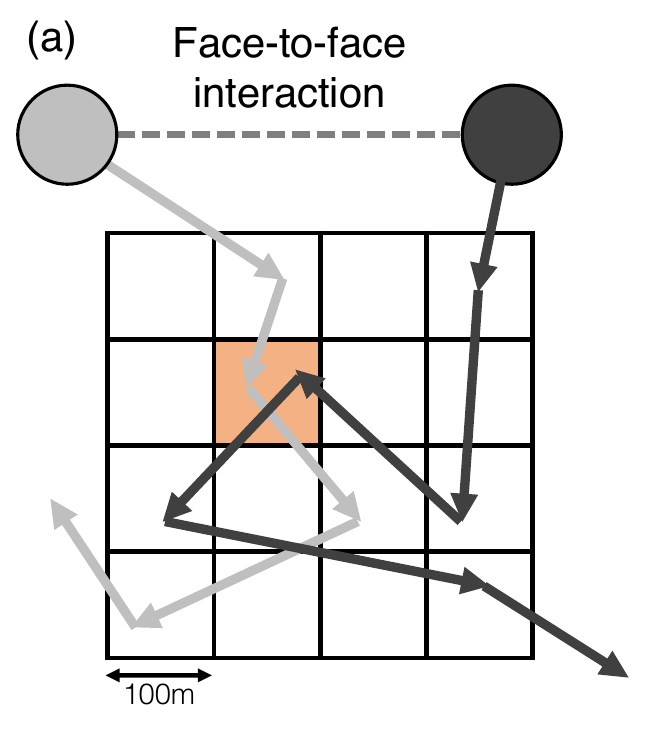} \\
 \includegraphics[width=60mm]{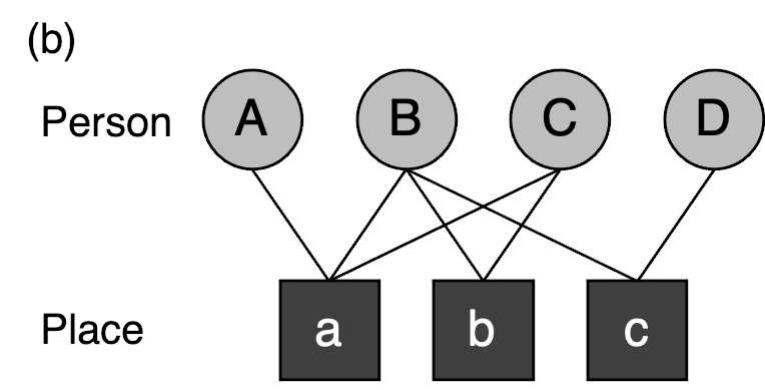} \\
 \includegraphics[width=60mm]{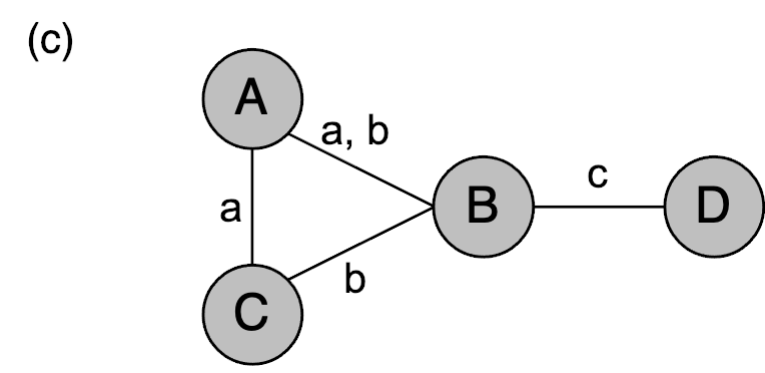}
 \caption{\label{figure2} Procedure for constructing a face-to-face interaction network: (a) Identify face-to-face interactions when two people stay in the same 100 m mesh $>1$ h, (b) Construct a person--place bipartite graph, (c) Project it into a person--person graph.}
\end{center}
\end{figure}

We constructed daily face-to-face interaction networks using this data. We set a condition to identify face-to-face interaction. When two people stayed in the same 100 m mesh for more than one hour, we regarded them as having face-to-face interaction. The procedure for constructing a face-to-face interaction network is as follows. First, we estimated 100 m mesh IDs from the data where people stayed every min. Second, we compared the trajectories of the two people and examined whether they stayed in the same 100 m mesh for $>1$ h, as shown in Fig. \ref{figure2}(a). This step was repeated for all combinations of people. Third, we connected two people and the place where they interacted and constructed a person--place bipartite graph, as shown in Fig. \ref{figure2}(b). Finally, we projected the person--place bipartite graph and constructed a person--person graph representing the face-to-face interaction network, as shown in Fig. \ref{figure2}(c).

\begin{figure}[h]
\begin{center}
 \includegraphics[width=70mm]{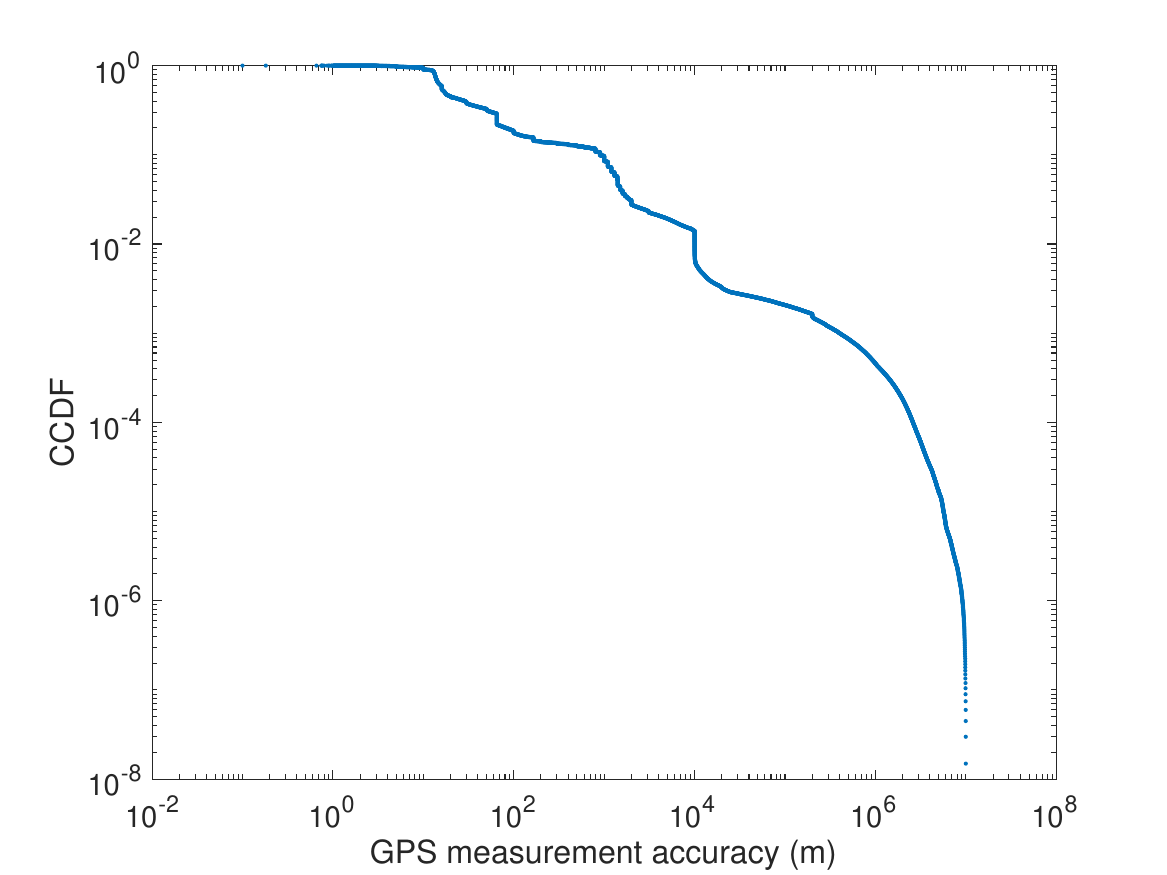}
 \caption{\label{figure3} Empirical complementary cumulative distribution functions of GPS measurement accuracy during the data period. We removed Data with an accuracy of 1,000 m or more, which is 9.7\% of the data. 81.5 \% of the data had an accuracy of 100 m or less.}
\end{center}
\end{figure}

We removed data for which the GPS measurement accuracy was much higher than 100 m. The distribution of the measurement accuracy during the data period is shown in Fig.  \ref{figure3}. The mean accuracy was 1,929 m. We removed data with an accuracy of 1,000 m or more, which was 9.7\% of the data. 81.5\% of the data had an accuracy of 100 m or less. This indicates that most of the data were accurate within the mesh size.

We also removed people with low log counts from the data analysis after removing the data due to poor measurement accuracy. We used data from people whose daily logs were recorded at a frequency of at least once every 30 min, because the raw data included people with very few daily logs. In addition, we used logs only during the activity period (09h00--16h59) because visitors tended to leave Kyoto City outside before and after this period and did not have any logs at night.

\begin{figure}[h]
\begin{center}
 \includegraphics[width=70mm]{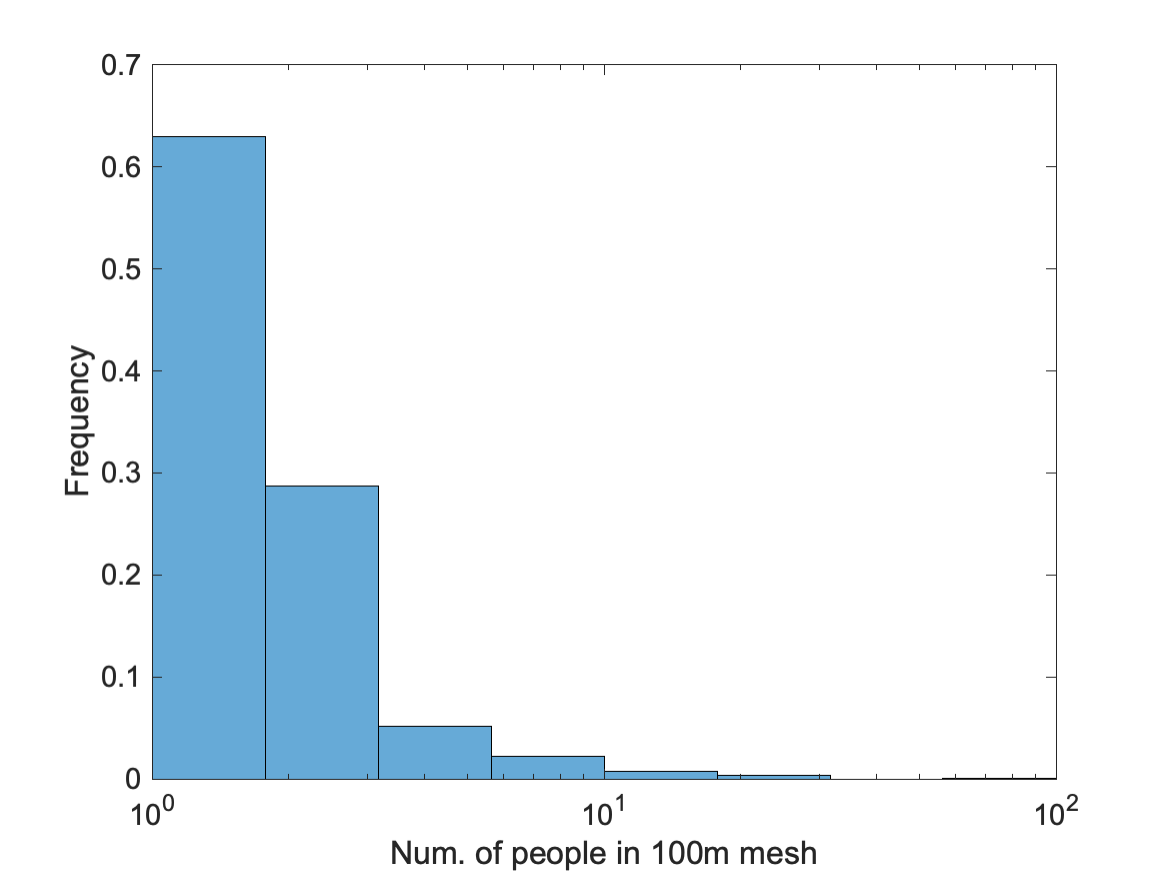}
 \caption{\label{figure4} Histogram of the number of people staying at the same time for each mesh on Tuesday, April 2, 2019, at 9:00 a.m.}
\end{center}
\end{figure}

Figure \ref{figure4} shows a histogram of the number of people staying at the same time for each mesh on Tuesday, April 2, 2019, at 9:00 a.m. We excluded meshes without any people staying from the distribution. This shows that approximately 60\% of the meshes have only one person staying, and approximately 95\% have four or fewer people staying simultaneously. Since only a few meshes showed a concentration of more than ten people, we consider that the mesh size is appropriate for identifying face-to-face interaction.

\subsection{Network features}

We calculated the average degree, distance, clustering coefficient, and assortativity as network features of face-to-face interaction networks. These features represent the overall network structure. We used these features to confirm the structural similarity of the constructed networks for all days.

\subsubsection*{Average degree}

The number of edges of a node $i$ is a degree $k_i$. The average degree $\langle k\rangle$ is the average of the degrees of all the nodes: $\langle k\rangle=\sum k_i/N=2L/N$ in a network with $N$ nodes and $L$ edges. The face-to-face interaction network represented the average number of interactions.
   
\subsubsection*{Average distance}

The distance $d_{ij}$ is the minimum number of edges between nodes $i$ and $j$. The average distance $\langle d \rangle=\sum_{i,j;i\neq j}d_{ij}/N(N-1)$ is the average of the distances among nodes. This means the average number of interactions that people connect through.

\subsubsection*{Average clustering coefficient}

The clustering coefficient of a node is the ratio of adjacent nodes with edges among them. When node $i$ has a degree $k_i$ and the number of edges among its adjacent nodes is $l_i$, the local clustering coefficient $C_i=2 l_i/k_i(k_i-1)$. The average value of $C_i$ for all nodes is the average clustering coefficient $\langle C\rangle=\sum C_i/N$. This indicates the tendency for when a person interacts with others; they also have interactions among themselves.
   
\subsubsection*{Assortativity}
  
Assortativity $r_\mathrm{ass}$ is the Pearson correlation coefficient between the degrees of both ends of the edges. When $r_\mathrm{ass}>0$, the face-to-face interaction network was an assortative network. However, when $r_\mathrm{ass}<0$, it was a disassortative network.

\subsection{Community analysis}
Infomap was used to detect the communities. This is a community analysis method that optimizes the map equation as an evaluation function\cite{RN51,Rosvall2009}. Modularity maximization is a widely used method for community analysis. However, it can fail to detect small communities owing to its resolution limit\cite{barabasi2016network}. Infomap has the advantage of a smaller resolution limit than modularity\cite{PhysRevE.91.012809}. We chose Infomap because of the merit of the resolution limit and the suitability for community analysis of large-scale networks.

The network was assumed to be partitioned into $n$ communities, and the trajectory of a random walker in this network was encoded in the most efficient manner. In other words, this trajectory was described by the smallest number of codes. In ideal encoding, the random walker remains in the communities and did not come out easily. We obtained this ideal code in the following manner. First, a codeword was assigned to each community. Second, one codeword was assigned to each node in each community. The same codeword can be used for different communities. Finally, an exit codeword was assigned to each community when a random walker left the community. The goal was to provide the shortest code for describing a random walk in a network. We determined the community structure by looking at the codebook, which has the code assigned to each community. 

The optimized code was obtained by minimizing the map equation:
\begin{eqnarray}
L_\mathrm{map}=q_{\curvearrowright}H_\mathrm{map}(\mathcal{Q})+\sum^{n}_{i=1}p^i_{\circlearrowright}H_\mathrm{map}(P_i) \label{mapequation}
\end{eqnarray}
where the first term was the expected value of the number of bits are needed to describe the random walker's movement between the communities. $q_{\curvearrowright}$ was the probability that the random walker moved between communities during a given step, and $H_\mathrm{map}(\mathcal{Q})$ was the entropy of the random walker's movement between the communities. The second term was the expected value of the number of bits required to describe a random walker's movement within communities. $p^i_{\circlearrowright}$ was the summation of the probability that the random walker moved within community $i$ and exited the community $i$ during a given step, and $H_\mathrm{map}(P_i)$  was the entropy of the random walker's movement within the community $i$.

$L_\mathrm{map}$ was the value obtained for a particular network partition. The best partition was determined by minimizing $L_\mathrm{map}$ across all possible partitions. Each node started as a separate community, and adjacent nodes were combined into a single node in random sequential order. If the integration of adjacent nodes into a single community reduced $L_\mathrm{map}$, then integration was adopted. After each integration, the value of $L_\mathrm{map}$ was updated using eq. (\ref{mapequation}). This was systematically repeated for all nodes. The resulting community structure was integrated into a new community to complete the path. We then repeated the algorithm for the new network with aggregated nodes.

\subsection{Identification of persistent communities}

We defined persistent communities as those that remained at the same location during the data period. We used the spatial distribution of interactions among nodes belonging to the community and clustering analysis of the distributions to identify persistent communities.

The edges of the face-to-face interaction network contained spatial information about the 100 m mesh ID where the interaction occurred. Community $i$ was characterized as the spatial distribution of interactions and was represented using the vector $\bm{x}_i$:
\begin{eqnarray}
\bm{x}_i=\left(f_{i1}, f_{i2}, \cdots, f_{il}, \cdots, f_{iM}\right),
\end{eqnarray}
where $f_{il}$ is the number of edges with a mesh $l$ in the community $i$, and $M$ is the total number of meshes where people have at least one interaction during the data period $d_\mathrm{all}$.

The distance $s_{ij}$ between communities $i$ and $j$ was calculated using the cosine distance ($1-$cosine similarity):
\begin{eqnarray}
s_{ij}=1-\frac{\bm{x}_i\bm{x}_j^T}{\sqrt{(\bm{x}_i\bm{x}_j^T)(\bm{x}_j\bm{x}_j^T})}.
\end{eqnarray}
The distance matrix $\bm{S}$ is a $n_\mathrm{all} \times n_\mathrm{all}$ matrix and its $i,j$ component is $s_{ij}$, where $n_{\mathrm{all}}$ is the total number of communities for all days. Let the number of communities of day $d$ be $n_d$, and $n_\mathrm{all}=\sum_{d=1}^{d_\mathrm{all}}n_d$. We classify $n_\mathrm{all}$ communities for $d_\mathrm{all}$ days into $k$ clusters using the distance matrix $\bm{S}$ by the $k$-means method.

From the results of the clustering analysis, we identified persistent communities. If communities belong to a cluster on more than 80\% of the days, the communities of the cluster are persistent.

\section{Theory of stable community structure} 

\begin{figure}[h]
\begin{center}
\includegraphics[width=80mm]{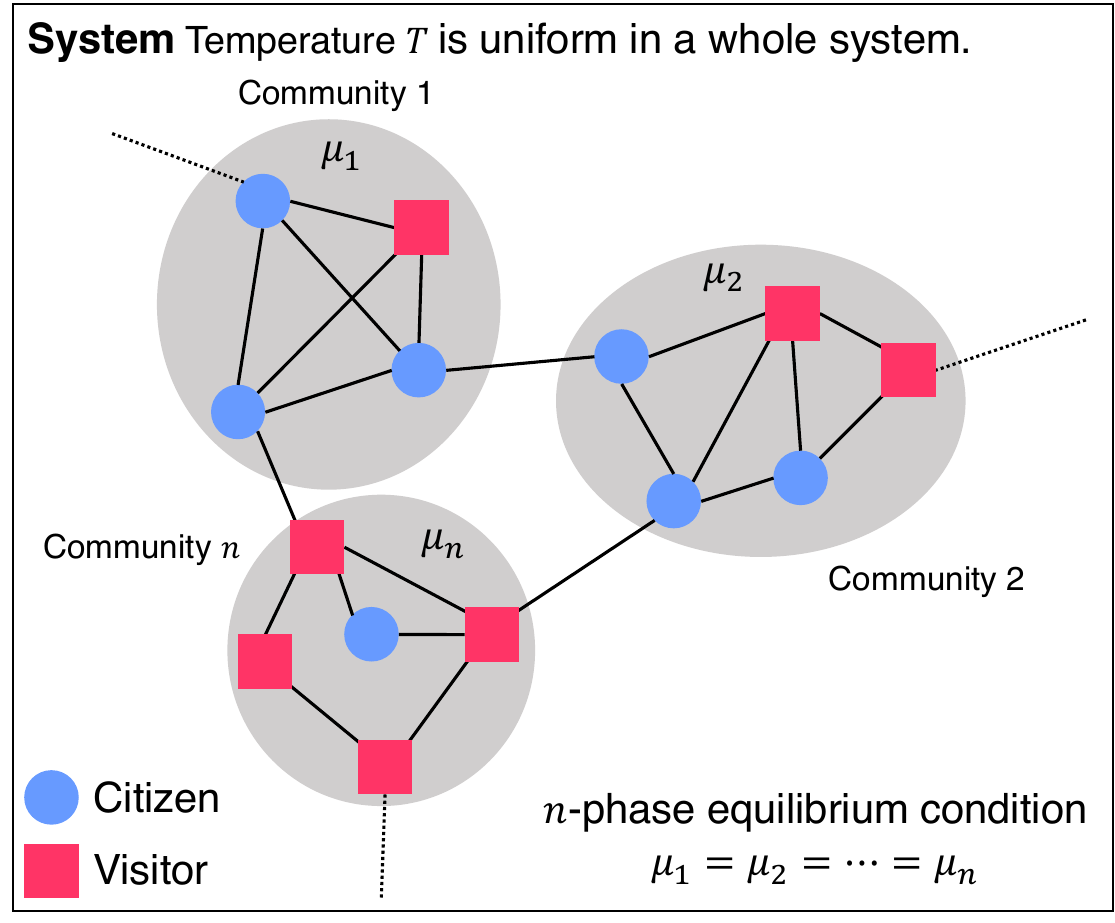}
\caption{\label{figure5}Model of the system in which the face-to-face interaction network is regarded as a system consisting of citizens and visitors and each community of the network as a phase in this system. $\mu$ is the chemical potential of each community. Values of $\mu$ for each community are equal under the equilibrium condition.}
\end{center}
\end{figure}

We developed a theory to examine stable community structures in terms of thermodynamics. A stable community structure indicates that relationships among communities satisfy the condition for thermodynamic stability. We regarded citizens and visitors as two types of particles and the community as a phase, and theorized the stability of the community structure using the equilibrium conditions among communities. Fig. \ref{figure5} illustrates the concept of this model.

\subsection{Basic assumption}

We considered Kyoto City as a two-dimensional plane and as system in which interactions occur between people moving around a location on the plane. When the system is in a bound state, some people remain in their neighborhood because of their interactions with neighboring people. Groups of people interact densely with each other in locations where multiple people are concentrated and form a cluster structure. Consequently, these clusters were detected as persistent communities in the face-to-face interaction network. Corresponding to persistent communities, we examined whether the equilibrium among persistent communities is satisfied as a function of thermodynamic stability. We assumed people move freely and interact with each other in a two-dimensional plane, and applied physical position and momentum to treat people's face-to-face interactions in a statistical mechanics model. We formulated an equilibrium relationship among communities in a face-to-face interaction network during the weekday activity period. This study focused on face-to-face interaction networks during the weekday activity period (09h00--16h59). Since many visitors flow into Kyoto City before the activity period starts and flow out of Kyoto City after the activity period ends, The number of people entering and leaving the system is sufficiently small during the activity period on weekdays. Therefore, it is appropriate to assume that the system is in equilibrium.

\subsubsection*{Unit system}

We defined the unit system using $\hbar,k_\mathrm{b},l_\mathrm{m}$, and $m_\mathrm{p}$ as the fundamental units. Unit of length $l_\mathrm{m}=100$ m to match the mesh size. $m_\mathrm{p}$ represents the mass of a person. We considered $m_\mathrm{p}=50$ kg as a uniform value for all people. We did not explicitly define the value of $k_\mathrm{b}$. Since $\beta=1/k_\mathrm{b} T$, we considered $T'=k_\mathrm{b} T$ as the thermodynamic temperature.

Since $\hbar$ is defined at the micro-scale $S_\mathrm{micro}$ (molecule-level), we need to correct the value of $\hbar$ for macro-scale $S_\mathrm{macro}$ (person-level). Table \ref{table2} shows the typical correspondence scales of the mass dimension M, length dimension L, and time dimension T on the macro- and micro-scales. The macro-scale quantities represent per-person quantities, whereas the micro-scale quantities represent per-molecule quantities. The scales of micro-scales are considered as follows: The mass dimension M scale is $10^{-23}$ g, because 1 mol of mass per molecule scales as the order of g. The length dimension L uses one $\AA$ as the scale of the size of the molecule. The time dimension T is one ns, which is the period of molecular vibration since the vibration frequency scale is of the order of GHz.

\begin{table}[h]
\caption{\label{table2}Correspondence scales on macro-scale $S_\mathrm{macro}$ and micro-scale $S_\mathrm{micro}$.}
\begin{tabular}{@{}llll}
\hline
Dimension & $S_\mathrm{macro}$ & $S_\mathrm{micro}$ & $S_\mathrm{macro}/S_\mathrm{micro}$ \\ \hline
Mass M & 10 kg & $10^{-23}$ g & $10^{27}$ \\
Length L & 1 m & 1 \AA & $10^{10}$ \\
Time T & 1 s & 1 ns & $10^{9}$ \\
\hline 
\end{tabular} \\
\end{table} 

If the quantity $x$ is approximately proportional to the scale, we can convert the micro-scale quantity $x_\mathrm{micro}$ to the macro-scale quantity $x_\mathrm{macro}$ according to the following equation:
\begin{eqnarray}
 x_\mathrm{macro} \simeq \frac{S_\mathrm{macro}[x]}{S_\mathrm{micro}[x]}x_\mathrm{micro}. \label{x}
\end{eqnarray}
Here $S_\mathrm{macro}[\cdot]$ and $S_\mathrm{micro}[\cdot]$ represent the macro- and micro-scales of a certain quantity, respectively. We examine the macro scale's value of $\hbar$ using this equation. We denoted $\hbar$ on the macro- and micro-scales as $\hbar_\mathrm{macro}$ and $\hbar_\mathrm{micro}$, respectively. Since $\hbar$ is a quantity with dimension $\mathrm{ML^2T^{-1}}$, the relationship between $\hbar_\mathrm{macro}$ and $\hbar_\mathrm{micro}$ is represented as follows using the relationship in eq. (\ref{x}): 
\begin{eqnarray}
 \hbar_\mathrm{macro}&\simeq&\frac{S_\mathrm{macro}[\hbar]}{S_\mathrm{micro}[\hbar]}\hbar_\mathrm{micro} \nonumber \\
 &=&\frac{S_\mathrm{macro}[m]}{S_\mathrm{micro}[m]} \cdot \left(\frac{S_\mathrm{macro}[l]}{S_\mathrm{micro}[l]}\right)^2 \cdot \left(\frac{S_\mathrm{macro}[t]}{S_\mathrm{micro}[t]}\right)^{-1} \hbar_\mathrm{micro}. \nonumber \\
 \label{scale}
\end{eqnarray}
Here $m$ is the mass, $l$ is the length, and $t$ is the time. Since $\hbar_\mathrm{micro}=1.054\times10^{-34} \ \mathrm{kg\cdot m^2/s}$ is used for the micro-scale system, we assigned this value and the values in Table \ref{table2} to the eq. (\ref{scale}) and obtained $\hbar_\mathrm{macro}\simeq1.054\times10^4 \ \mathrm{kg\cdot m^2/s}$ as the Dirac's constant for the macro-scale system. Hereafter, $\hbar_\mathrm{macro}$ is simply denoted as $\hbar$ and used as one of the fundamental units. 

These fundamental units entirely determine the unit of each dimension: the mass dimension M, length dimension L, time dimension T, and thermodynamic temperature dimension $\Theta$. All of the units treated in this theory were assembled using these fundamental units.

\subsubsection*{Kinetic energy}

Kinetic energy is the energy due to the movement of people and represents the activity of movement of a person at a certain location on a two-dimensional plane moving around its neighborhood. The kinetic energy $\mathcal{H}_i^{\mathrm{id}}$ of person $i$ is represented by the following equation:
\begin{eqnarray}
 \mathcal{H}_i^{\mathrm{id}}=\frac{1}{2m_\mathrm{p}}\bm{p}_i^2. \label{h_id}
\end{eqnarray}
Here $\bm{p}_i$ denotes the momentum of $i$. People move around randomly on a two-dimensional plane when there is no interaction with other people.

\subsubsection*{Interaction energy}

The interaction energy represents the potential energy acting between two people. Since the face-to-face interaction between people takes place at an approximately constant distance, we assumed there is a stable distance for the interaction of face-to-face interaction. We introduced the interaction energy to represent the stability of face-to-face interaction. The interaction energy $\mathcal{H}_{ij}^{\mathrm{int}}$ between persons $i$ and $j$ is expressed as
\begin{eqnarray}
\mathcal{H}_{ij}^{\mathrm{int}}=\phi_{gh}(|\bm{r}_i-\bm{r}_j|) \quad (g,h=\mathrm{c},\mathrm{v}). \label{h_int}
\end{eqnarray}
Here, subscripts c and v denote citizens and visitors, respectively and distinguish interactions between people of the same and different types. $\bm{r}_i$ and $\bm{r}_j$ denote the positions of persons $i$ and $j$. We assumed the interaction potential $\phi$ to be Lennard-Jones potential. Since face-to-face interactions between people are temporary, we used the Lennard-Jones potential, which represents a weak interaction, such as van der Waals forces. Therefore, $\phi(r)$ can be expressed as follows:
\begin{eqnarray}
\phi_{gh}(r)=4\epsilon_{gh}\left[\left(\frac{\sigma_{gh}}{r}\right)^{12}-\left(\frac{\sigma_{gh}}{r}\right)^6\right] \quad (g,h=\mathrm{c},\mathrm{v}). \label{LJ}
\end{eqnarray}
The Leonard-Jones potential has parameters $\epsilon$ and $\sigma$. We assumed that these parameters have different values for each combination of citizen and visitor.

\subsubsection*{Average kinetic energy and inverse temperature $\beta$}

According to the physical picture, we need to set $\beta$ as the value for which the root-mean-square speed $\sqrt{\langle v^2 \rangle}$ is an appropriate value for the velocity of movement of a person. In addition, it should be satisfied that the system is in a bound state under the value of $\beta$.

First, we considered the root-mean-square speed $\sqrt{\langle v^2 \rangle}$. The inverse temperature $\beta$ represents the average kinetic energy as in the following equation: 
\begin{eqnarray}
\frac{1}{2m_\mathrm{p}}\langle \bm{p}^2 \rangle=\frac{1}{2}m_\mathrm{p}\langle v^2\rangle =\frac{1}{\beta}. \label{relation}
\end{eqnarray}
Equation (\ref{relation}) holds when both members are calculated using the same unit system. When $\beta$ is given under the unit system described in this section, we must convert the kinetic energy on the left member into a value in this unit system. We obtained a unit of energy derived from the fundamental units as follows:
\begin{eqnarray}
\frac{\hbar^2}{m_\mathrm{p}l_\mathrm{m}^2}=1[\mathrm{E}],
\end{eqnarray}
where E represents the unit of energy in the unit system. Using this quantity of one unit of energy, eq. (\ref{relation}) can be rewritten as follows:
\begin{eqnarray}
 \frac{1}{2}m_\mathrm{p}\langle v^2\rangle \cdot \frac{m_\mathrm{p}l_\mathrm{m}^2}{\hbar^2}=\frac{1}{\beta},
\end{eqnarray}
where all quantities on the left member represent quantities under the SI unit system. From the above equation, we obtained $\sqrt{\langle v^2 \rangle}$ as following equation:
\begin{eqnarray}
 \sqrt{\langle v^2 \rangle}=\sqrt{\frac{2}{\beta}}\frac{\hbar}{m_\mathrm{p}l_\mathrm{m}} \label{v}
\end{eqnarray}
Since this value represents the scale of the velocity of a person's movement, we need to check that $\sqrt{\langle v^2 \rangle}$ is the appropriate size for a person's movement.

Next, we considered the relationship between the depth of the potential and the average kinetic energy. The bound state of the system means that the interaction energy between people in close proximity is larger than the average kinetic energy. Therefore, it is necessary to confirm $\epsilon>1/\beta$.

\subsection{Formulation of chemical potentials using spatial coordinates}

Communities are thermodynamically stable when in equilibrium. When $n$ communities exist in the network, the equilibrium condition is that the chemical potential $\mu$ of each community is equal. The chemical potential indicates the energy required for one person to transfer between communities. The following equation shows the equilibrium conditions:
\begin{eqnarray}
\mu_1=\mu_2=\cdots=\mu_n. \label{peq}
\end{eqnarray}
Thus, if the chemical potential for each community was formulated, the equilibrium conditions among the communities could be examined. The detail of this derivation is described in Appendix A. The total energy $\mathcal{H}$ of a community in this system is expressed as follows:  
\begin{eqnarray}
\mathcal{H}=(\mathcal{H}_\mathrm{c}^{\mathrm{id}}+\mathcal{H}_{\mathrm{cc}}^{\mathrm{int}})+(\mathcal{H}_\mathrm{v}^\mathrm{id}+\mathcal{H}_{\mathrm{vv}}^{\mathrm{int}})+(\mathcal{H}_{\mathrm{cv}}^{\mathrm{int}}). \label{H}
\end{eqnarray}
$\mathcal{H}_\mathrm{c}^{\mathrm{id}}$ and $\mathcal{H}_\mathrm{v}^\mathrm{id}$ represent the total kinetic energy of citizens and visitors, respectively. And, $\mathcal{H}_{\mathrm{cc }}^{\mathrm{int}}$, $\mathcal{H}_{\mathrm{vv}}^{\mathrm{int}}$, and $\mathcal{H}_{\mathrm{cv}}^{\mathrm{int}}$ represent the total interaction energy between citizen--citizen, visitor--visitor, and citizen--visitor, respectively. These quantities are represented using eqs. (\ref{h_id}) and (\ref{h_int}) as follows: 
\begin{eqnarray}
\mathcal{H}^{\mathrm{id}}_g&=&\frac{1}{2m_\mathrm{p}}\sum_{i=1}^{N_g} \bm{p}_i^2 \quad (g=\mathrm{c},\mathrm{v}) \\
\mathcal{H}_{gh}^\mathrm{int}&=&\sum_{i,j}\phi_{gh}(|\bm{r}_i-\bm{r}_j|) \quad (g,h=\mathrm{c},\mathrm{v}),
\end{eqnarray}
where $N_\mathrm{c}$ and $N_\mathrm{v}$ denote the number of citizens and visitors, respectively.

A canonical ensemble was used to formulate the chemical potential. We consider a situation in which the temperature is given by the inverse temperature $\beta$ and is maintained constant. The temperature indicates the average activity level of the movement of people. The total partition function $Z^{\mathrm{tot}}$ is given by,
\begin{eqnarray}
Z^{\mathrm{tot}}=(Z_{\mathrm{c}}^{\mathrm{id}} \times Z_{\mathrm{cc}}^{\mathrm{int}})\cdot(Z_{\mathrm{v}}^{\mathrm{id}} \times Z_{\mathrm{vv}}^{\mathrm{int}})\cdot(Z_{\mathrm{cv}}^{\mathrm{int}}).
\end{eqnarray}

First, the partition function for an ideal gas, $Z^{\mathrm{id}}$ is calculated. It is calculated as follows, considering the momenta of citizens and visitors are independent:
\begin{eqnarray}
Z^{\mathrm{id}}=\frac{1}{N_\mathrm{c}! N_\mathrm{v}!} \left( \frac{m_\mathrm{p}}{2 \pi \hbar^2 \beta} \right)^{N_\mathrm{c}+N_\mathrm{v}}
\end{eqnarray}
Since we defined $m_\mathrm{p}$ and $\hbar$ as the fundamental units, let $m_\mathrm{p}=1,\hbar=1$ and we obtain the following equation:
\begin{eqnarray}
Z^{\mathrm{id}}=\frac{1}{N_\mathrm{c}! N_\mathrm{v}!} \left( \frac{1}{2 \pi \beta} \right)^{N_\mathrm{c}+N_\mathrm{v}} \label{id}.
\end{eqnarray}

Next, the partition function for the interaction $Z^{\mathrm{int}}$ was calculated. The interaction potentials $\mathcal{H}^{\mathrm{int}}$ is derived from $\mathcal{H}_{\mathrm{cc}}^{\mathrm{int}}$, $\mathcal{H}_{\mathrm{vv}}^{\mathrm{int}}$ and $\mathcal{H}_{\mathrm{cv}}^{\mathrm{int}}$. Following eq. (\ref{LJ}), we assume that $\phi(r)$ is the Lennard-Jones potential. $Z^\mathrm{int}$ is calculated as the following equation: 
\begin{eqnarray}
Z^{\mathrm{int}}=S_\mathrm{c}^{N_\mathrm{c}}S_\mathrm{v}^{N_\mathrm{v}}\exp(-N_\mathrm{c}^2B_{\mathrm{cc}}-N_\mathrm{v}^2B_{\mathrm{vv}}-N_\mathrm{c}N_\mathrm{v}B_{\mathrm{cv}}). \label{int}
\end{eqnarray}
Here, $S_\mathrm{c}$ and $S_\mathrm{v}$ represent areas where citizens and visitors stay at least once, respectively. $B_{\mathrm{cc}},B_{\mathrm{cv}}$, and $B_{\mathrm{cv}}$ can be expressed as follows:
\begin{eqnarray}
B_{\mathrm{cc}}&=&\frac{1-\bm{q}_\mathrm{c}\bm{\Phi}_\mathrm{cc}\bm{q}_\mathrm{c}^{T}}{2S_\mathrm{c}^2} \\
B_{\mathrm{vv}}&=&\frac{1-\bm{q}_\mathrm{v}\bm{\Phi}_\mathrm{vv}\bm{q}_\mathrm{v}^{T}}{2S_\mathrm{v}^2} \\
B_{\mathrm{cv}}&=&\frac{1-\bm{q}_\mathrm{c}\bm{\Phi}_\mathrm{cv}\bm{q}_\mathrm{v}^{T}}{S_\mathrm{c}S_\mathrm{v}}.
\end{eqnarray}
The vector $\bm{q}_\mathrm{c},\bm{q}_\mathrm{v}$ in the above equation are,
\begin{eqnarray}
\bm{q_\mathrm{c}}&=&\left(q_{\mathrm{c}1}, q_{\mathrm{c}2}, \cdots, q_{\mathrm{c}M_\mathrm{c}}\right) \label{q_c}\\
\bm{q_\mathrm{v}}&=&\left(q_{\mathrm{v}1}, q_{\mathrm{v}2}, \cdots, q_{\mathrm{v}M_\mathrm{v}}\right) \label{q_v}.
\end{eqnarray}
These represent the fractions of time spent in each mesh by citizens and visitors, respectively. $M_c$ and $M_v$ are the numbers of meshes where citizens and visitors stay at least once, respectively. They were calculated from the mobility data. Matrices $\bm{\Phi}_\mathrm{cc},\bm{\Phi}_\mathrm{vv}$, and $\bm{\Phi}_\mathrm{cv}$ are as follows: 
\begin{eqnarray}
\bm{\Phi}_{gh}=\left[\exp\left\{-\beta\phi_{gh}\left(\sqrt{(x_k-x_l)^2+(y_k-y_l)^2}\right)\right\}\right] \\ \nonumber
(g,h=\mathrm{c},\mathrm{v}).
\end{eqnarray}
The $k,l$ components of $\bm{\Phi}$ are the Boltzmann factors of the interactions of the distance between meshes $k$ and $l$. Here $(x_k,y_k)$ and $(x_l,y_l)$ denote the positions of meshes $k$ and $l$.

Because the Helmholtz free energy $F=-1/\beta\ln Z$, $F$ is obtained from eqs. (\ref{id}) and (\ref{int}) as,
\begin{eqnarray}
F&=&\frac{N_\mathrm{c}}{\beta}\left[\ln\left(\frac{2\pi\beta N_\mathrm{c}}{S_\mathrm{c}}\right)-1\right]+\frac{N_\mathrm{v}}{\beta}\left[\ln\left(\frac{2\pi\beta N_\mathrm{v}}{S_\mathrm{v}}\right)-1\right] \nonumber \\
&&+N_\mathrm{c}^2B_\mathrm{cc}+N_\mathrm{v}^2B_\mathrm{vv}+N_\mathrm{c}N_\mathrm{v}B_\mathrm{cv}.
\end{eqnarray}
Finally, the chemical potential $\mu$ is formulated from $F$ as follows:
\begin{eqnarray}
\mu&=&\frac{1}{\beta(N_\mathrm{c}+N_\mathrm{v})}\Biggl[N_\mathrm{c}\ln\left(\frac{2\pi\beta N_\mathrm{c}}{S_\mathrm{c}}\right)+N_\mathrm{v}\ln\left(\frac{2\pi\beta N_\mathrm{v}}{S_\mathrm{v}}\right) \nonumber \\
&&+2N_\mathrm{c}^2B_\mathrm{cc}+2N_\mathrm{v}^2B_\mathrm{vv}+2N_\mathrm{c}N_\mathrm{v}B_\mathrm{cv}\Biggr] \label{mu}.
\end{eqnarray}
We use this equation to examine the thermodynamic stability of the community structure of the face-to-face interaction network. In the next section, we describe a formulation of the chemical potential using the adjacency matrix to demonstrate the applicability of the theory to general networks without spatial coordinates. 

\subsection{Formulation of chemical potentials using adjacency matrix without spatial coordinates}

We use the adjacency matrix to calculate the partition function $Z^\mathrm{int}$ to formulate chemical potentials without spatial coordinates. The detail of the derivation is described in Appendix B. Let the interaction potential $\mathcal{H}_{ij}^\mathrm{int}$ between nodes $i$ and $j$ be as,
\begin{eqnarray}
 \mathcal{H}_{ij}^\mathrm{int} = \begin{cases}
  \epsilon_\mathrm{cc} & (\text{Citizen--citizen interaction}) \\
  \epsilon_\mathrm{vv} & (\text{Visitor--visitor interaction}) \\
  \epsilon_\mathrm{cv} & (\text{Citizen--visitor interaction}) \\
  0 & (\text{No interaction}) \\
 \end{cases}.
\end{eqnarray}
We consider that an interaction occurs when there is an edge exists between $i$ and $j$. Partition function of the interaction $Z^\mathrm{int}$ is formulated using the adjacency matrix as,
\begin{eqnarray}
Z^\mathrm{int}&=&S_\mathrm{c}^{N_\mathrm{c}}S_\mathrm{v}^{N_\mathrm{v}}\left[1+\frac{\exp(-\beta\epsilon_\mathrm{cc})-1}{2S_\mathrm{c}^{2}}\bm{I}_\mathrm{c}\bm{A}_\mathrm{cc}\bm{I}_\mathrm{c}^T \right. \nonumber \\
&-&\left. \frac{\exp(-\beta\epsilon_\mathrm{vv})-1}{2S_\mathrm{v}^{2}}\bm{I}_\mathrm{v}\bm{A}_\mathrm{vv}\bm{I}_\mathrm{v}^T-\frac{\exp(-\beta\epsilon_\mathrm{cv})-1}{S_\mathrm{c}S_\mathrm{v}}\bm{I}_\mathrm{c}\bm{A}_\mathrm{cv}\bm{I}_\mathrm{v}^T \right].\nonumber \\
\end{eqnarray}
$\bm{A}_\mathrm{cc},\bm{A}_\mathrm{vv}$, and $\bm{A}_\mathrm{cv}$ represent citizen--citizen, visitor--visitor, and citizen--visitor adjacency matrices, respectively. $\bm{I}_\mathrm{c}$ and $\bm{I}_\mathrm{v}$ are column vectors with all components are one in the $N_\mathrm{c}$ and $N_\mathrm{v}$ columns. We consider $\ln S_\mathrm{c} \propto \ln N_\mathrm{c}$ and  $\ln S_\mathrm{v} \propto \ln N_\mathrm{v}$ and replace $S_\mathrm{c}$ and $S_\mathrm{v}$ with $N_\mathrm{c}$ and $N_\mathrm{v}$, respectively:
\begin{eqnarray}
S_\mathrm{c}=b_\mathrm{c}N_\mathrm{c}^{a_\mathrm{c}}, \ S_\mathrm{v}=b_\mathrm{v}N_\mathrm{v}^{a_\mathrm{v}}. \label{prop}
\end{eqnarray}
Using eq. (\ref{prop}), the partition function is represented without spatial coordinates as follows:
\begin{eqnarray}
Z^\mathrm{int}&=&(b_\mathrm{c}N_\mathrm{c}^{a_\mathrm{c}})^{N_\mathrm{c}} (b_\mathrm{v}N_\mathrm{v}^{a_\mathrm{v}})^{N_\mathrm{v}} \left[1+\frac{\exp(-\beta\epsilon_\mathrm{cc})-1}{2(b_\mathrm{c}N_\mathrm{c}^{a_\mathrm{c}})^{2}}\bm{I}_\mathrm{c}\bm{A}_\mathrm{cc}\bm{I}_\mathrm{c}^T \right. \nonumber \\
&+& \frac{\exp(-\beta\epsilon_\mathrm{vv})-1}{2(b_\mathrm{v}N_\mathrm{v}^{a_\mathrm{v}})^{2}}\bm{I}_\mathrm{v}\bm{A}_\mathrm{vv}\bm{I}_\mathrm{v}^T \nonumber \\
&+&\left.\frac{\exp(-\beta\epsilon_\mathrm{cv})-1}{(b_\mathrm{c}N_\mathrm{c}^{a_\mathrm{c}})(b_\mathrm{v}N_\mathrm{v}^{a_\mathrm{v}})}\bm{I}_\mathrm{c}\bm{A}_\mathrm{cv}\bm{I}_\mathrm{v}^T \right].
\end{eqnarray}

Since the Helmholtz free energy $F=-1/\beta\ln Z$, $F$ is obtained as follows:
\begin{eqnarray}
F&=&\frac{N_\mathrm{c}}{\beta}\left[\ln\left(\frac{2\pi\beta}{b_\mathrm{c} N_\mathrm{c}^{a_\mathrm{c}-1}}\right)-1\right] +\frac{N_\mathrm{v}}{\beta}\left[\ln\left(\frac{2\pi\beta}{b_\mathrm{v} N_\mathrm{v}^{a_\mathrm{v}-1}}\right)-1\right] \nonumber \\
 &+&\frac{1}{\beta}\left(B_\mathrm{cc}\bm{I}_\mathrm{c}\bm{A}_\mathrm{cc}\bm{I}_\mathrm{c}^T+B_\mathrm{vv}\bm{I}_\mathrm{v}\bm{A}_\mathrm{vv}\bm{I}_\mathrm{v}^T+B_\mathrm{cv}\bm{I}_\mathrm{c}\bm{A}_\mathrm{cv}\bm{I}_\mathrm{v}^T \right).
\end{eqnarray}
Here, $B_\mathrm{cc},B_\mathrm{vv}$, and $B_\mathrm{cv}$ are as follows:
\begin{eqnarray}
B_\mathrm{cc}=\frac{1-\exp(-\beta\epsilon_\mathrm{cc})}{2b_\mathrm{c}^2N_\mathrm{c}^{2a_\mathrm{c}}} \label{B_cc} \\
B_\mathrm{vv}=\frac{1-\exp(-\beta\epsilon_\mathrm{vv})}{2b_\mathrm{v}^2N_\mathrm{v}^{2a_\mathrm{v}}} \\
B_\mathrm{cv}=\frac{1-\exp(-\beta\epsilon_\mathrm{cv})}{b_\mathrm{c}b_\mathrm{v}N_\mathrm{c}^{a_\mathrm{c}}N_\mathrm{v}^{a_\mathrm{v}}} \label{B_cv}.
\end{eqnarray}
Finally, the chemical potential $\mu$ is derived as the following equation:
\begin{eqnarray}
 \mu&=&\frac{1}{\beta(N_\mathrm{c}+N_\mathrm{v})}\left\{N_\mathrm{c}\left[\ln \left( \frac{2 \pi \beta}{b_\mathrm{c} N_\mathrm{c}^{a_\mathrm{c}-1}} \right)-a_\mathrm{c}\right]\right. \nonumber \\ 
 &+&\left.N_\mathrm{v}\left[\ln \left( \frac{2 \pi \beta}{b_\mathrm{v} N_\mathrm{v}^{a_\mathrm{v}-1}} \right) -a_\mathrm{v} \right] \right\}-\left[2a_\mathrm{c}B_\mathrm{cc}\bm{I}_\mathrm{c}\bm{A}_\mathrm{cc}\bm{I}_\mathrm{c}^T\right. \nonumber \\
 &+&\left.2a_\mathrm{v}B_\mathrm{vv}\bm{I}_\mathrm{v}\bm{A}_\mathrm{vv}\bm{I}_\mathrm{v}^T+(a_\mathrm{c}+a_\mathrm{v})B_\mathrm{cv}\bm{I}_\mathrm{c}\bm{A}_\mathrm{cv}\bm{I}_\mathrm{v}^T \right],
\end{eqnarray}
which is associated with the formulation using spatial coordinates.

\subsection{Parameter estimation}

The model parameters must be estimated to calculate the chemical potential using eq. (\ref{mu}), which is a formulation that uses spatial coordinates. We solved the optimization problem to obtain the values of the model parameter that satisfy eq. (\ref{peq}), which is the condition for equilibrium among $n$ communities. The parameters included in $\mu$ are $\beta,\epsilon_\mathrm{cc},\epsilon_\mathrm{vv},\epsilon_\mathrm{cv},\sigma_\mathrm{cc},\sigma_\mathrm{vv}$, and $\sigma _\mathrm{cv}$. 

Considering the average kinetic energy and $\beta$ in Sect 3.1, we set $\beta$ as the value at which the root-mean-square speed takes the appropriate value as a scale of velocity for a person, and the system is in a bound state (see Sect. 4.3). In addition, we assume $\epsilon_\mathrm{cv}$ and $\sigma_\mathrm{cv}$ are determined according to the Lorentz--Berthelot combining rules:
\begin{eqnarray}
\epsilon_\mathrm{cv}=\sqrt{\epsilon_\mathrm{cc}\epsilon_\mathrm{vv}}, \ \sigma_\mathrm{cv}=\frac{\sigma_\mathrm{cc}+\sigma_\mathrm{vv}}{2}. \label{sigma_cv}
\end{eqnarray}
Thus, we must estimate $\epsilon_\mathrm{cc},\epsilon_\mathrm{vv},\sigma_\mathrm{cc}$, and $\sigma_\mathrm{vv}$.

The loss function $\mathcal{L}$ is defined as follows:
\begin{eqnarray}
\mathcal{L}(\epsilon_\mathrm{cc},\epsilon_\mathrm{vv},\sigma_\mathrm{cc},\sigma_\mathrm{vv}):=\frac{2}{n(n-1)}\sum_{i<j}(\mu_i-\mu_j)^2. \label{loss}
\end{eqnarray}
The model parameters were obtained by solving the minimization problem of $\mathcal{L}$. $\mu$ in eq. (\ref{loss}) was calculated using eq. (\ref{mu}) derived from the chemical potential formulation using spatial coordinates. Since $\mu$ is composed of the chemical potential due to the ideal gas $\mu^\mathrm{id}$ and the chemical potential due to the interaction $\mu^\mathrm{int}$, we can represent it as follows:
\begin{eqnarray}
\mu&=&\mu^\mathrm{id}+\mu^\mathrm{int}\\
\mu^\mathrm{id}&=&\frac{1}{\beta(N_\mathrm{c}+N_\mathrm{v})}\left[N_\mathrm{c}\ln\left(\frac{2\pi\beta N_\mathrm{c}}{S_\mathrm{c}}\right)+N_\mathrm{v}\ln\left(\frac{2\pi\beta N_\mathrm{v}}{S_\mathrm{v}}\right) \right] \nonumber \\
\label{mu_id} \\
\mu^\mathrm{int}&=&\frac{1}{\beta(N_\mathrm{c}+N_\mathrm{v})}\left(2N_\mathrm{c}^2B_\mathrm{cc}+2N_\mathrm{v}^2B_\mathrm{vv}+2N_\mathrm{c}N_\mathrm{v}B_\mathrm{cv}\right) \label{mu_int1} \nonumber \\ 
&=&\mu_\mathrm{c}^\mathrm{int}+\mu_\mathrm{v}^\mathrm{int}+\mu_\mathrm{cv}^\mathrm{int}. \label{mu_int2} 
\end{eqnarray}
Here, $\mu_\mathrm{c}^\mathrm{int}$,$\mu_\mathrm{v}^\mathrm{int}$, and $\mu_\mathrm{cv}^\mathrm{int}$ represent the chemical potentials of the citizen--citizen, visitor--visitor and citizen--visitor interactions, respectively. For the calculation, we considered the distance between adjacent meshes as $l_\mathrm{m}=100$ m. However, the actual mesh is based on latitude and longitude, and the size of the length and width were not exactly 100 m.

\section{Result}

The features of the constructed networks and the results of the community analysis were shown to identify persistent communities from real data analysis. The model parameters were then optimized and used to calculate the chemical potentials to satisfy the equilibrium condition among the persistent communities.

\subsection{Network features of constructed networks}

\begin{figure}[h]
 (a) \\
 \vspace{-5mm}
 \begin{center}
 \includegraphics[width=70mm]{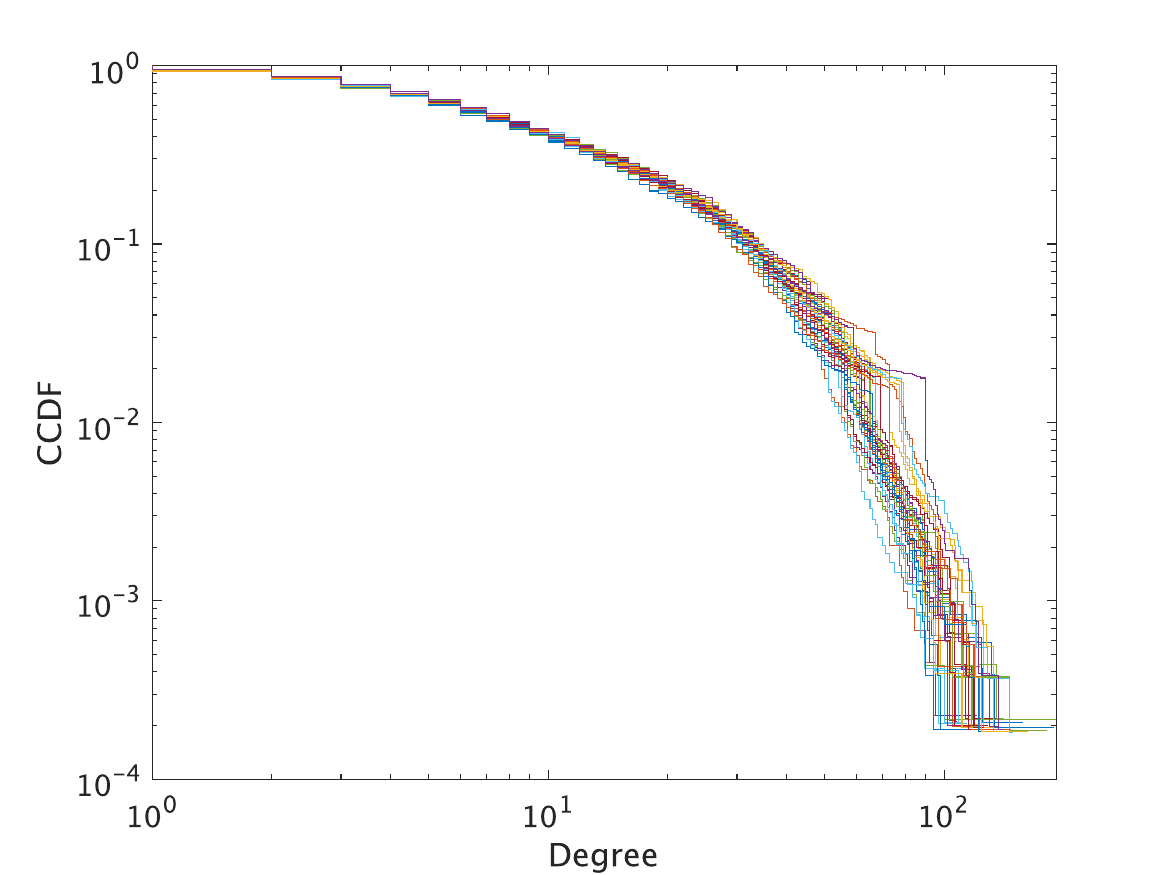}
 \end{center}
 (b) \\
 \vspace{-5mm}
 \begin{center}
 \includegraphics[width=70mm]{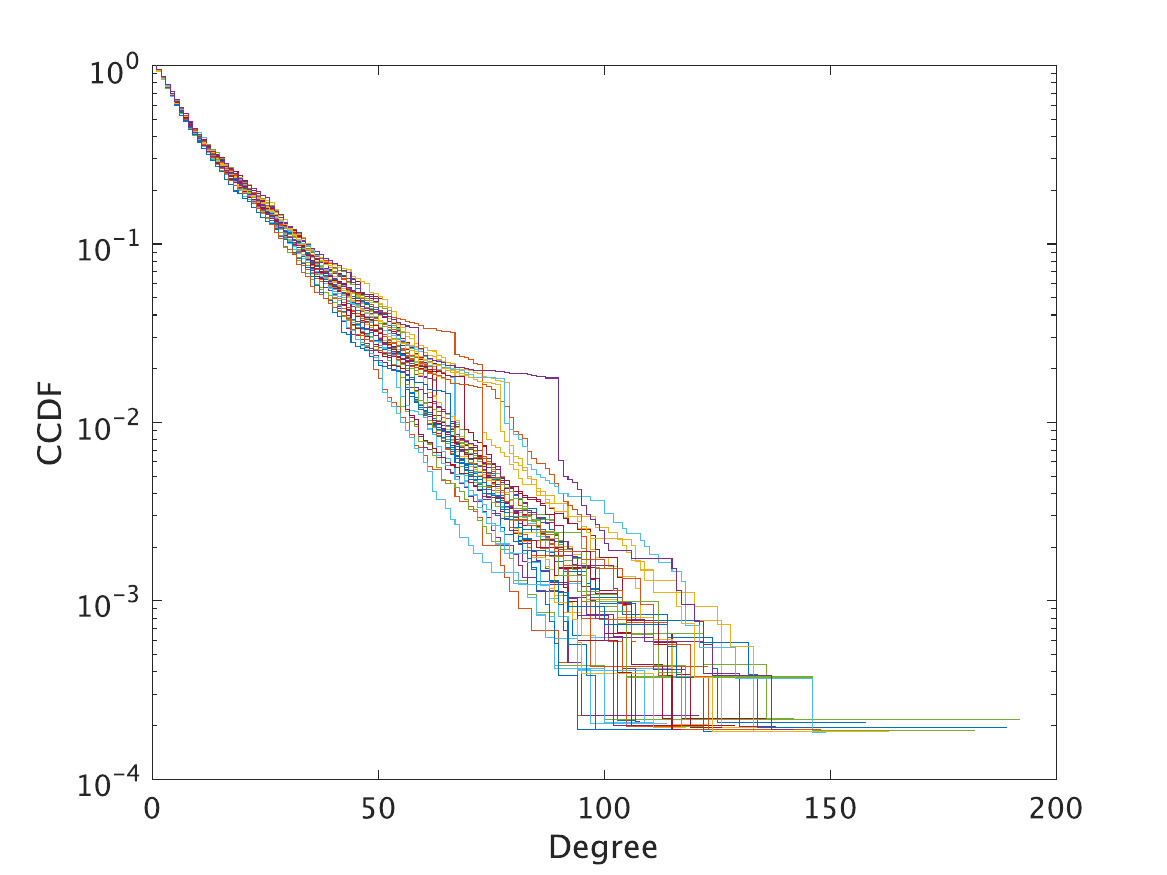}
\caption{\label{figure6}Empirical complementary cumulative distribution functions of degrees in which each color represents the distribution for each day: (a) double logarithmic plot, and (b) single logarithmic plot.}
 \end{center}
\end{figure}

\begin{figure}[h]
\begin{center}
\includegraphics[width=7cm]{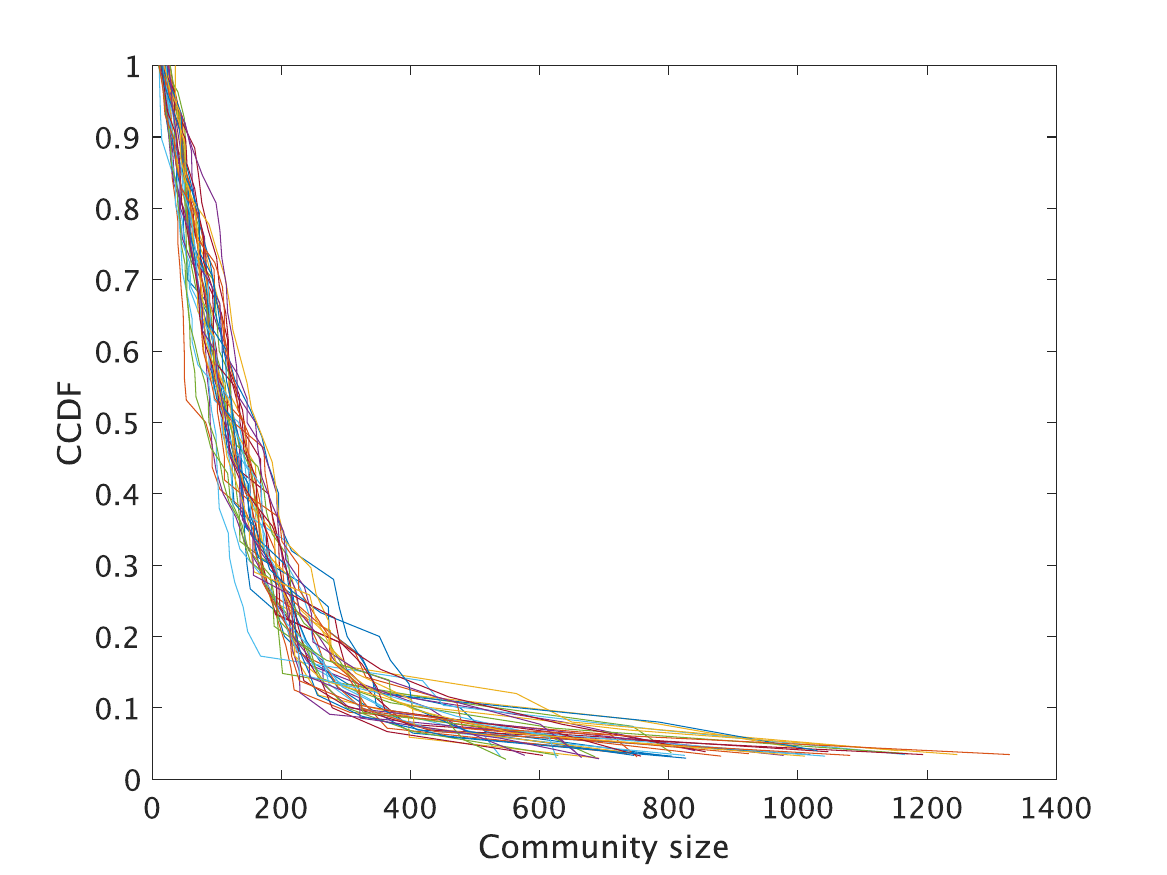}
\caption{\label{figure7}Empirical complementary cumulative distribution functions of community sizes. Each color represents the distribution for each day.}
\end{center}
\end{figure}

Daily face-to-face interaction networks were constructed using mobility data, and the maximum connected components of the networks were analyzed. The network features of each daily network were then calculated, and the communities were detected. The average network features over the data period and the number of communities are listed in Table \ref{table3}. Multi-level Infomap was used for community analysis, and the first level of communities was used for further analysis. The coefficients of variation were less than 10\%. Complementary empirical cumulative distribution functions of degrees and community sizes are shown in Figs. \ref{figure6} and \ref{figure7}, respectively. Based on these results, the network structures of the constructed networks were confirmed to be similar over the data period.

\begin{table}[h]
\caption{\label{table3}Average values of network features and number of communities of daily face-to-face interaction networks.}
\begin{tabular}{@{}llll}
\hline
& Number & Fraction & Number\\
& of nodes & of citizens & of edges \\
\hline
Mean & 4974 & 0.641 & 32850 \\
SD & 298 & 0.010 & 2387 \\
CV & 0.060 & 0.015 & 0.073 \\ 
\hline
& & Shortest & Clustering \\
& Degree & path length & coefficient \\
\hline
Mean & 13.2 & 6.33 & 0.779 \\
SD & 0.4 & 0.15 & 0.005 \\
CV & 0.033 & 0.024 & 0.006 \\ 
\hline
& & Number of & \\
& Assortativity & communities & \\
\hline
Mean & 0.667 & 29.7 &\\
SD & 0.044 & 2.9 &\\
CV & 0.067 & 0.099 &\\ 
\hline 
\end{tabular} \\
SD: standard deviation, CV: coefficient of variation.
\end{table} 

\begin{figure}[h]
\begin{center}
\includegraphics[width=7cm]{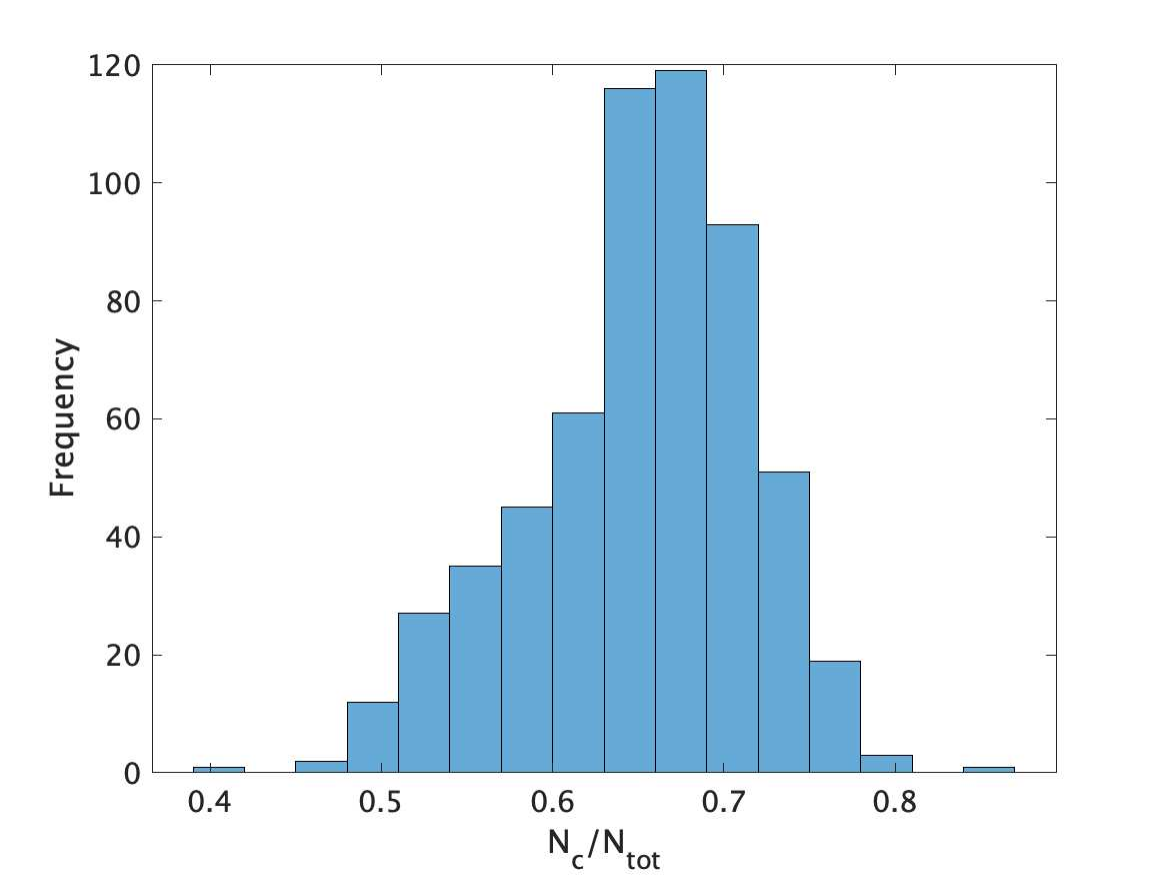}
\caption{\label{figure8}Histogram of the proportion of citizens $N_\mathrm{c}/N_\mathrm{tot}$ in the top 15 communities with the largest number of components per day.}
\end{center}
\end{figure}

Figure \ref{figure8} shows a histogram of the percentage of citizens $N_\mathrm{c}/N_{\mathrm{tot}}$ for the top 15 communities with the most components for each day, where $N_{\mathrm{tot}}=N_\mathrm{c}+N_\mathrm{v}$. The peak was in the range of 65--70\%, and the distribution was around the peak. The overall proportion of citizens in the network is consistent with the percentage of the peak. All communities contain both citizens and visitors, which implies that they coexist in each community.

\subsection{Identifying persistent communities from data analysis}

We used clustering analysis to detect persistent communitiess. The total number of communities for all days of the data period $n_\mathrm{all}$ was 1,157. We classified them into $k=40$ clusters using the $k$-means method. To determine the number of clusters, clustering analysis was performed for $k=30$, $40$, and $50$. We used the result of $k=40$, because the geospatial boundaries of each cluster were clear compared to the results of $k=30$ and $50$. 

The communities in the seven clusters satisfy the conditions for a persistent community. The communities classified in each of the seven clusters were present for 80\% or more of the days during the data period. The center of gravity of the spatial distribution of the interactions of each persistent community is shown in Fig. 9. From this result, we confirmed that each of the seven clusters was characterized by a specific location in Kyoto City. The total number of persistent communities on all days was 284, averaging of 7.3 communities/day. A total of 49.8\% of people belonged to persistent communities. These clusters include communities in the urban areas of Kyoto City (\textit{Shijo Karasuma} and \textit{Karasuma Oike}) and those around individual objects, such as large train stations (\textit{Kyoto Station} and \textit{Katsura Station}), city hall (\textit{Kyoto City office}), hospitals (\textit{Kyoto University Hospital}), and workplaces (\textit{ROHM Co., Ltd.}). The areas and objects where persistent communities were observed attracted large populations during weekday daytime.

The network features of persistent communities in each cluster are listed in Tables \ref{table4}--\ref{table6}. The formulations of the network features are described in Sect. 2.3. If more than one community belongs to a cluster on the same day, it is treated as a single network, and if it is disconnected, the average distance is calculated from the largest connected component of the network. The coefficient of variation, which is the standard deviation divided by the mean, was used to evaluate the variation in community features (Table \ref{table6}). The coefficients of variation for the number of nodes and the number of edges are relatively large, varying from 50 to 60\% at the maximum. However, the average degree, distance, and assortativity varied from 10 to 30\%, and the average clustering coefficient varied by less than 10\%. Although the variation is large compared to the overall network features shown in Table \ref{table3}, the coefficients of variation are all smaller than one and can be considered to vary within a certain range. Figure \ref{figure10} illustrates the transition of each network feature for the top two clusters with a large number of components. On days when no communities belong to a cluster, the previous day's values were used for imputation of the missing values. These figures indicate that there is a similarity in the network structure of communities that occur persistently in the same location.

\begin{figure}[h]
\begin{center}
\includegraphics[width=8cm]{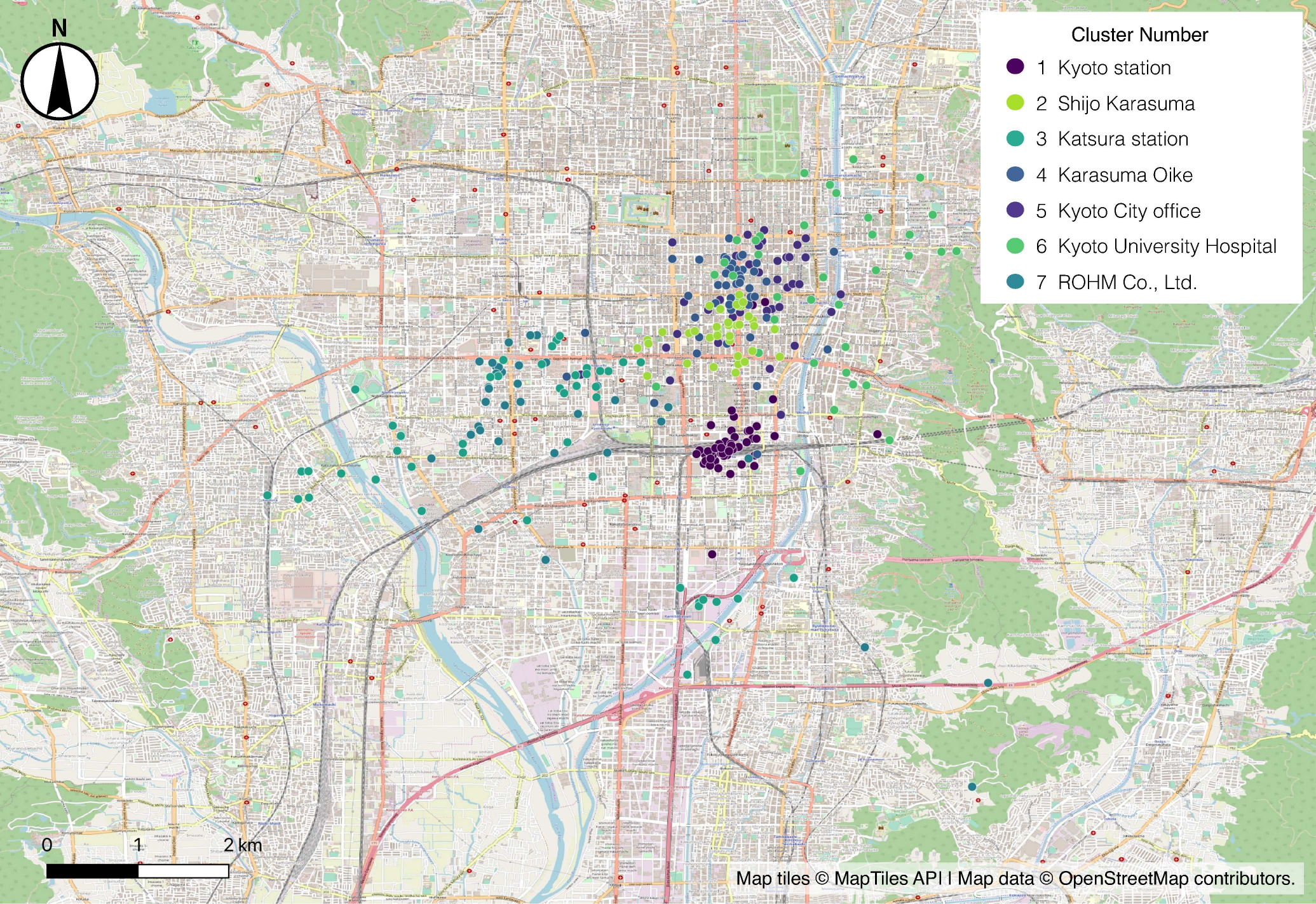}
\caption{\label{figure9}Diagram of the centers of gravity of communities belonging to persistent clusters.}
\end{center}
\end{figure}

\begin{table}[h]
\caption{\label{table4}Average values of network features of persistent communities belonging to seven clusters.}
\begin{tabular}{@{}llll}
\hline
 & Number & Number & \\
No. & of nodes & of edges & Degree \\
\hline
1 & 904 & 9732 & 21.8 \\
2 & 498 & 3840 & 15.3 \\
3 & 344 & 2770 & 16.3 \\
4 & 319 & 1911 & 12.0 \\
5 & 253 & 1269 & 10.2 \\
6 & 229 & 1065 & 9.3 \\
7 & 124 & 556 & 9.1 \\
\hline
 & Shortest & Clustering & \\
No. & path length & coefficient & Assortativity \\
\hline
1 & 4.18 & 0.823 & 0.578 \\
2 & 4.17 & 0.813 & 0.442 \\
3 & 4.14 & 0.831 & 0.654 \\
4 & 4.57 & 0.798 & 0.572 \\
5 & 4.68 & 0.787 & 0.507 \\
6 & 4.66 & 0.786 & 0.480 \\
7 & 4.24 & 0.801 & 0.535 \\
\hline
\end{tabular}
\end{table}

\begin{table}[h]
\caption{\label{table5}Standard deviations of network features of persistent communities belonging to seven clusters.}
\begin{tabular}{@{}llll}
\hline
 & Number & Number & \\
No. & of nodes & of edges & Degree \\
\hline
1 & 202 & 1958 & 2.64 \\
2 & 181 & 1543 & 2.20 \\
3 & 193 & 1628 & 3.77 \\
4 & 114 & 708 & 1.55 \\
5 & 68 & 294 & 1.46 \\
6 & 77 & 448 & 1.52 \\
7 & 59 & 278 & 1.65 \\
\hline
 & Shortest & Clustering & \\
No. & path length & coefficient & Assortativity \\
\hline
1 & 0.36 & 0.013 & 0.099 \\
2 & 0.49 & 0.023 & 0.070 \\
3 & 0.99 & 0.035 & 0.173 \\
4 & 0.65 & 0.023 & 0.076 \\
5 & 0.68 & 0.028 & 0.073 \\
6 & 0.70 & 0.022 & 0.100 \\
7 & 1.10 & 0.032 & 0.147 \\
\hline
\end{tabular}
\end{table}

\begin{table}[h]
\caption{\label{table6}Coefficients of variation in network features of persistent communities belonging to seven clusters.}
\begin{tabular}{@{}llll}
\hline
 & Number & Number & \\
No. & of nodes & of edges & Degree \\
\hline
1 & 0.223 & 0.201 & 0.121 \\
2 & 0.364 & 0.402 & 0.144 \\
3 & 0.559 & 0.588 & 0.232 \\
4 & 0.357 & 0.370 & 0.129 \\
5 & 0.269 & 0.232 & 0.143 \\
6 & 0.336 & 0.420 & 0.164 \\
7 & 0.472 & 0.500 & 0.183 \\
\hline
 & Shortest & Clustering & \\
No. & path length & coefficient & Assortativity \\
\hline
1 & 0.086 & 0.016 & 0.171 \\
2 & 0.117 & 0.028 & 0.158 \\
3 & 0.239 & 0.042 & 0.265 \\
4 & 0.142 & 0.029 & 0.133 \\
5 & 0.146 & 0.036 & 0.145 \\
6 & 0.150 & 0.029 & 0.207 \\
7 & 0.259 & 0.040 & 0.275 \\
\hline
\end{tabular}
\end{table}

\begin{figure*}[h]
\begin{center}
 \begin{minipage}{0.49\hsize}
 (a) 
 \vspace{-6mm}
 \begin{center}
  \includegraphics[width=70mm]{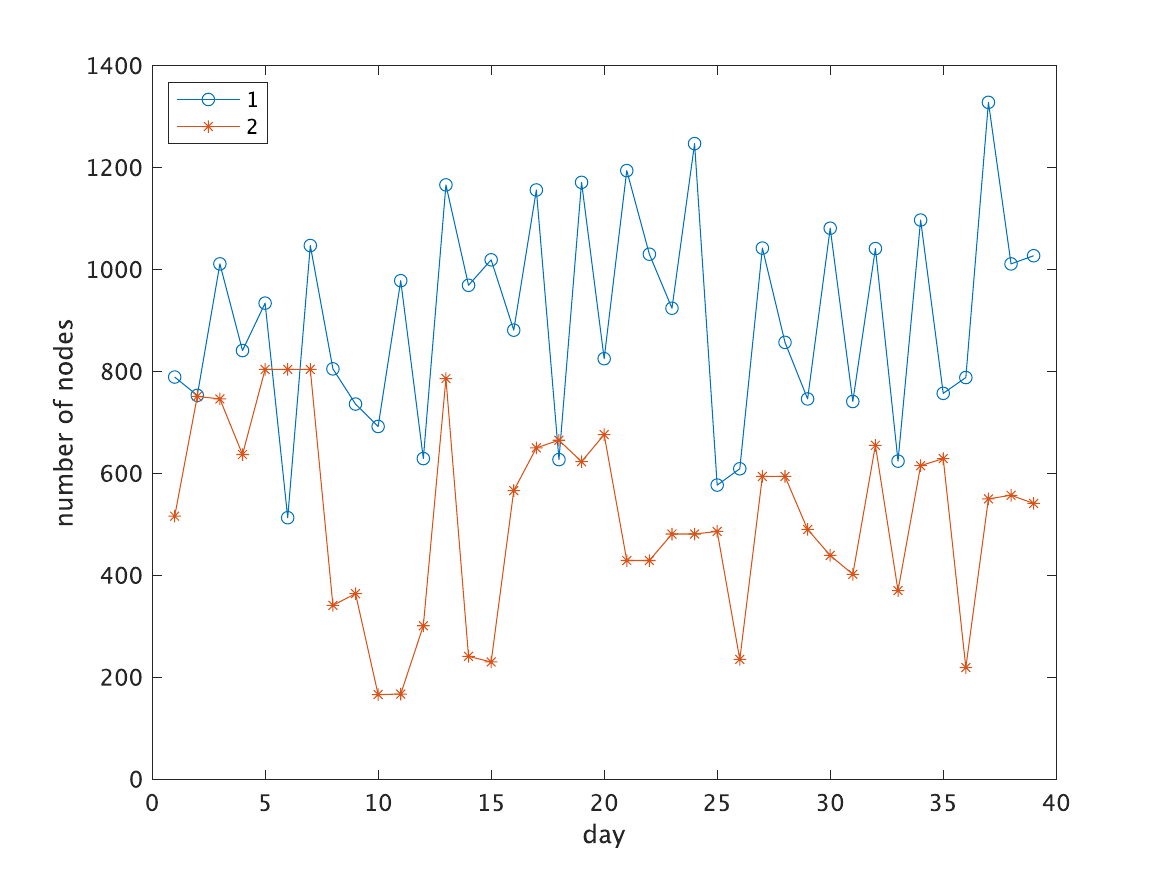}
 \end{center}
 \end{minipage}
 \begin{minipage}{0.49\hsize}
 (b) 
 \vspace{-6mm}
 \begin{center}
  \includegraphics[width=70mm]{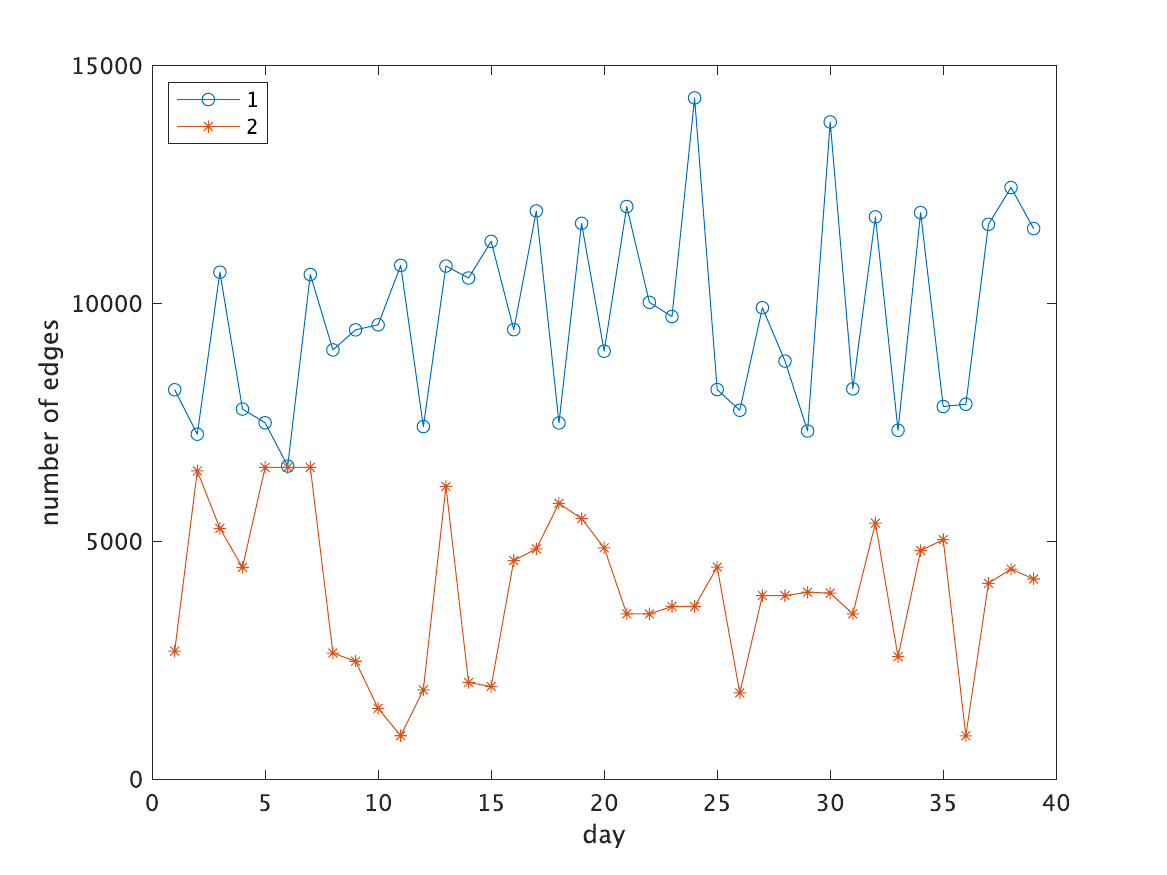}
 \end{center}
 \end{minipage} \\
  \begin{minipage}{0.49\hsize}
 (c) 
 \vspace{-6mm}
 \begin{center}
  \includegraphics[width=70mm]{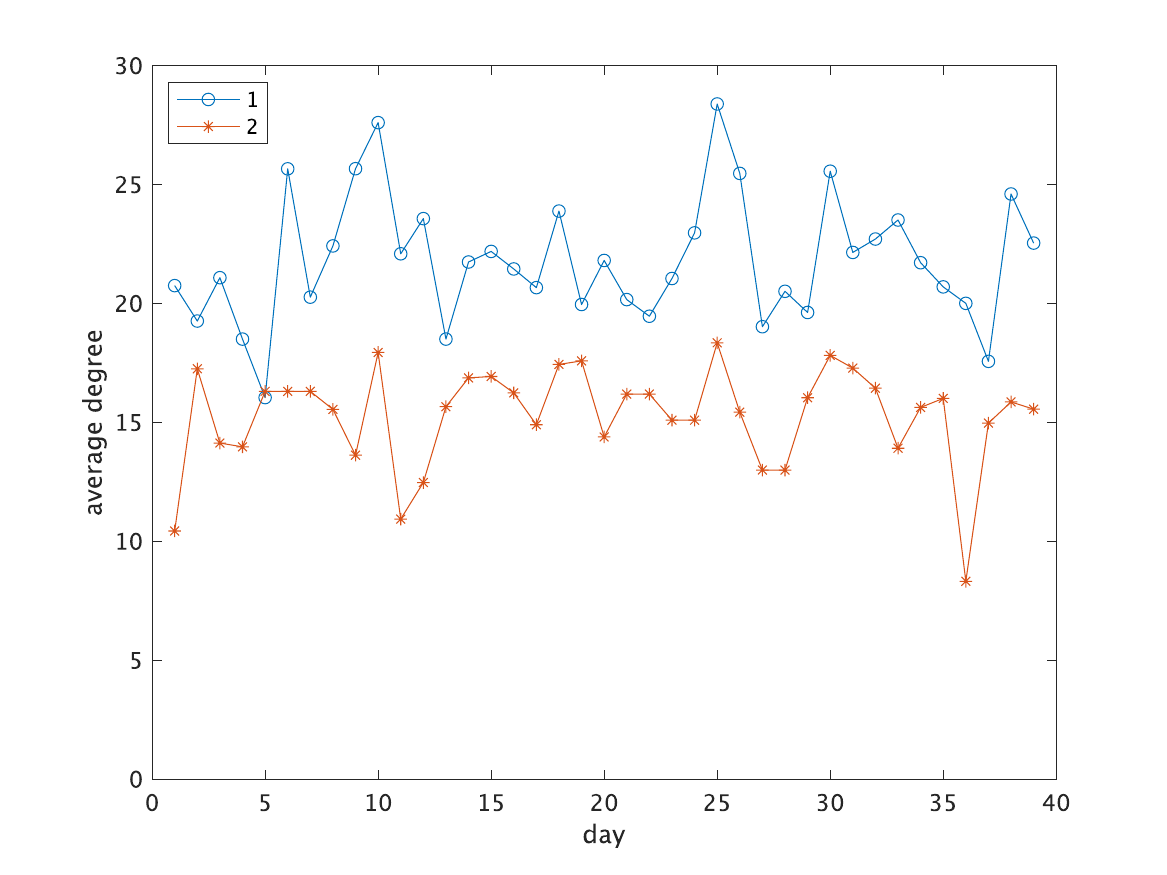}
 \end{center}
 \end{minipage}
 \begin{minipage}{0.49\hsize}
 (d) 
 \vspace{-6mm}
 \begin{center}
  \includegraphics[width=70mm]{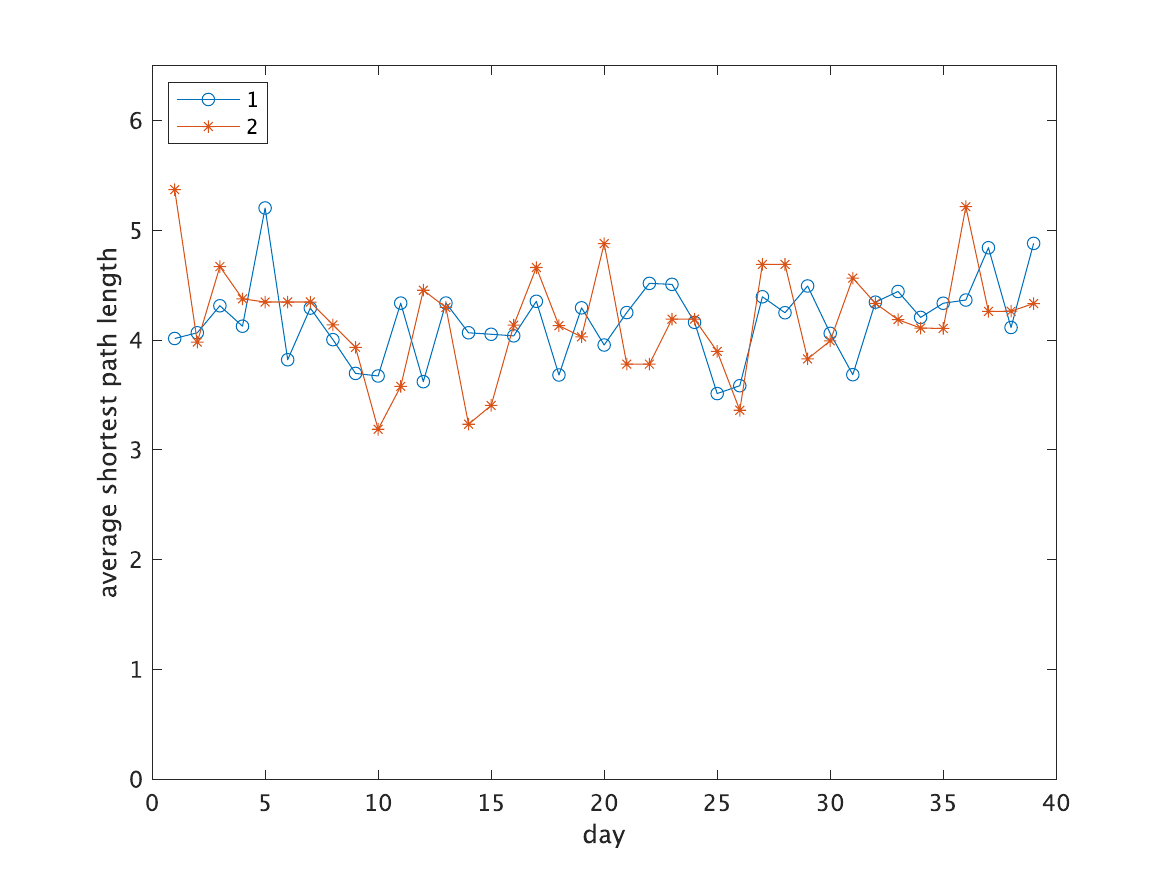}
 \end{center}
 \end{minipage} \\
  \begin{minipage}{0.49\hsize}
 (e) 
 \vspace{-6mm}
 \begin{center}
  \includegraphics[width=70mm]{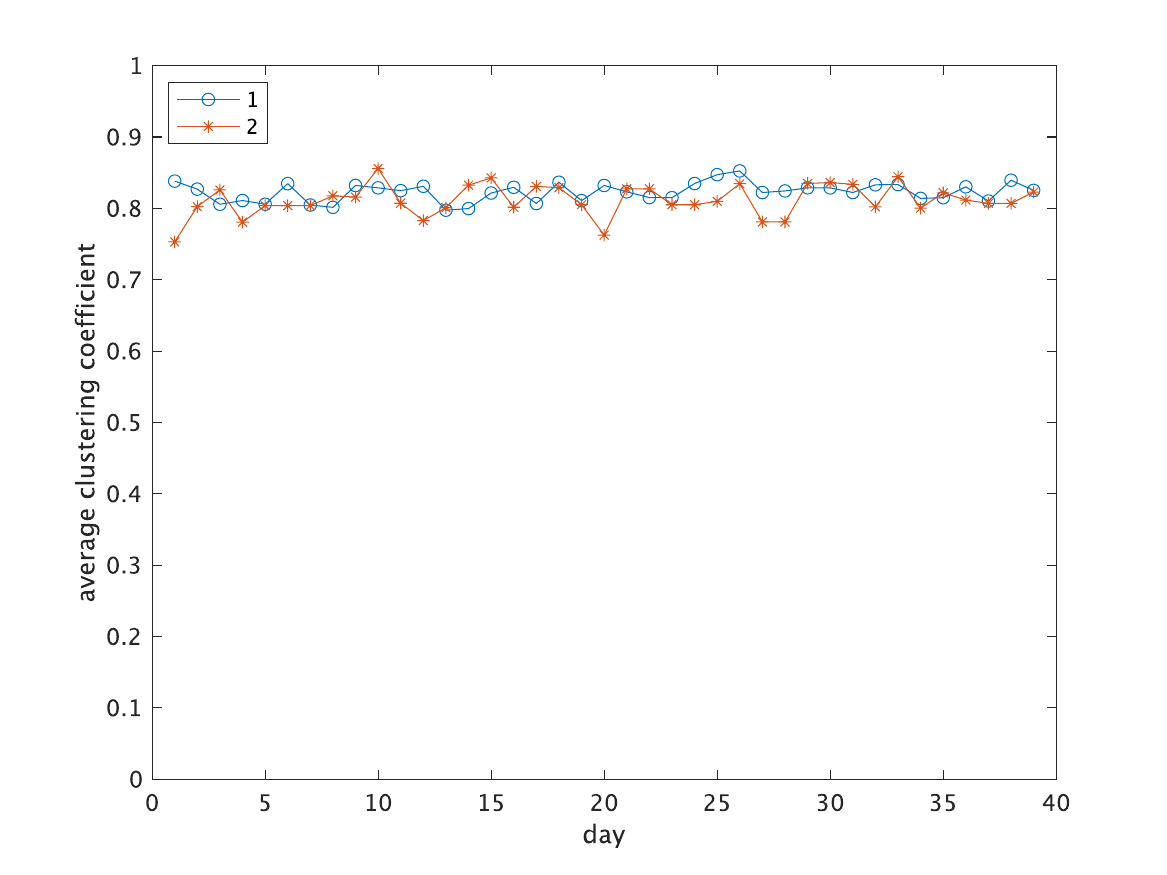}
 \end{center}
 \end{minipage}
 \begin{minipage}{0.49\hsize}
 (f) 
 \vspace{-6mm}
 \begin{center}
  \includegraphics[width=70mm]{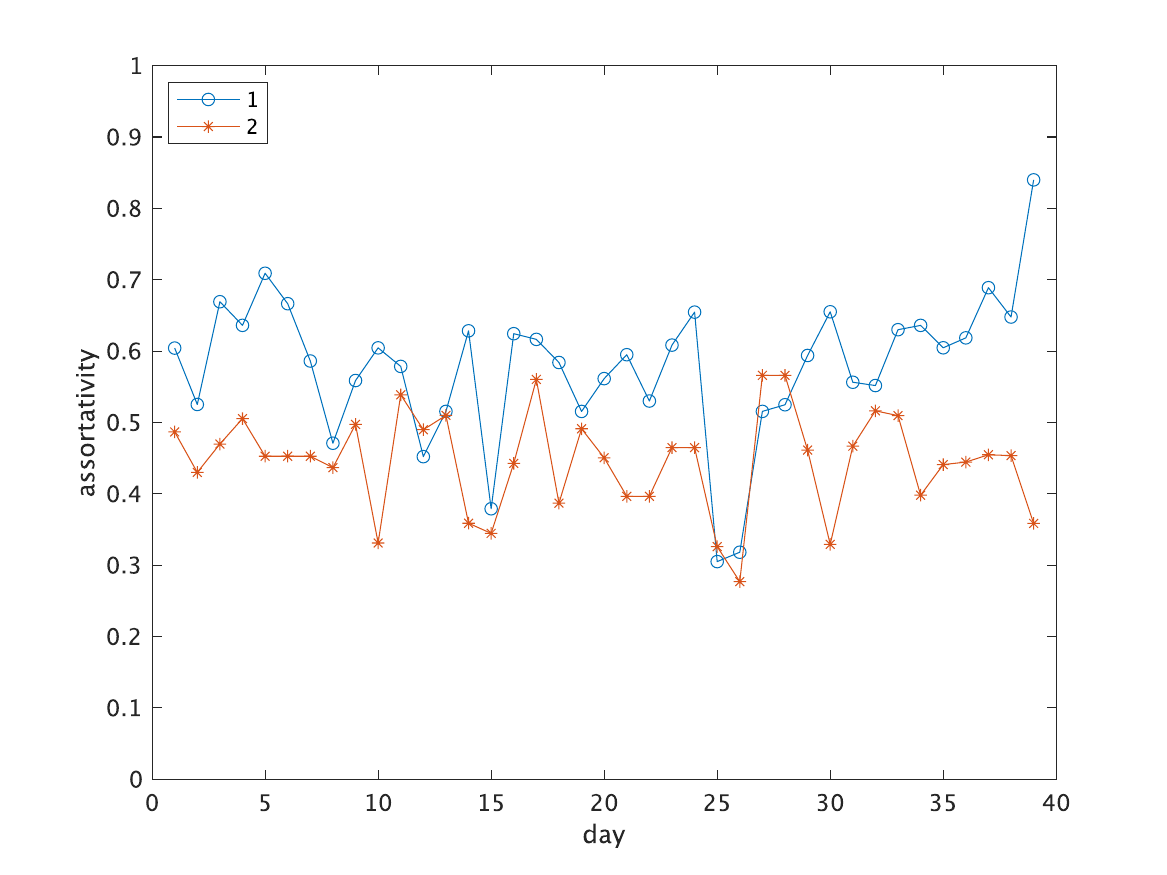}
 \end{center}
 \end{minipage} 
\caption{\label{figure10}Transition of network features for the networks of the top two clusters with the largest number of components among the clusters to which persistent communities belong. Missing values on days when the communities belonging to the cluster do not exist are complemented with the previous day's values. (a) The number of nodes. (b) The number of edges. (c) Average degree. (d) Average distance. (e) Average clustering coefficient. (f) Assortativity.}
\end{center}
\end{figure*}

\subsection{Estimation of chemical potential} 

We solved the optimization problem to estimate the chemical potential $\mu$ of the persistent communities for each day. We targeted the persistent communities in the top 15 with the largest number of components for each day to limit the community size to a certain level. The number of target communities was 241, with an average of 6.2 communities/day. These communities accounted for 48.2\% of the population.

\begin{figure}[h]
\begin{center}
\includegraphics[width=8cm]{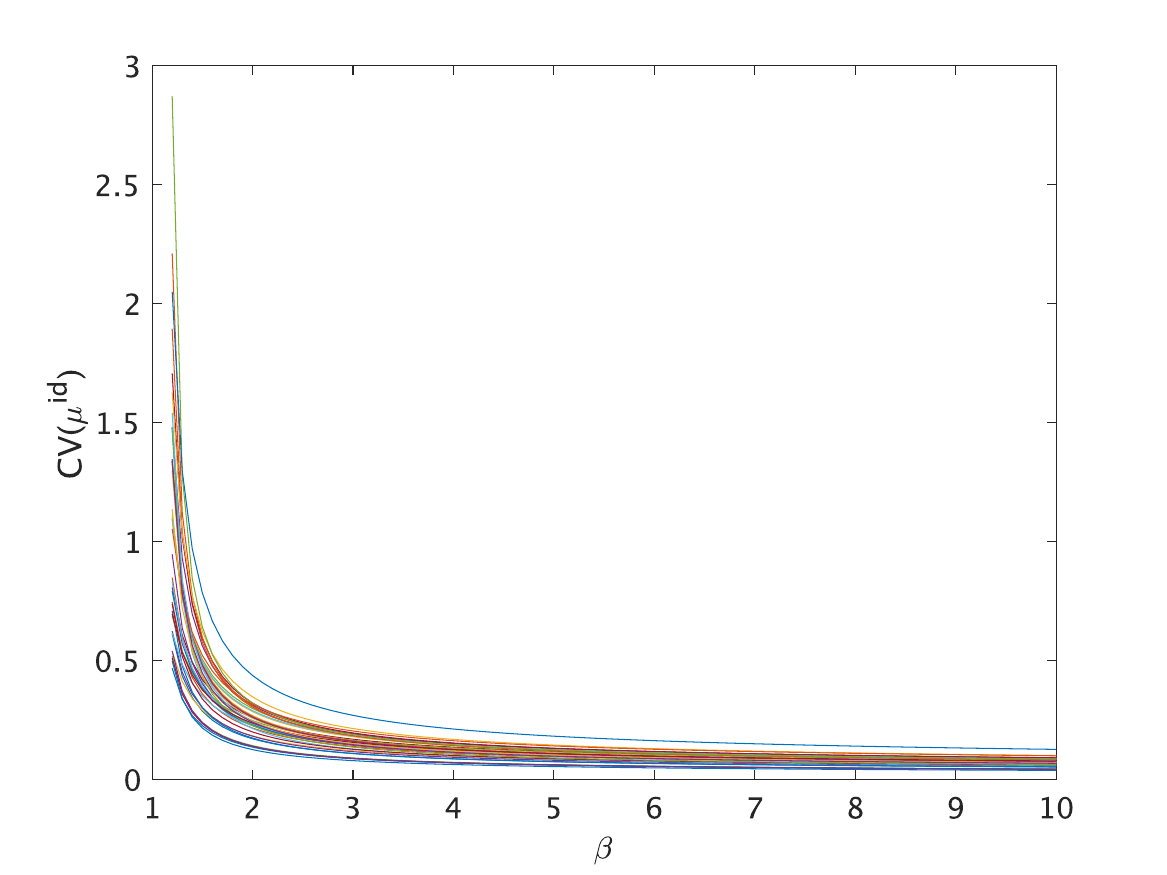}
\caption{\label{figure11}Coefficient of variation of the ideal gas component of chemical potential $\mu^\mathrm{id}$.}
\end{center}
\end{figure}

First, we determined the value of $\beta$ to determine $\mu$. We set $\beta=5$ as the value at which the root-mean-square speed is appropriate as the velocity of a person's movement, and the system is in a bound state. In this case, the error in $\mu^\mathrm{id}$ is approximately 10\%, which is sufficiently large to estimate the proper model parameters, as shown in Fig. \ref{figure11}.

Second, we considered the constraints of the optimization problem. We constrained $\epsilon>0$ and $\sigma>0$ because all had positive values. We also constrained $\mu^\mathrm{int}<0$ because it had a negative value, assuming a real gas. The optimization problem considering these two constraints is as follows:
\begin{eqnarray}
\begin{aligned}
& {\text{minimize}} &&\mathcal{L}(\epsilon_\mathrm{cc},\epsilon_\mathrm{vv},\sigma_\mathrm{cc},\sigma_\mathrm{vv})\\
&\text{subject to} && \mu_i^\mathrm{int}<0 &&& i=\{1,2,\cdots,n\} \\
& &&\epsilon_\mathrm{cc},\epsilon_\mathrm{vv}>0&&& \\
& &&\sigma_\mathrm{cc},\sigma_\mathrm{vv}>0.&&&
\end{aligned} \label{o1}
\end{eqnarray}
We used an interior point method algorithm to solve the constrained nonlinear optimization problem.

\begin{figure}[h]
\begin{center}
\includegraphics[width=8cm]{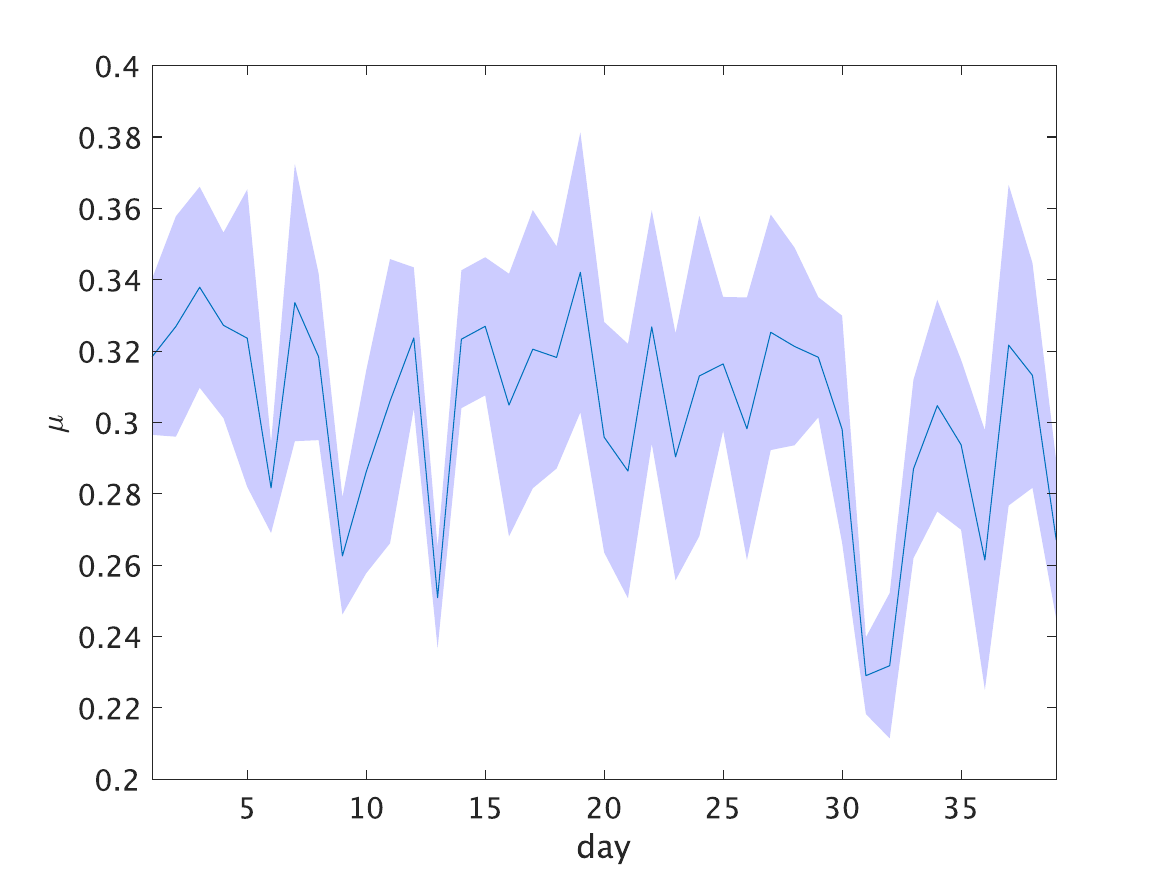}
\caption{\label{figure12}Average value of the estimated chemical potential $\mu$. The standard deviation is shown above and below.}
\end{center}
\end{figure}

\begin{figure}[h]
 (a) \\
 \vspace{-5mm}
 \begin{center}
 \includegraphics[width=70mm]{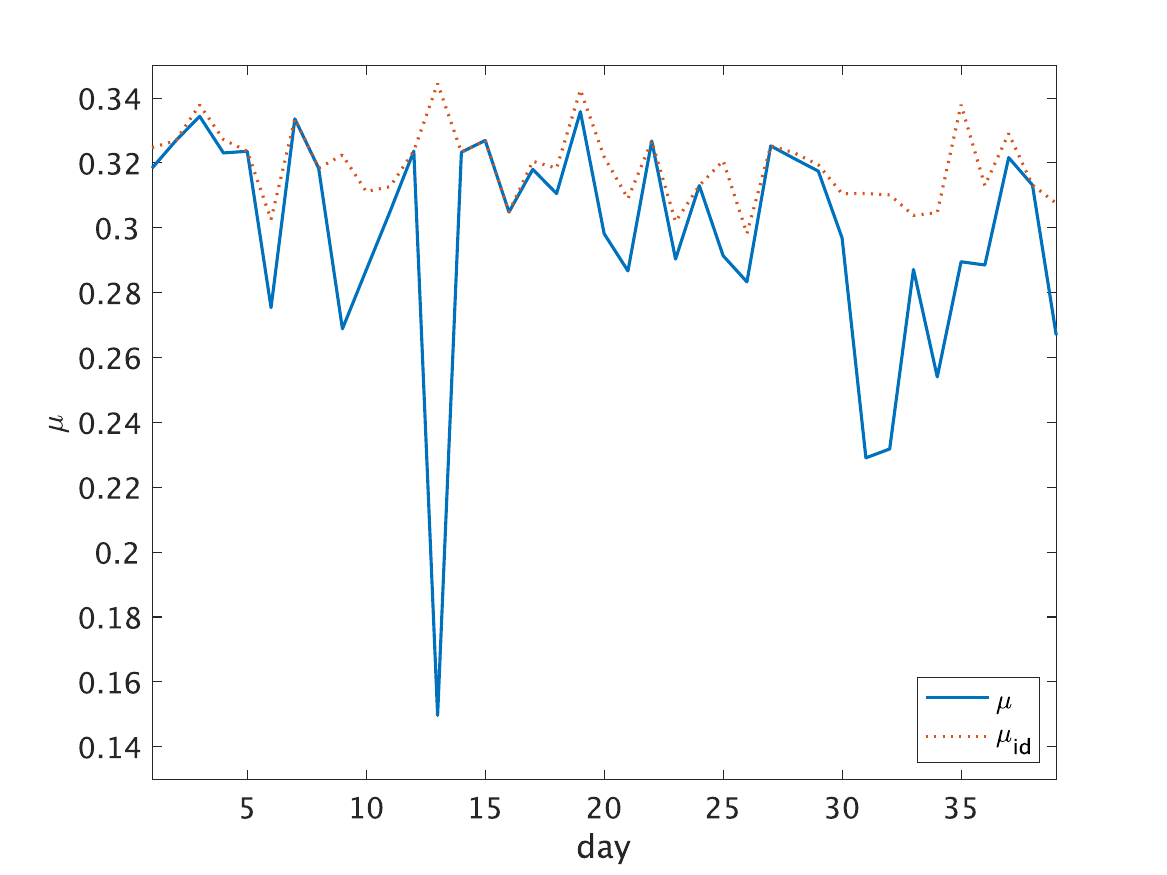}
 \end{center}
 (b) \\
 \vspace{-5mm}
 \begin{center}
 \includegraphics[width=70mm]{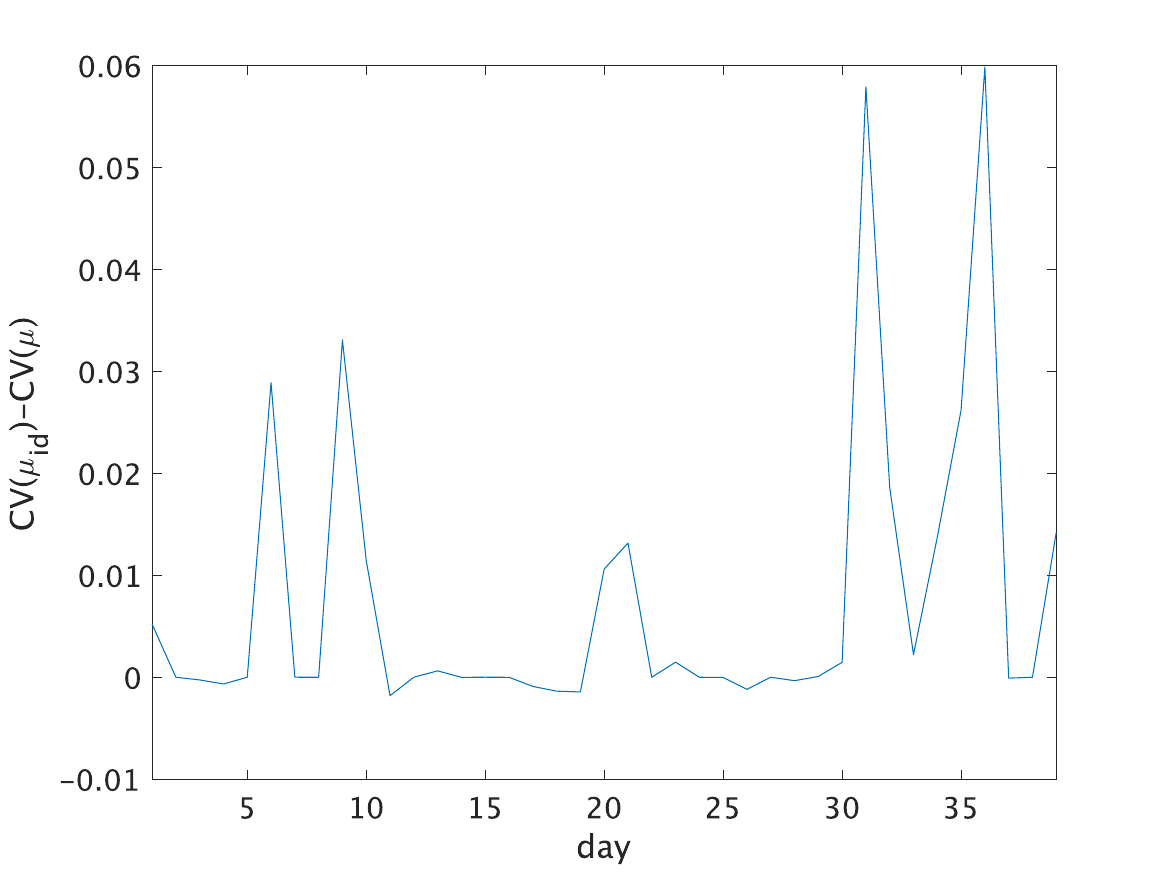}
 \caption{\label{figure13}Comparison of chemical potential $\mu$ and its ideal gas component $\mu^\mathrm{id}$: (a) Average values of $\mu$ and $\mu^\mathrm{id}$, and (b) Difference in coefficient of variation between $\mu$ and $\mu^\mathrm{id}$.}
\end{center}
\end{figure}

We estimated $\mu$ of persistent communities for each day and obtained model parameters. Figure \ref{figure12} shows the results of the estimated $\mu$ values. Standard deviations are shown above and below the mean values. Under conditions in which it is expected to be estimated with an error of approximately10\%, the value of the chemical potential was estimated with a smaller error than expected due to the contribution of the interaction of people. We compared $\mu$ and $\mu^\mathrm{id}$ in Fig. \ref{figure13}(a). $\mu$ was estimated to be smaller than $\mu^\mathrm{id}$. We also confirmed that the contribution of $\mu^\mathrm{int}$ in satisfying the equilibrium condition varied from day to day. The maximum effect of $\mu^\mathrm{int}$ was approximately 6\% from Fig.  \ref{figure13}(b), which is the difference between the coefficients of variation of $\mu$ and $\mu^\mathrm{id}$.

\begin{figure}[h]
 (a) \\
 \vspace{-5mm}
 \begin{center}
 \includegraphics[width=70mm]{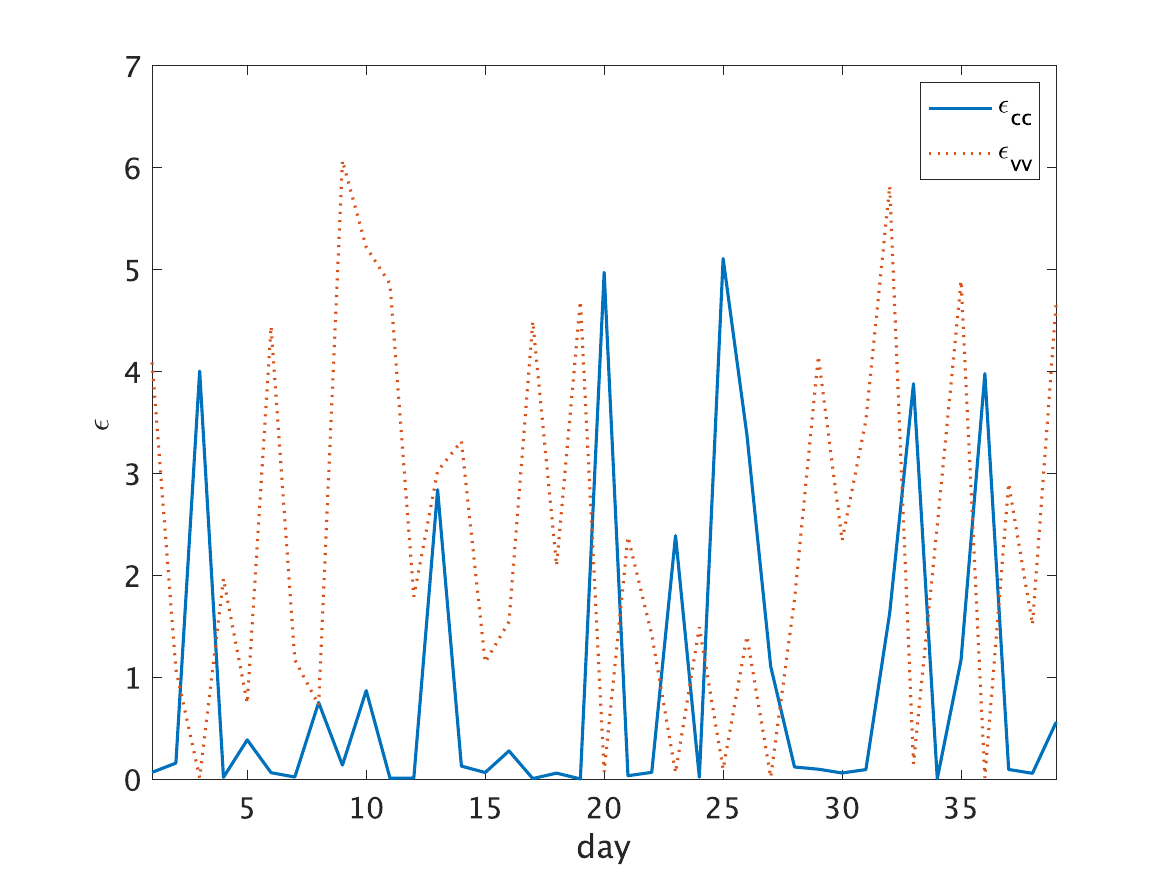}
 \end{center}
 (b) 
 \vspace{-5mm}
 \begin{center}
 \includegraphics[width=70mm]{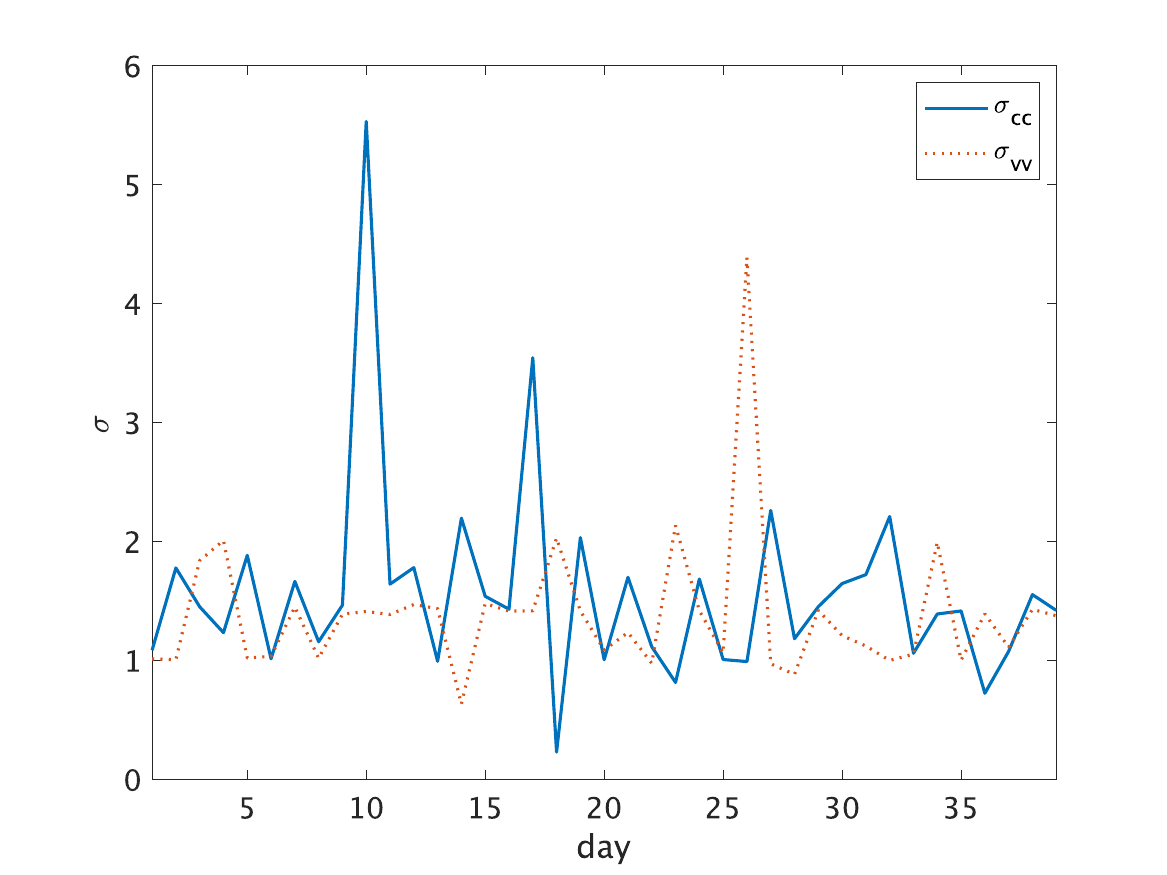}
 \caption{\label{figure14}Estimated values of model parameters. $\epsilon$ is a parameter of the depth of interactions. $\sigma$ is a parameter of the range of interactions: (a) $\epsilon$, and (b) $\sigma$.}
\end{center}
\end{figure}

Table \ref{table7} and Fig. \ref{figure14} show the values of the model parameters. The values were within a certain range for the data period, although some variation was observed. The average $\sigma$ value was approximately one. This indicated that the distance of the interaction was related to the mesh size.

\begin{table}[h]
\caption{\label{table7}Estimated values of model parameters.}
\begin{tabular}{@{}lllllll}
\hline
 & $\epsilon_\mathrm{cc}$ & $\epsilon_\mathrm{vv}$ & $\epsilon_\mathrm{cv}$ & $\sigma_\mathrm{cc}$ & $\sigma_\mathrm{vv}$ & $\sigma_\mathrm{cc}$ \\ \hline
Mean & 1.12 & 2.31 & 1.07 & 1.05 & 1.04 & 1.05 \\
SD & 2.08 & 1.89 & 1.49 & 0.55 & 0.27 & 0.28 \\
CV & 1.85 & 0.82 & 1.40 & 0.52 & 0.26 & 0.27 \\
\hline
\end{tabular} \\
SD: standard deviation, CV: coefficient of variation.
\end{table}

Finally, we checked the consistency between the given condition of $\beta$, $\beta=5$, and the physical picture. The root-mean-square velocity was calculated using eq. (\ref{v}), and is found to be $\sqrt{\langle v^2 \rangle}\simeq1.3$m/s. We confirmed that this value of  $\sqrt{\langle v^2 \rangle}$ is valid for the velocity at which a person moves. In addition, it was confirmed $\epsilon>1/\beta$ comparing the average kinetic energy $1/\beta$ and $\epsilon$, which means that the system is in a bound state. Therefore, we considered $\beta=5$ appropriate for this analysis.

\section{Discussion}

We discuss these results in terms of the goals of this study. First, we constructed a face-to-face interaction network using the mobility data with approximately 5,000 people in the maximum connected component. The sizes of the networks were much larger than networks constructed by conventional methods using Bluetooth, cell phones, and RFID sensors with a few hundred people\cite{RN54,RN43}. In addition, when we analyzed mobility data for more people in a wider area, we constructed a larger network. We also found that the network structures were similar during the data period. This implies that the proposed method for constructing face-to-face interaction networks is valid. We succeeded in proposing a method for constructing large-scale face-to-face interaction networks using mobility data.

However, the features of the constructed network were different from those of other social networks. Scale-free properties have been revealed as a characteristic of social networks \cite{Xu2019,Muchnik2013}. However, the degree distribution of the constructed face-to-face interaction networks was similar to an exponential distribution and did not exhibit a scale-free property. We confirmed that the degree distributions are similar to an exponential distribution because the tail is linear when plotted in a one-logarithmic form, as shown in Fig. \ref{figure6}. Further analysis is required to understand whether this is a unique feature of face-to-face interaction networks.

Second, we identified persistent communities using clustering analysis of daily face-to-face interaction networks over the data period. We used clustering analysis of the spatial distribution of each community to define persistent communities without information regarding individual node identification. We found that communities belonging to seven clusters were persistent, and each cluster was characterized by a specific location in Kyoto City. This indicates that communication patterns between citizens and visitors are stationary on weekdays because of the commuting flow in and out of Kyoto City. We infer that this communication pattern is common in many cities and that the persistent community structure may be a universal feature of face-to-face interaction networks in cities.

Third, we developed the theory of stable community structure to formulate the chemical potential of each community and adapted the theory to the results of the data analysis. Consequently, we estimated the chemical potential $\mu$ of the communities each day with errors of approximately 10\% under $\beta=5$. This suggests that the statistical mechanics model can explain the emergence of persistent communities because of their stable community structure.

Although the estimated values of $\mu$ matched over the data period, the level of agreement varied from day to day. We analyzed the reason for this variance in $\mu$ by separating $\mu^\mathrm{id}$ and $\mu^\mathrm{int}$. We focused on the particle density and configuration for each community because the value of $\mu^\mathrm{id}$ is determined by the particle density, and $\mu^\mathrm{int}$ is determined by the second-order particle density and configuration.

\begin{figure}[h]
 (a) \\
 \vspace{-5mm}
 \begin{center}
 \includegraphics[width=70mm]{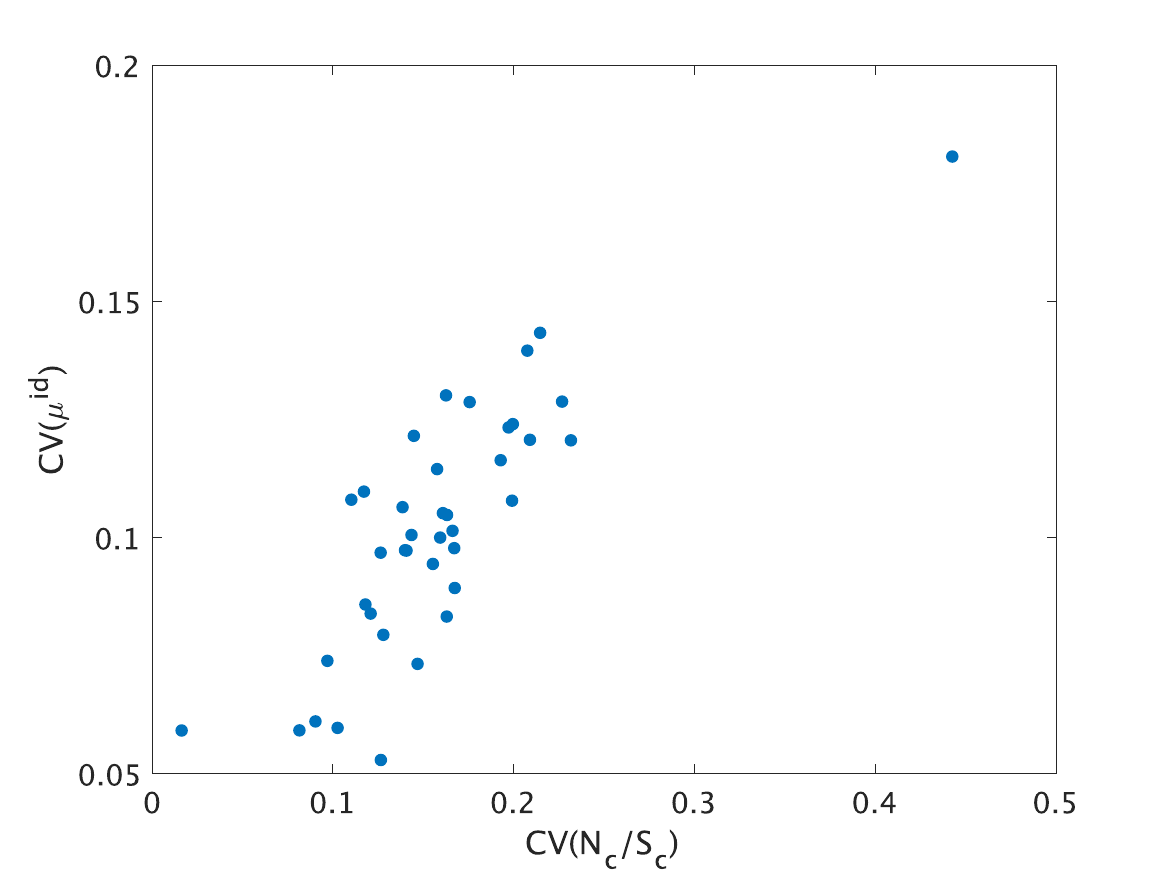}
 \end{center}
 (b) \\
 \vspace{-5mm}
 \begin{center}
 \includegraphics[width=70mm]{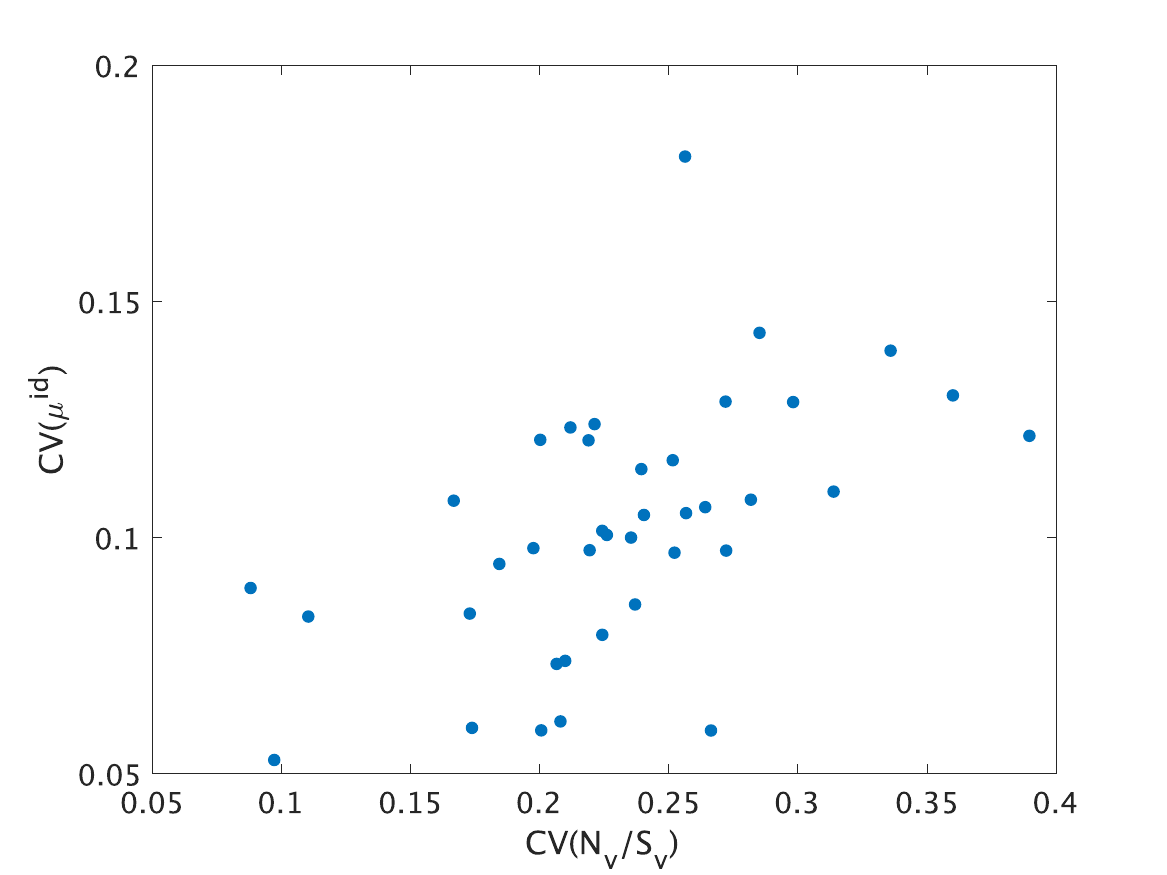}
 \caption{\label{figure15} Scatter plots of the ideal gas component of the chemical potential $\mu^\mathrm{id}$ and particle densities of citizens and visitors: (a) $\mu^\mathrm{id}$ and the particle density of citizens $N_\mathrm{c}/V_\mathrm{c}$, and (b) $\mu^\mathrm{id}$ and the particle density of visitors $N_\mathrm{v}/V_\mathrm{v}$.}
\end{center}
\end{figure}

From eq. (\ref{mu_id}), the value of $\mu^\mathrm{id}$ was determined by the particle density, $N_\mathrm{c}/S_\mathrm{c}$ and $N_\mathrm{v}/S_\mathrm{v}$. Therefore, the variance in $\mu^\mathrm{id}$ reflects the difference in particle density among the communities. $\mu^\mathrm{id}$ and $N_\mathrm{c}/S_\mathrm{c},N_\mathrm{v}/S_\mathrm{v}$ are positively related to each other, as shown in Fig. \ref{figure15} (the correlation coefficient $R=0.821$ and $0.537$, respectively). This means that forces act on particles from communities with high particle density to communities with low particle density owing to $\mu^\mathrm{id}$.

From the results shown in Fig. \ref{figure13}(b), $\mu^\mathrm{int}$ corrects the instability caused by the variance of $\mu^\mathrm{id}$ and stabilizes the community structures on some days. When $\mu^\mathrm{int}$ contributes to community stability, it has an opposite effect on the action of $\mu^\mathrm{id}$. This suggests that the community structure is stable if each particle is configured close to one another, and acts in communities with a higher particle density.

We used the Shannon entropy $H(p)=-\sum_{r}p_r \log p_r$ to confirm this consideration, which indicates the localization of the particle configurations. Here, $r$ is the particle distance, and $p_r$ is the frequency at which the distance between particles is $r$. Particles tend to be configured in a more concentrated manner if $H$ is higher in the community. We obtained the distributions of distances between citizens $p_\mathrm{c}$, visitors $p_\mathrm{v}$, and citizens--visitors $p_\mathrm{cv}$, from vectors $\bm{q}_\mathrm{c}$ and $\bm{q}_\mathrm{v}$ as shown in eqs. (\ref{q_c}) and (\ref{q_v}), respectively. The probability $p_r$ of particle distance $r$ is calculated as $p_r=\sum_{i,j\in R_{ij}} q_i\cdot q_j$, where $R_{ij}:=\left\{i,j | \sqrt{(x_i-x_j)^2+(y_i-y_j)^2}=r\right\}$, the set for which the distance between meshes $i$ and $j$ is equal to $r$.  

\begin{figure}[h]
 (a) \\
 \vspace{-5mm}
 \begin{center}
 \includegraphics[width=70mm]{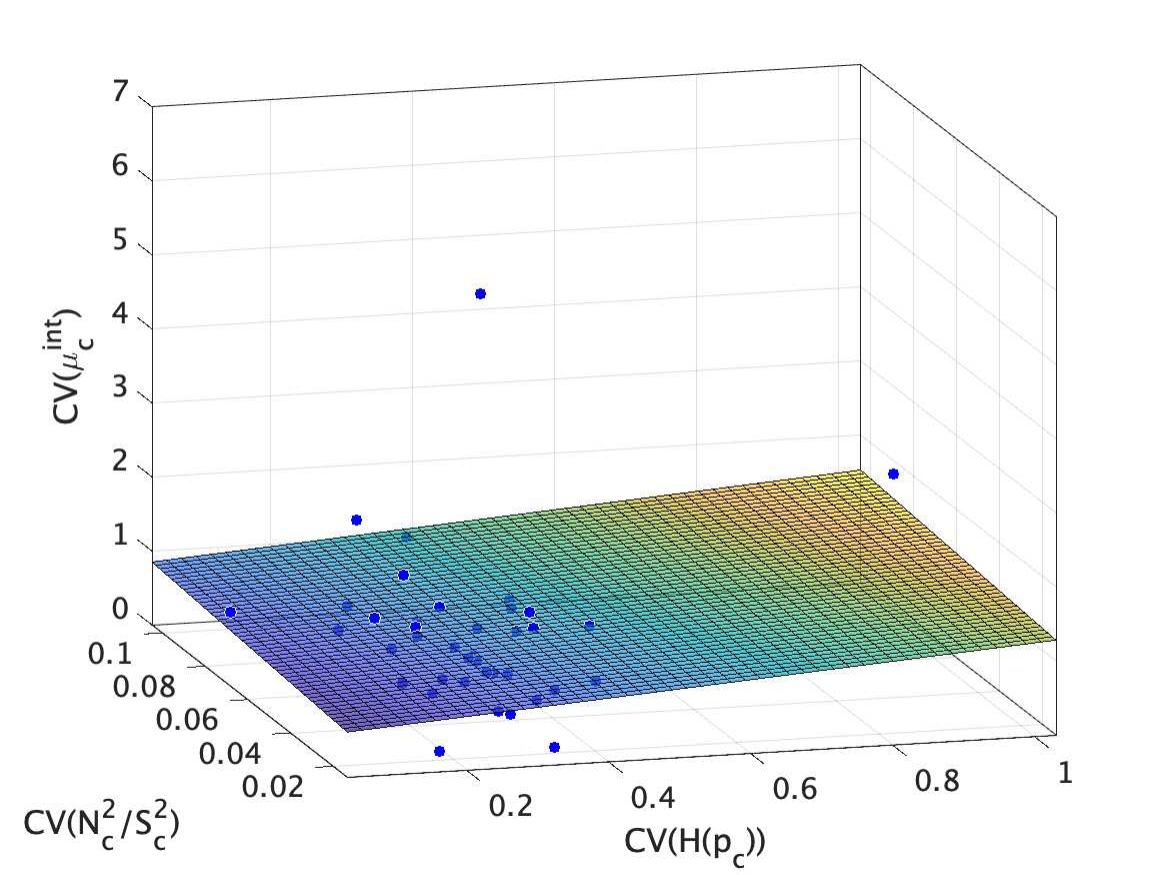}
 \end{center}
 (b) \\
 \vspace{-5mm}
 \begin{center}
 \includegraphics[width=70mm]{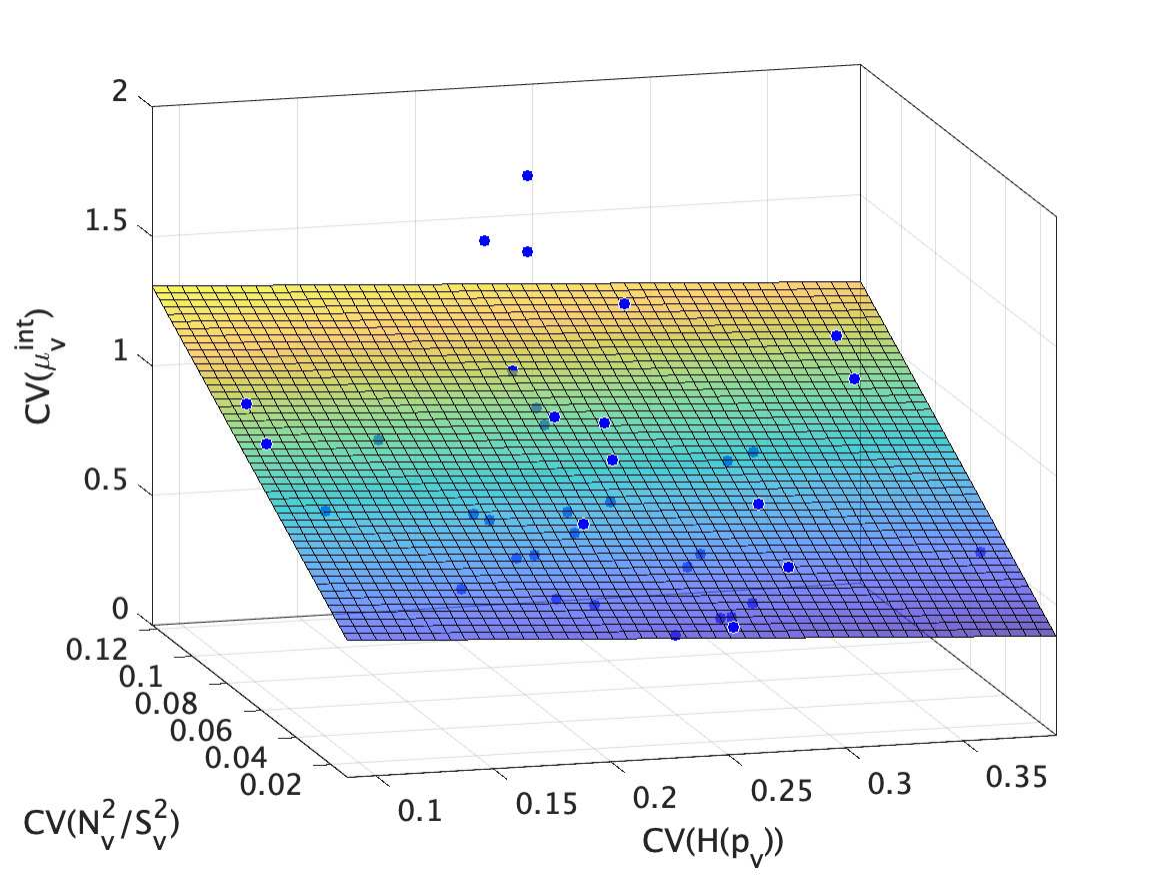}
 \end{center}
 (c) \\
 \vspace{-5mm}
 \begin{center}
 \includegraphics[width=70mm]{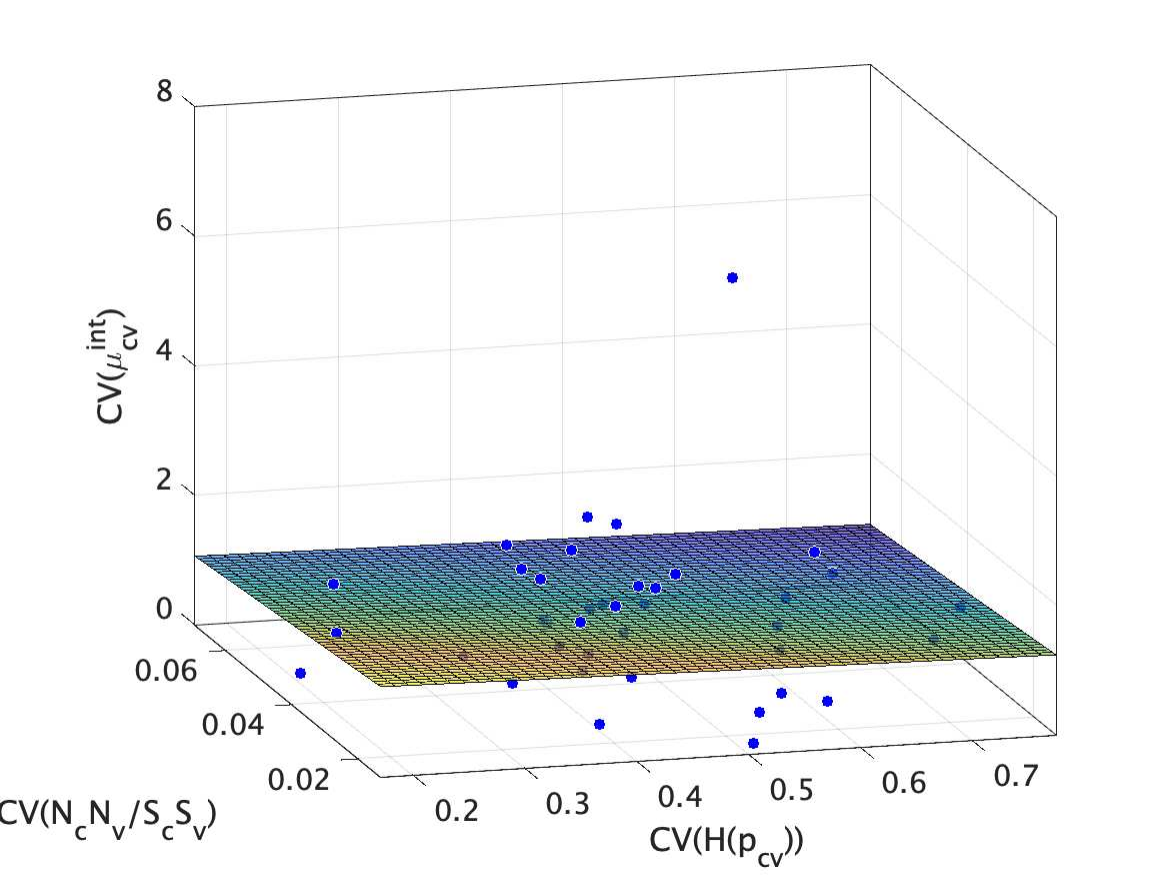}
 \caption{\label{figure16}Scatter plots of the interaction component of the chemical potential $\mu^\mathrm{int}$, second-order particle densities, and Shannon entropy of the distribution of particle distances. (a) Citizen--citizen relationships: $\mu^\mathrm{int}_\mathrm{c}$, $N_\mathrm{c}^2/S_\mathrm{c}^2$, and $H(p_\mathrm{c})$, (b) Visitor--visitor relationships: $\mu^\mathrm{int}_\mathrm{v}$, $N_\mathrm{v}^2/S_\mathrm{v}^2$, and $H(p_\mathrm{v})$, (c) Citizen--visitor relationships: $\mu^\mathrm{int}_\mathrm{cv}$, $N_\mathrm{c}N_\mathrm{v}/S_\mathrm{c}S_\mathrm{v}$, and $H(p_\mathrm{cv})$. The linear regression planes are plotted together.}
 \end{center}
\end{figure}

We analyzed the relationship among the coefficient of variation of $\mu^\mathrm{int}$, Shannon entropy $H(p_\mathrm{c}),H(p_\mathrm{v})$, and $H(p_\mathrm{cv})$, and second-order particle densities $N_\mathrm{c}^2/S_\mathrm{c}^2,N_\mathrm{v}^2/S_\mathrm{v}^2$, and $N_\mathrm{c}N_\mathrm{v}/S_\mathrm{c}S_\mathrm{v}$. Figure \ref{figure16} shows that the difference in density and the configuration of particles leads to a variance in the estimated value of $\mu^\mathrm{int}$. When the variances are large, $\mu^\mathrm{int}$ contributes to stabilizing the community structure. From eq. (\ref{mu_int1}), each component of $\mu^\mathrm{int}$ is the product of the second-order particle density and the value determined by the particle configuration because the denominators of $B_\mathrm{cc}, B_\mathrm{vv}$, and $B_\mathrm{cv}$ shown in eqs. (\ref{B_cc}--\ref{B_cv}) are calculated from the vectors $\bm{q}_\mathrm{c}$ and $\bm{q}_\mathrm{v}$. However, the chemical potential $\mu_\mathrm{v}^\mathrm{int}$ among visitors is negatively related to the variation in $H(p_\mathrm{v})$. We also consider that the citizen-visitor parameters $\epsilon_\mathrm{cv}$ and $\sigma_\mathrm{cv}$ are not independently determined, so that no relationship appears at all in $\mu_\mathrm{cv}^\mathrm{int}$.

As a result of these considerations, it seems that there are two requirements for $\mu^\mathrm{int}$ to act as a correction for the variation in $\mu^\mathrm{id}$: (i) a tendency toward structural instability with large variations in particle density, and (ii) large variations in the localization of the particle configuration. If (i) is not satisfied, the contribution of $\mu^\mathrm{int}$ is unnecessary because the structure is already stable. If (ii) is not satisfied, the contribution of $\mu^\mathrm{int}$ is almost uniform for all communities, and it does not act to stabilize the structure. Therefore, when both (i) and (ii) satisfy, $\mu^\mathrm{int}$ is estimated to stabilize the structure.

\begin{figure}[h]
 (a) \\
 \vspace{-5mm}
 \begin{center}
 \includegraphics[width=70mm]{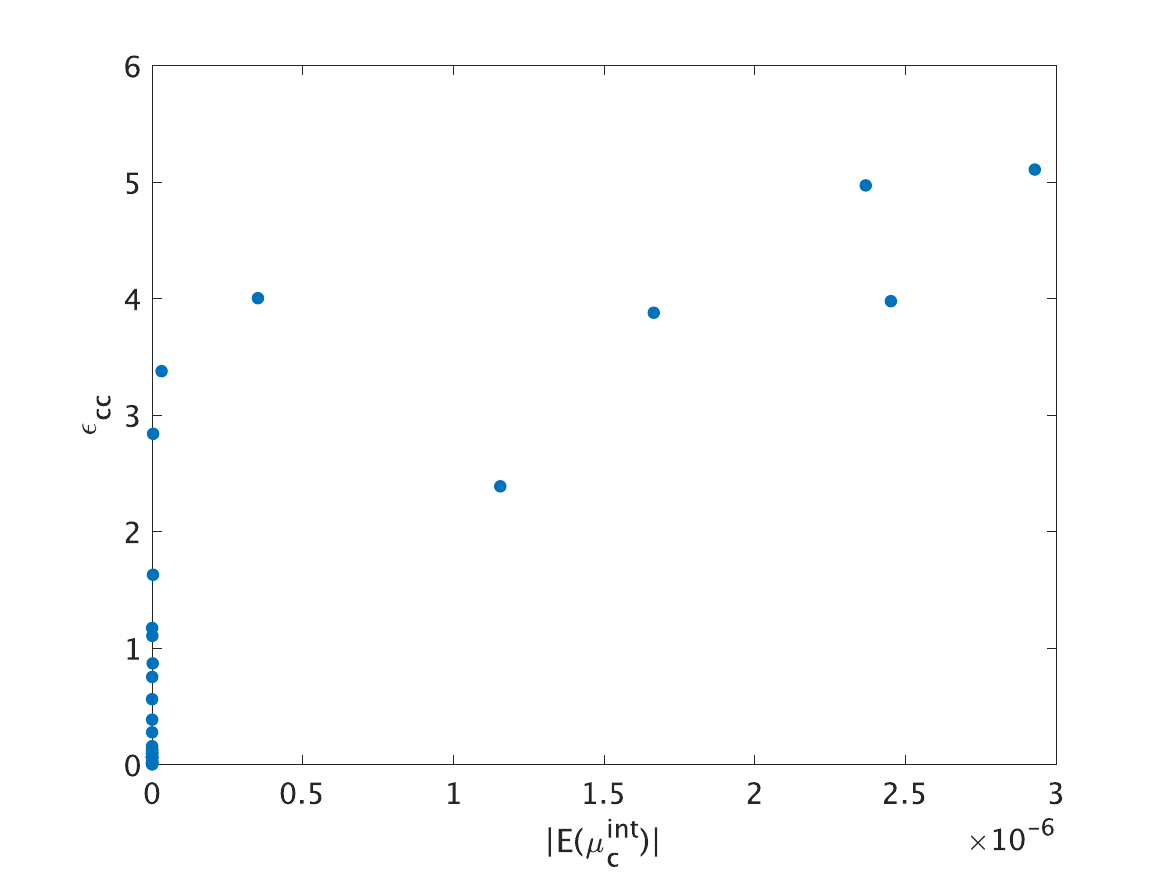}
 \end{center}
 (b) \\
 \vspace{-5mm}
 \begin{center}
 \includegraphics[width=70mm]{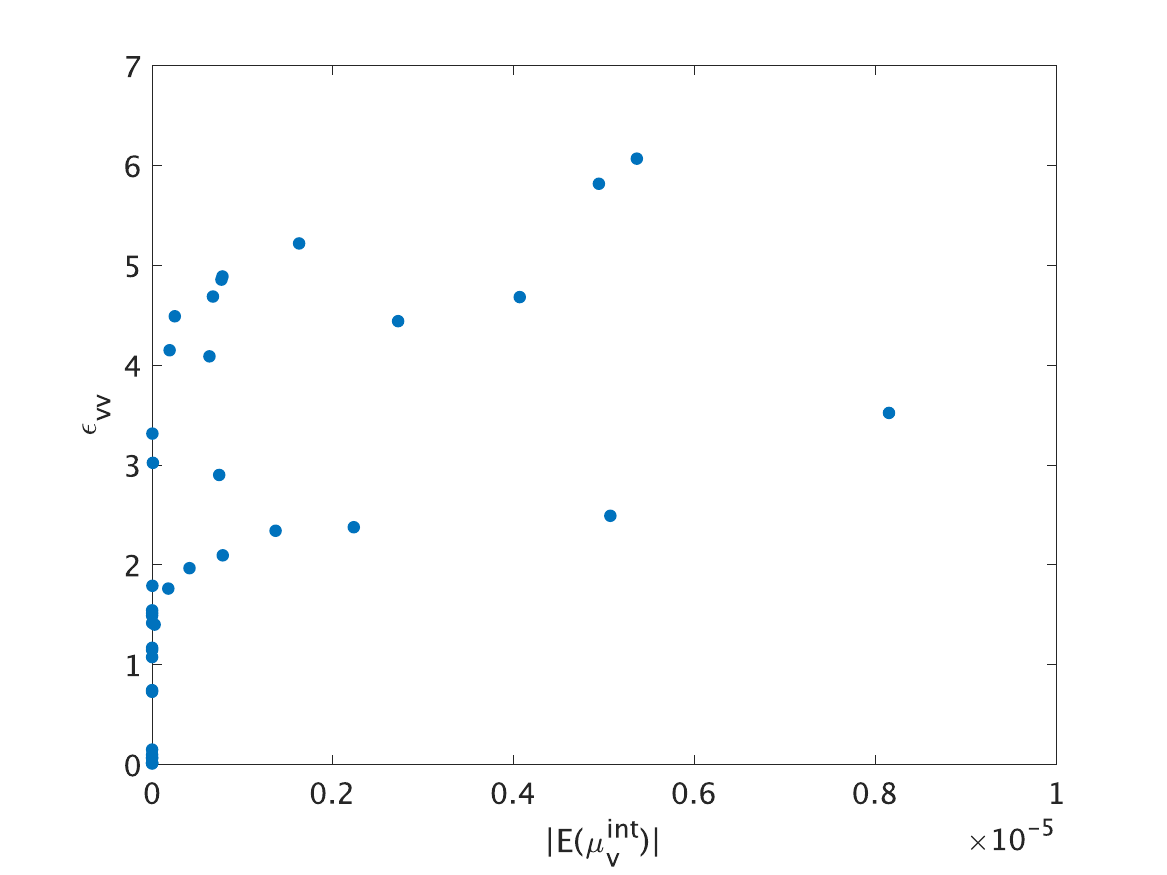}
 \end{center}
 (c) \\
 \vspace{-5mm}
 \begin{center}
 \includegraphics[width=70mm]{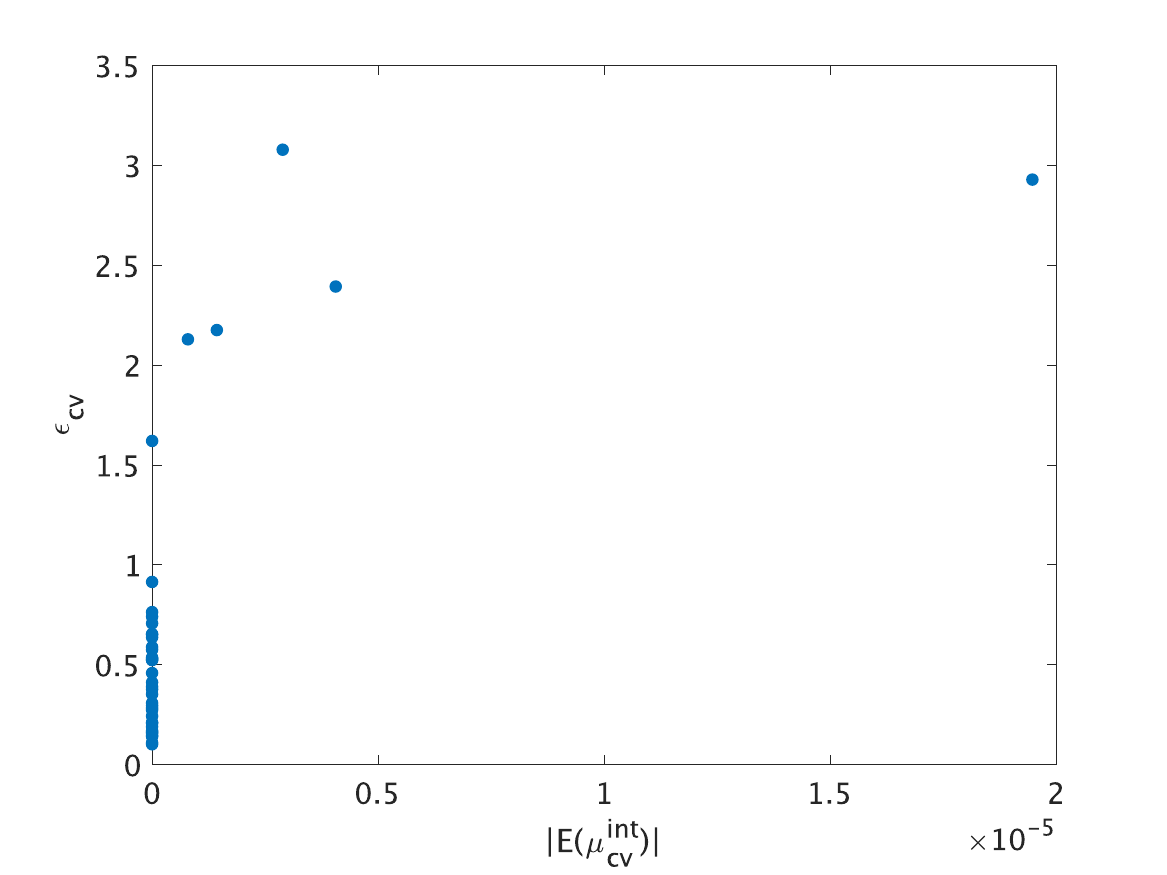}
 \caption{\label{figure17} Scatter plots of the model parameter $\epsilon$ and the interaction component of the chemical potential $\mu^\mathrm{int}$. $\epsilon_{cc}$ and $\mu^\mathrm{int}_\mathrm{c}$. (b) $\epsilon_\mathrm{vv}$ and $\mu^\mathrm{int}_\mathrm{v}$. (c) $\epsilon_\mathrm{cv}$ and $\mu^\mathrm{int}_\mathrm{cv}$. $\epsilon$ represents the depth of the interaction.}
 \end{center}
\end{figure}

We also discuss the estimated values of the model parameters. When $\epsilon_\mathrm{cc},\epsilon_\mathrm{vv}$, and $\epsilon_\mathrm{cv}$ are estimated with large values, the contributions of $\mu_\mathrm{c}^\mathrm{int},\mu_\mathrm{v}^\mathrm{int}$, and $\mu_\mathrm{cv}^\mathrm{int}$ tends to be large as shown in Fig. \ref{figure17} (the correlation coefficient $R=0.822,0.546$, and $0.638$, respectively). Since $\epsilon$ is a parameter of the depth of interaction, in case the value of $\epsilon$ is larger, the effect of $\mu^\mathrm{int}$ is larger too. When the particle density and configuration within communities satisfied the situation in which $\mu^\mathrm{int}$ acted to stabilize the community structure, the value of $\epsilon$ was estimated to be higher.

\begin{figure}[h]
 (a) \\
 \vspace{-5mm}
 \begin{center}
 \includegraphics[width=70mm]{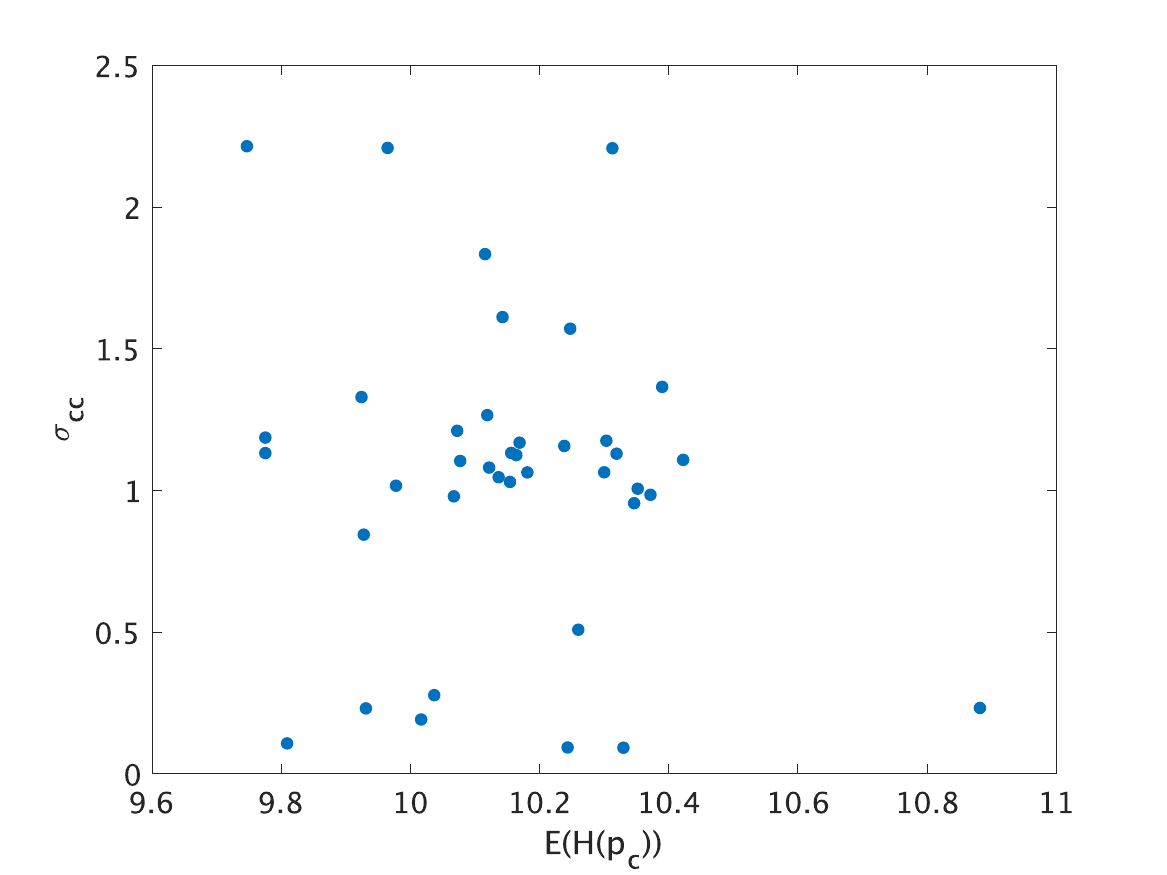}
 \end{center}
 (b) 
 \vspace{-5mm}
 \begin{center}
 \includegraphics[width=70mm]{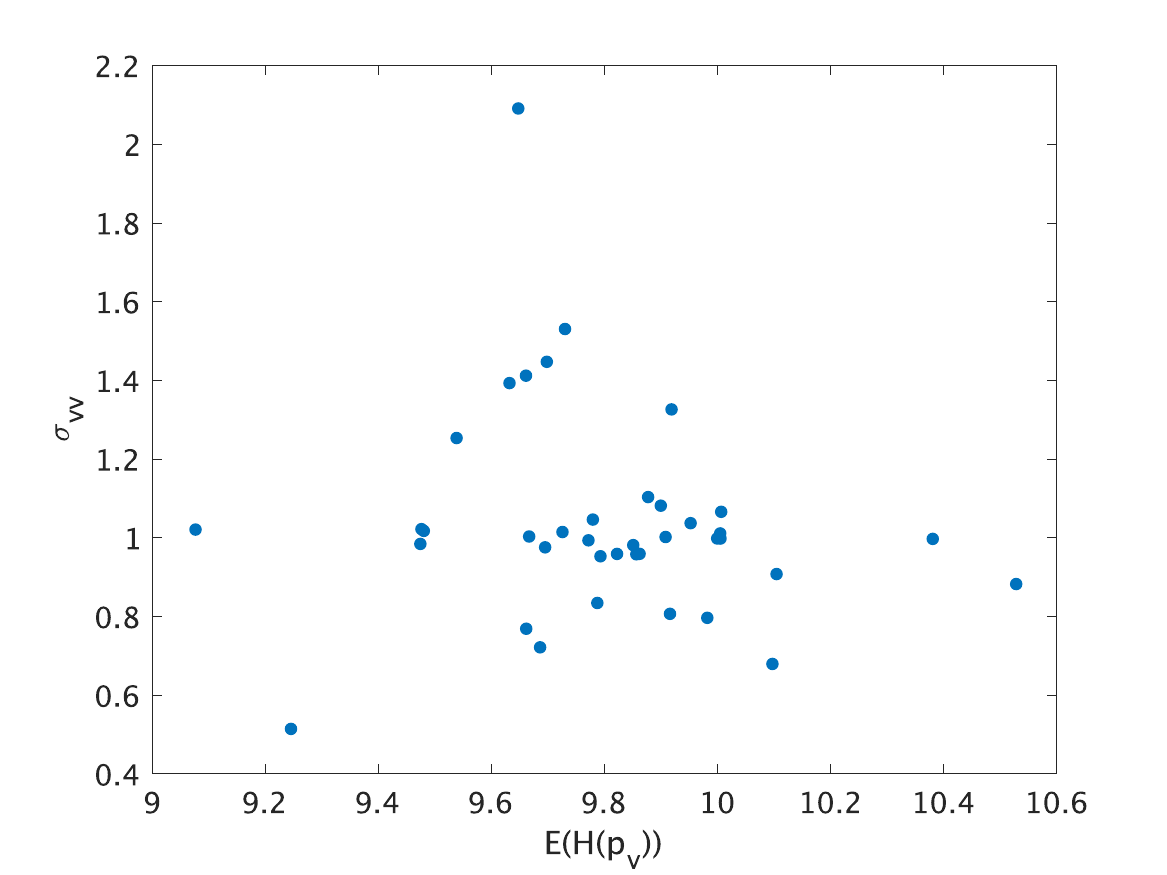}
 \end{center}
 (c) \\
 \vspace{-5mm}
 \begin{center}
 \includegraphics[width=70mm]{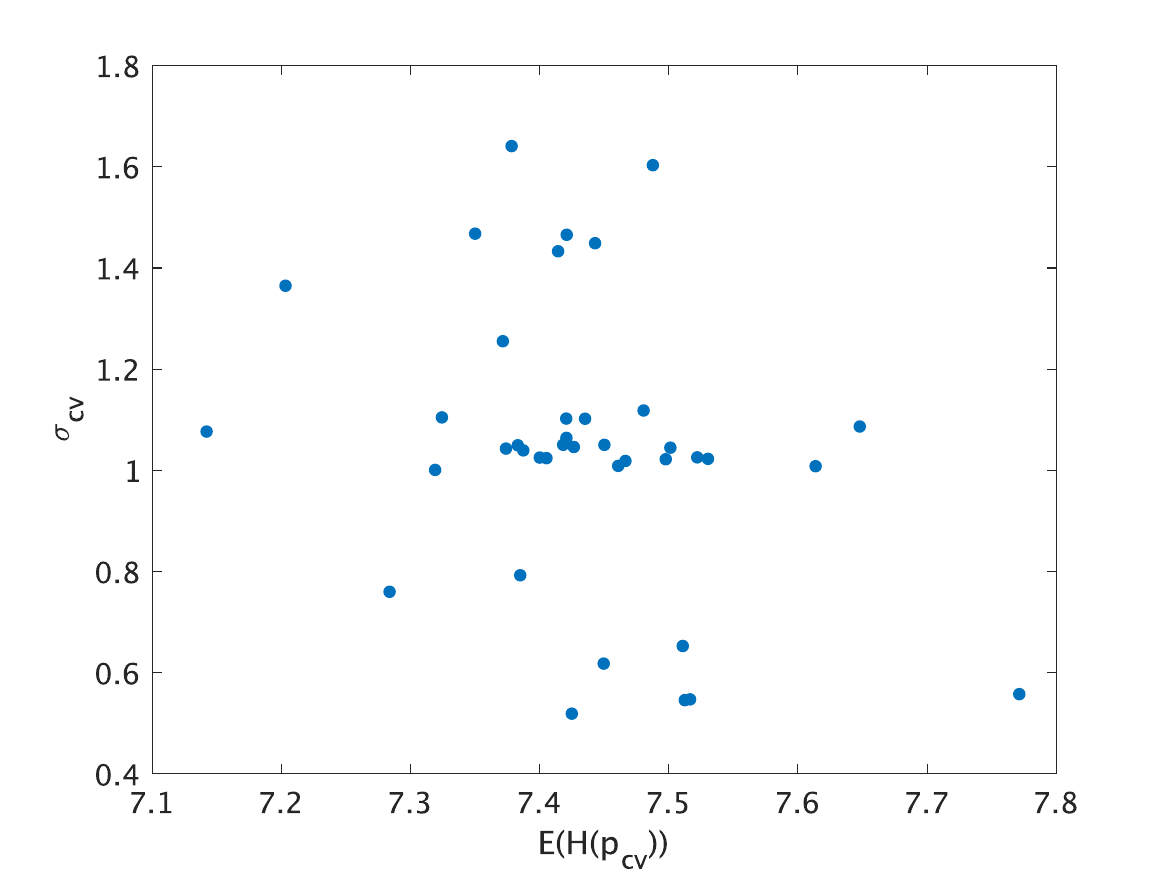}
\caption{\label{figure18}Scatter plots of the model parameter $\sigma$ and Shannon entropy of the distribution of particle distances. (a) $\sigma_\mathrm{cc}$ and $H(p_\mathrm{c})$. (b) $\sigma_\mathrm{vv}$ and $H(p_\mathrm{v})$. (c) $\sigma_\mathrm{cv}$ and $H(p_\mathrm{cv})$. $\sigma$ represents the range of interactions.}
 \end{center}
\end{figure}

The value of $\sigma$ represents the range of the interactions. We considered that the distributions of the distance between particles in the communities were related to $\sigma$. When the proportion of particles in the proximity distance was high, it was expected to be small. Shannon entropy $H(p_\mathrm{c}),H(p_\mathrm{v})$, and $H(p_\mathrm{cv})$ were used as the values for the particle localization. Figure \ref{figure18} shows the relationship between $\sigma$ and $H$. The larger the average value of $H$, the smaller is the value of $\sigma$ that tends to be estimated (the correlation coefficient $R=-0.157,-0.120$, and $-0.314$, respectively). This indicates that if the distribution of particles within a community tends to be more locally concentrated, the estimated value of $\sigma$ decreases.

The variance of  $\epsilon$ was larger than that of $\sigma$. This suggests that $\sigma$ is less sensitive to differences in the particle density and configuration among communities. Although there are variations in particle density and configuration among communities, the distance of interaction is constant to some extent. In addition, the variances of $\epsilon_\mathrm{cc}$ and $\sigma_\mathrm{cc}$ were larger than $\epsilon_\mathrm{vv}$ and $\sigma_\mathrm{vv}$. This implies that the communication pattern among visitors is more stationary than that among citizens.

From the above discussion, we found that the variation in the estimated value of $\mu$ resulted from the differences in particle density and configuration among communities in each daily face-to-face interaction network. The variations of $\epsilon$ and $\sigma$ are also explained in the same way. Based on the developed theory, we discuss the conditions of a stable community structure in terms of particle density and configuration. This result demonstrates that communities in the face-to-face interaction network are stable, and this stability leads to persistent communities in the face-to-face interaction network.

\section{Conclusion}

Social behavior is an important aspect of people's lives and is projected as a social network formed by means of communication. Face-to-face interaction networks are one of them and are worth analyzing to understand social behavior. However, it has been difficult to construct large-scale face-to-face interaction networks despite their importance. Because of the limitations, previous studies have not revealed the characteristics of the communication pattern in a city using a network science approach. We intended this research to accomplish three goals: (i) to construct a large-scale face-to-face interaction network using mobility data; (ii) to find persistent communities consisting of citizens and visitors; (iii) to explain the cause of persistent communities according to the theory of stable community structure using a statistical mechanics model. 

We used mobility data for 39 weekdays in February and April 2019 in Kyoto City and analyzed daily face-to-face interaction networks to identify persistent communities. We found persistent communities that existed over the data period belonging to seven clusters, and each cluster was characterized by a specific location in Kyoto City. This result indicates that persistent communities of face-to-face interaction networks are formed by the communication pattern between citizens and visitors in a specific location in Kyoto City. Moreover, the chemical potential of each persistent community was calculated to explain the thermodynamically stable community structure. We found that the values of the chemical potentials were matched with approximately 10\% error for each day. This means that persistent communities satisfy the condition of a stable community structure. This indicates that a stable community structure creates stent communities.

In this study, we revealed that persistent communities were formed by stationary communication patterns between cities and visitors on weekdays as one of the features of face-to-face interaction networks. We have developed a theory to explain the cause of persistent communities as a stable community structure in terms of thermodynamics. Future studies are planned to adapt this theory to other networks without spatial coordinates. We used it to investigate the stable community structures of universal networks, including social networks via cell phones or SNS, and to understand the characteristics of people's communication patterns.

\section*{Acknowledgments}

This study was supported by JSPS KAKENHI (grant number JP17KT0034). YO thanks JST for the support of the establishment of university fellowships for the creation of science technology innovation (grant number JPMJFS2123).

\appendix

\section{Deriving of chemical potentials using spatial coordinates}

If we denote citizens and visitors by the subscripts c and v, the total energy $\mathcal{H}$ of a phase of this system is as follows:
\begin{eqnarray}
\mathcal{H}=(\mathcal{H}_\mathrm{c}^\mathrm{id}+\mathcal{H}_\mathrm{cc}^\mathrm{int})+(\mathcal{H}_\mathrm{v}^\mathrm{id}+\mathcal{H}_\mathrm{vv}^\mathrm{int})+(\mathcal{H}_\mathrm{cv}^\mathrm{int}).
\end{eqnarray}
$\mathcal{H}^\mathrm{id}$ represents the energy of an ideal gas, that is, without interaction, and $\mathcal{H}^\mathrm{int}$ represents the energy of the interaction. $\mathcal{H}_\mathrm{c}^{\mathrm{id}}$ and $\mathcal{H}_\mathrm{v}^\mathrm{id}$ represent the total kinetic energy of citizens and visitors, respectively. And, $\mathcal{H}_{\mathrm{cc }}^{\mathrm{int}}$, $\mathcal{H}_{\mathrm{vv}}^{\mathrm{int}}$, and $\mathcal{H}_{\mathrm{cv}}^{\mathrm{int}}$ represent the total interaction energy between citizen--citizen, visitor--visitor, and citizen--visitor, respectively. These quantities are represented by eqs. (\ref{h_id}) and (\ref{h_int}) as follows:
\begin{eqnarray}
\mathcal{H}^{\mathrm{id}}_g&=&\frac{1}{2m_\mathrm{p}}\sum_{i=1}^{N_g} \bm{p}_i^2 \quad (g=\mathrm{c},\mathrm{v}) \\
\mathcal{H}_{gh}^\mathrm{int}&=&\sum_{i,j}\phi_{gh}(|\bm{r}_i-\bm{r}_j|) \quad (g,h=\mathrm{c},\mathrm{v}).
\end{eqnarray}
where $N_\mathrm{c}$ and $N_\mathrm{v}$ denote the number of citizens and visitors, respectively.
   
The partition function was calculated to detemine the chemical potential. The total partition function $Z^\mathrm{tot}$ can be derived as follows:
\begin{eqnarray}
Z^\mathrm{tot}=(Z_{c}^\mathrm{id} \times Z_\mathrm{cc}^\mathrm{int})\cdot(Z_{v}^\mathrm{id} \times Z_\mathrm{vv}^\mathrm{int})\cdot(Z_\mathrm{cv}^\mathrm{int})
\end{eqnarray}
$\mathcal{H}^\mathrm{id}$ is a function of momentum $\bm{p}$, and $\mathcal{H}^\mathrm{int}$ is a function of coordinates $\bm{x}$. Replace $d\bm{\Gamma}_p=d\bm{p}_1\ldots d\bm{p}_N,d\bm{\Gamma}_x=d\bm{p}_1\ldots d\bm{x}_N$ with $m_\mathrm{p}$ and $\hbar=1$, $Z^\mathrm{tot}$ is as follows: 
\begin{eqnarray}
 Z^\mathrm{\mathrm{tot}}&=&\frac{1}{N_\mathrm{c}!N_\mathrm{v}! (2\pi)^{2N_\mathrm{tot}}} \nonumber\\ 
 &\times& \int\int d\bm{\Gamma}_{p\mathrm{c}} d\bm{\Gamma}_{p\mathrm{v}} \exp\left[ -\beta \left( \sum_{i=1,\ldots,N_\mathrm{c}} \mathcal{H}_{\mathrm{c}i}^\mathrm{id}+\sum_{i=1,\ldots,N_\mathrm{v}} \mathcal{H}_{\mathrm{v}i}^\mathrm{id} \right) \right] \nonumber \\ 
 &\times& \int\int d\bm{\Gamma}_{x\mathrm{c}} d\bm{\Gamma}_{x\mathrm{v}} \exp\left[ -\beta \left( \sum_{\substack{i<j \\ i,j=1,\ldots,N_\mathrm{c}}} \mathcal{H}_{\mathrm{c}i\mathrm{c}j}^\mathrm{int} \right. \right. \nonumber\\ 
 &+&\left. \left.\sum_{\substack{i<j \\ i,j=1,\ldots,N_\mathrm{v}}}\mathcal{H}_{\mathrm{v}i\mathrm{v}j}^\mathrm{int}+\sum_{\substack{i=1,\ldots,N_\mathrm{c} \\ j=1,\ldots,N_\mathrm{v}}} \mathcal{H}_{\mathrm{c}i\mathrm{v}j}^\mathrm{int} \right\}\right] . \label{pf}
\end{eqnarray}
$N_\mathrm{tot}=N_\mathrm{c}+N_\mathrm{v}$, where $N_\mathrm{c}$ is the number of citizens and $N_\mathrm{v}$ is the number of visitors. The partition function is defined as the integral of the Boltzmann factor $e^{-\beta E}$ over the phase space divided by $2\pi\hbar$. We assume that the quantum states are distributed in the phase space at a rate of one per area, $2\pi\hbar$. It is divided into $N_\mathrm{c}!$ and $N_\mathrm{v}!$ to account for the overlap of homogeneous particles.

We set the first and second terms as $Z^\mathrm{id}$, and the third term as $Z^\mathrm{int}$:
\begin{eqnarray}
 Z^\mathrm{id}&=&\frac{1}{N_\mathrm{c}!N_\mathrm{v}! (2\pi)^{2N_{\mathrm{tot}}}} \int\int d\bm{\Gamma}_{p\mathrm{c}} d\bm{\Gamma}_{p\mathrm{v}} \nonumber \\
 &&\exp\left[ -\beta \left( \sum_{i=1,\ldots,N_\mathrm{c}} \mathcal{H}_{\mathrm{c}i}^\mathrm{id}+\sum_{i=1,\ldots,N_\mathrm{v}} \mathcal{H}_{\mathrm{v}i}^\mathrm{id} \right) \right] \label{pfid} \\
 Z^\mathrm{int}&=&\int\int d\bm{\Gamma}_{x\mathrm{c}} d\bm{\Gamma}_{x\mathrm{v}} \exp \left[ -\beta \left( \sum_{\substack{i<j \\ i,j=1,\ldots,N_\mathrm{c}}} \mathcal{H}_{\mathrm{c}i\mathrm{c}j}^\mathrm{int} \right.\right.\nonumber \\
 &&+\left.\left.\sum_{\substack{i<j \\ i,j=1,\ldots,N_\mathrm{v}}} \mathcal{H}_{\mathrm{v}i\mathrm{v}j}^\mathrm{int}+\sum_{\substack{i=1,\ldots,N_\mathrm{c} \\ j=1,\ldots,N_\mathrm{v}}} \mathcal{H}_{\mathrm{c}i\mathrm{v}j}^\mathrm{int} \right) \right]. 
\label{pfint}
\end{eqnarray}
First, we calculated the partition function of the ideal gas $Z^\mathrm{id}$ as follows:
\begin{eqnarray}
Z^\mathrm{id}&=&\frac{1}{N_\mathrm{c}!N_\mathrm{v}! (2\pi)^{2N_\mathrm{tot}}} \left( \frac{2\pi}{\beta}\right)^{N_\mathrm{c}}\left( \frac{2\pi}{\beta}\right)^{N_\mathrm{v}} \nonumber\\
&=&\frac{1}{N_\mathrm{c}! N_\mathrm{v}!} \left( \frac{1}{2 \pi \beta} \right)^{N_\mathrm{c}+N_\mathrm{v}}.
\end{eqnarray}

Next, we calculated the partition function for the interaction $Z^\mathrm{int}$. The interaction potential $\mathcal{H}^{\mathrm{int}}$ acting between the particles is respectively given by,
\begin{eqnarray}
\mathcal{H}_{\mathrm{cc}}^{\mathrm{int}}&=\sum_{i,j}\phi_{\mathrm{cc}}(r_{ij}) \ (\text{Citizen--citizen interaction}) \\
\mathcal{H}_{\mathrm{vv}}^{\mathrm{int}}&=\sum_{i,j}\phi_{\mathrm{vv}}(r_{ij}) \ (\text{Visitor-visitor interaction}) \\
\mathcal{H}_{\mathrm{cv}}^{\mathrm{int}}&=\sum_{i,j}\phi_{\mathrm{cv}}(r_{ij}) \ (\text{Citizen-visitor interaction}) .
\end{eqnarray}
$\phi(r)$ is assumed to be the Lennard-Jones potential:
\begin{eqnarray}
\phi_{gh}(r)=4\epsilon_{gh}\left[\left(\frac{\sigma_{gh}}{r}\right)^{12}-\left(\frac{\sigma_{gh}}{r}\right)^6\right] \quad (g,h=\mathrm{c},\mathrm{v}).
\end{eqnarray}

$Z^\mathrm{int}$ is calculated as follows: 
\begin{eqnarray}
 Z^\mathrm{int}&=&\int\int d\bm{\Gamma}_{x\mathrm{c}} d\bm{\Gamma}_{x\mathrm{v}} \prod_{\substack{i<j \\ i,j=1,\ldots,N_\mathrm{c}}} (1+f_{ij}^\mathrm{c}) \prod_{\substack{i<j \\ i,j=1,\ldots,N_\mathrm{v}}} (1+f_{ij}^\mathrm{v}) \nonumber \\
 &&\prod_{\substack{i=1,\ldots,N_\mathrm{c} \\ j=1,\ldots,N_\mathrm{v}}} (1+f_{ij}^\mathrm{cv}). \label{hint}
\end{eqnarray}
Note that $f_{ij}^\mathrm{c},f_{ij}^\mathrm{v}$, and $f_{ij}^\mathrm{cv}$ are respectively defined as,
\begin{eqnarray}
1+f_{ij}^\mathrm{c}&=&\exp\{-\beta\phi_\mathrm{cc}(|\bm{r}_i-\bm{r}_j|)\} \\
1+f_{ij}^\mathrm{v}&=&\exp\{-\beta\phi_\mathrm{vv}(|\bm{r}_i-\bm{r}_j|)\} \\
1+f_{ij}^\mathrm{cv}&=&\exp\{-\beta\phi_\mathrm{cv}(|\bm{r}_i-\bm{r}_j|)\}.
\end{eqnarray}

We used a discrete probability distribution using the proportion of time people spent at mesh $m$, $q_{m}$, to represent The probability $p_i(x_m,y_m)$ that a person $i$ exists in a certain mesh $m(x_m,y_m)$:
\begin{eqnarray}
q_{m}&=&p_i(x_m,y_m) \\
\sum_{m=1}^{M}q_{m}&=&\sum_{m=1}^{M}p_i(x_m,y_m)=1.
\end{eqnarray}
$M$ denotes the total number of meshes. We represent the existence probability $p_j(x_m,y_m)$ of another person $j$ at mesh $m$ using $q_m$ in the same manner. We consider $q_m=p_i(x_m,y_m)=p_j(x_m,y_m)$. We treat the discrete probability density functions $p_i(x,y)$ and $p_j(x,y)$ as continuous probability density functions using the delta function and obtain the following equation:
\begin{eqnarray}
p_i(x,y)=p_j(x,y)=\sum_{m=1}^M q_{m} \delta(x-x_m) \delta(y-y_m). \label{pfun2}
\end{eqnarray}

We use this probability distribution to calculate $Z^\mathrm{int}$ considering the approximation for a real gas. Equation (\ref{hint}) expands to terms representing the interaction of $i$ pairs of particles. Since the maximum number of combinations for $i$-particle pair interactions occurs when all the particles in the particle pairs are different, we obtain the following equation leaving only the largest contribution as the approximation: 
\begin{eqnarray}
Z^\mathrm{int}\simeq S_\mathrm{c}^{N_\mathrm{c}}S_\mathrm{v}^{N_\mathrm{v}}\sum_{i=0}\frac{1}{i!}(-N_\mathrm{c}^2B_\mathrm{cc}-N_\mathrm{v}^2B_\mathrm{vv}-N_\mathrm{c}N_\mathrm{v}B_\mathrm{cv})^i. \label{A19} 
\end{eqnarray}
Since eq. (\ref{A19}) is in the form of a Taylor series of exponential functions, and $Z^\mathrm{int}$ is represented as follows: 
\begin{eqnarray}
Z^\mathrm{int}=S_\mathrm{c}^{N_\mathrm{c}}S_\mathrm{v}^{N_\mathrm{v}}\exp(-N_\mathrm{c}^2B_\mathrm{cc}-N_\mathrm{v}^2B_\mathrm{vv}-N_\mathrm{c}N_\mathrm{v}B_\mathrm{cv}) \label{int2}
\end{eqnarray}
In eq. (\ref{int2}), $S_\mathrm{c}$ and $S_\mathrm{v}$ represent areas where citizens and visitors stay at least once, respectively. And $B_{\mathrm{cc}},B_{\mathrm{cv}}$, and $B_{\mathrm{cv}}$ are expressed as follows:
\begin{eqnarray}
B_{\mathrm{cc}}&=&\frac{1-\bm{q}_\mathrm{c}\bm{\Phi}_\mathrm{cc}\bm{q}_\mathrm{c}^{T}}{2S_\mathrm{c}^2} \\
B_{\mathrm{vv}}&=&\frac{1-\bm{q}_\mathrm{v}\bm{\Phi}_\mathrm{vv}\bm{q}_\mathrm{v}^{T}}{2S_\mathrm{v}^2} \\
B_{\mathrm{cv}}&=&\frac{1-\bm{q}_\mathrm{c}\bm{\Phi}_\mathrm{cv}\bm{q}_\mathrm{v}^{T}}{S_\mathrm{c}S_\mathrm{v}}.
\end{eqnarray}
The vector $\bm{q}_\mathrm{c},\bm{q}_\mathrm{v}$ in the above equation are,
\begin{eqnarray}
\bm{q_\mathrm{c}}&=&\left[
  \begin{array}{cccc}
   q_{\mathrm{c}1} & q_{\mathrm{c}2} & \cdots & q_{\mathrm{c}M_\mathrm{c}}
  \end{array}
\right] \\
\bm{q_\mathrm{v}}&=&\left[
  \begin{array}{cccc}
   q_{\mathrm{v}1} & q_{\mathrm{v}2} & \cdots & q_{\mathrm{v}M_\mathrm{v}}
  \end{array}
\right].
\end{eqnarray}
These represent the fractions of time spent in each mesh by citizens and visitors, respectively. $M_\mathrm{c}$ and $M_\mathrm{v}$ are the number of meshes where citizens and visitors stay at least once, respectively. We calculated them by examining the time spent in each mesh of citizens and visitors from the mobility data. Matrices $\bm{\Phi}_\mathrm{cc},\bm{\Phi}_\mathrm{vv}$, and $\bm{\Phi}_\mathrm{cv}$ are as follows: 
\begin{eqnarray}
\bm{\Phi}_{gh}&=&\left[\exp\left\{-\beta\phi_{gh}\left(\sqrt{(x_k-x_l)^2+(y_k-y_l)^2}\right)\right\}\right] \nonumber \\
&&(g,h=\mathrm{c},\mathrm{v}).
\end{eqnarray}
The $k,l$ components of Boltzmann factor are citizen--citizen, visitor--visitor, and citizen--visitor interactions of a distance between meshes $k$ and $l$, respectively.

Because the Helmholtz free energy is $F=-1/\beta\ln Z$, we obtain $F$ as,
\begin{eqnarray}
F&=&\frac{N_\mathrm{c}}{\beta}\left\{\ln\left(\frac{2\pi\beta N_\mathrm{c}}{S_\mathrm{c}}\right)-1\right\}+\frac{N_\mathrm{v}}{\beta}\left\{\ln\left(\frac{2\pi\beta N_\mathrm{v}}{S_\mathrm{v}}\right)-1\right\} \nonumber \\ 
&&+N_\mathrm{c}^2B_\mathrm{cc}+N_\mathrm{v}^2B_\mathrm{vv}+N_\mathrm{c}N_\mathrm{v}B_\mathrm{cv}.
\end{eqnarray}
We obtain the chemical potential of the citizen $\mu_\mathrm{c}=\partial F/\partial N_\mathrm{c}$ as,
\begin{eqnarray}
\mu_\mathrm{c}=\frac{1}{\beta}\ln\left(\frac{2\pi\beta N_\mathrm{c}}{S_\mathrm{c}}\right)+\frac{1}{\beta}(2N_\mathrm{c}B_\mathrm{cc}+N_\mathrm{v}B_\mathrm{cv}).
\end{eqnarray}
We also obtain the chemical potential of the visitor $\mu_\mathrm{v}=\partial F/\partial N_\mathrm{v}$ as,
\begin{eqnarray}
\mu_\mathrm{v}=\frac{1}{\beta}\ln\left(\frac{2\pi\beta N_\mathrm{v}}{S_\mathrm{v}}\right)+\frac{1}{\beta}(2N_\mathrm{v}B_\mathrm{vv}+N_\mathrm{c}B_\mathrm{cv}).
\end{eqnarray}
Therefore, the chemical potential $\mu$ is formulated as follows:
\begin{eqnarray}
\mu&=&\frac{1}{\beta(N_\mathrm{c}+N_\mathrm{v})}\Biggl\{N_\mathrm{c}\ln\left(\frac{2\pi\beta N_\mathrm{c}}{S_\mathrm{c}}\right)+N_\mathrm{v}\ln\left(\frac{2\pi\beta N_\mathrm{v}}{S_\mathrm{v}}\right)  \nonumber \\ 
&&+2N_\mathrm{c}^2B_\mathrm{cc}+2N_\mathrm{v}^2B_\mathrm{vv}+2N_\mathrm{c}N_\mathrm{v}B_\mathrm{cv}\Biggr\}. \label{mu_a}
\end{eqnarray}

\section{Deriving chemical potentials using adjacency matrix without spatial coordinates}

We computed the $Z^\mathrm{int}$ using an adjacency matrix. We obtain the following equation from eq. (\ref{hint}):
\begin{eqnarray}
 Z^\mathrm{int}&\simeq&\int\int d\bm{\Gamma}_{xc} d\bm{\Gamma}_{x\mathrm{v}} \nonumber \\
&& \left(1+\sum_{\substack{i<j\\ i,j=1,\ldots,N_\mathrm{c}}} f_{ij}^\mathrm{c}+\sum_{\substack{i<j\\ i,j=1,\ldots,N_\mathrm{v}}} f_{ij}^\mathrm{v}+\sum_{\substack{i=1,\ldots,N_\mathrm{c}\\ j=1,\ldots,N_\mathrm{v}}}
 f_{ij}^\mathrm{cv}\right) . \nonumber \\
 \label{zint2}
\end{eqnarray}
Here, we assumed that the second and subsequent terms of $f_{ij}$ are small and approximated it. 

Let the interaction potential $\mathcal{H}_{ij}^\mathrm{int}$ between nodes $i,j$ be,
\begin{eqnarray}
 \mathcal{H}_{ij}^\mathrm{int} = \begin{cases}
  \epsilon_\mathrm{cc} & (\text{Citizen--citizen interaction}) \\
  \epsilon_\mathrm{vv} & (\text{Visitor--visitor interaction}) \\
  \epsilon_\mathrm{cv} & (\text{Citizen--visitor interaction}) \\
  0 & (\text{No interaction}) \\
 \end{cases}.
\end{eqnarray}
We consider the interaction to occur when there is an edge between $i$ and $j$. At this time, $f_{ij}^\mathrm{c},f_{ij}^\mathrm{v}$, and $f_{ij}^\mathrm{cv}$ are respectively calculated as follows:
\begin{eqnarray}
1+f_{ij}^\mathrm{c}=\begin{cases}
  \exp(-\beta\epsilon_\mathrm{cc}) &(\text{Interaction between $i,j$)}\\
  1 & (\text{No interaction between $i,j$})\\
  \end{cases} \nonumber \\
  \\
1+f_{ij}^\mathrm{v}=\begin{cases}
  \exp(-\beta\epsilon_\mathrm{vv}) & (\text{Interaction between $i,j$)}\\
  1 & (\text{No interaction between $i,j$})\\
  \end{cases} \nonumber \\
  \\
1+f_{ij}^\mathrm{cv}=\begin{cases}
  \exp(-\beta\epsilon_\mathrm{cv}) & (\text{Interaction between $i,j$)}\\
  1 & (\text{No interaction between $i,j$}).\\
  \end{cases} \nonumber \\
\end{eqnarray}
We use these to calculate each term of eq. (\ref{zint2}).

Sorting the adjacency matrix, $\bm{A}$ such that $N_\mathrm{c}$ rows and columns are citizens, and the rest are visitors, and we can represent $\bm{A}$ as,
\begin{eqnarray}
\bm{A}&=&\left[
  \begin{array}{c|c}
   \bm{A}_\mathrm{cc} & \bm{A}_\mathrm{cv}\\ \hline
   \bm{A}_{vc} & \bm{A}_\mathrm{vv}\\
  \end{array}
\right]
\end{eqnarray}
where $\bm{A}_\mathrm{cc},\bm{A}_\mathrm{vv}$, and $\bm{A}_\mathrm{cv}$ represent the citizen--citizen, visitor--visitor, and citizen--visitor adjacency matrices, respectively. $\bm{I}_\mathrm{c}$ and $\bm{I}_\mathrm{v}$are column vectors with all components set to one in the $N_\mathrm{c}$ and $N_\mathrm{v}$ columns. Using $\bm{A}_\mathrm{cc},\bm{A}_\mathrm{vv},\bm{A}_\mathrm{cv},\bm{I}_{c}$, and $\bm{I}_{v}$, we formulate the partition function of the interaction $Z^\mathrm{int}$ using the adjacency matrix:
\begin{eqnarray}
Z^\mathrm{int}&=&S_\mathrm{c}^{N_\mathrm{c}}S_\mathrm{v}^{N_\mathrm{v}}\left[1+\frac{\exp(-\beta\epsilon_\mathrm{cc})-1}{2S_\mathrm{c}^{2}}\bm{I}_\mathrm{c}\bm{A}_\mathrm{cc}\bm{I}_\mathrm{c}^T \right. \nonumber \\
&-&\left. \frac{\exp(-\beta\epsilon_\mathrm{vv})-1}{2S_\mathrm{v}^{2}}\bm{I}_\mathrm{v}\bm{A}_\mathrm{vv}\bm{I}_\mathrm{v}^T \right. \nonumber \\
&-&\left.\frac{\exp(-\beta\epsilon_\mathrm{cv})-1}{S_\mathrm{c}S_\mathrm{v}}\bm{I}_\mathrm{c}\bm{A}_\mathrm{cv}\bm{I}_\mathrm{v}^T \right].
\end{eqnarray}
We consider $\ln S_\mathrm{c} \propto \ln N_\mathrm{c}$ and  $\ln S_\mathrm{v} \propto \ln N_\mathrm{v}$ and replace $S_\mathrm{c}$ and $S_\mathrm{v}$ with $N_\mathrm{c}$ and $N_\mathrm{v}$, respectively:
\begin{eqnarray}
S_\mathrm{c}=b_\mathrm{c}N_\mathrm{c}^{a_\mathrm{c}}, \ S_\mathrm{v}=b_\mathrm{v}N_\mathrm{v}^{a_\mathrm{v}}. \label{prop2}
\end{eqnarray}
Using eq. (\ref{prop2}), the partition function is represented without spatial coordinates as follows:
\begin{eqnarray}
Z^\mathrm{int}&=&(b_\mathrm{c}N_\mathrm{c}^{a_\mathrm{c}})^{N_\mathrm{c}} (b_\mathrm{v}N_\mathrm{v}^{a_\mathrm{v}})^{N_\mathrm{v}} \left[1+\frac{\exp(-\beta\epsilon_\mathrm{cc})-1}{2(b_\mathrm{c}N_\mathrm{c}^{a_\mathrm{c}})^{2}}\bm{I}_\mathrm{c}\bm{A}_\mathrm{cc}\bm{I}_\mathrm{c}^T \right. \nonumber \\
&+& \frac{\exp(-\beta\epsilon_\mathrm{vv})-1}{2(b_\mathrm{v}N_\mathrm{v}^{a_\mathrm{v}})^{2}}\bm{I}_\mathrm{v}\bm{A}_\mathrm{vv}\bm{I}_\mathrm{v}^T \nonumber \\
&+&\left.\frac{\exp(-\beta\epsilon_\mathrm{cv})-1}{(b_\mathrm{c}N_\mathrm{c}^{a_\mathrm{c}})(b_\mathrm{v}N_\mathrm{v}^{a_\mathrm{v}})}\bm{I}_\mathrm{c}\bm{A}_\mathrm{cv}\bm{I}_\mathrm{v}^T \right].
\end{eqnarray}

Since the Helmholtz free energy $F=-1/\beta\ln Z$, we obtain $F$ as,
\begin{eqnarray}
F&=&\frac{N_\mathrm{c}}{\beta}\left[\ln\left(\frac{2\pi\beta}{b_\mathrm{c} N_\mathrm{c}^{a_\mathrm{c}-1}}\right)-1\right] +\frac{N_\mathrm{v}}{\beta}\left[\ln\left(\frac{2\pi\beta}{b_\mathrm{v} N_\mathrm{v}^{a_\mathrm{v}-1}}\right)-1\right] \nonumber \\
 &&+\frac{1}{\beta}\left(B_\mathrm{cc}\bm{I}_\mathrm{c}\bm{A}_\mathrm{cc}\bm{I}_\mathrm{c}^T+B_\mathrm{vv}\bm{I}_\mathrm{v}\bm{A}_\mathrm{vv}\bm{I}_\mathrm{v}^T+B_\mathrm{cv}\bm{I}_\mathrm{c}\bm{A}_\mathrm{cv}\bm{I}_\mathrm{v}^T \right). \nonumber \\
\end{eqnarray}
$B_\mathrm{cc},B_\mathrm{vv}$, and $B_\mathrm{cv}$ are as follows:
\begin{eqnarray}
B_\mathrm{cc}=\frac{1-\exp(-\beta\epsilon_\mathrm{cc})}{2b_\mathrm{c}^2N_\mathrm{c}^{2a_\mathrm{c}}} \\
B_\mathrm{vv}=\frac{1-\exp(-\beta\epsilon_\mathrm{vv})}{2b_\mathrm{v}^2N_\mathrm{v}^{2a_\mathrm{v}}} \\
B_\mathrm{cv}=\frac{1-\exp(-\beta\epsilon_\mathrm{cv})}{b_\mathrm{c}b_\mathrm{v}N_\mathrm{c}^{a_\mathrm{c}}N_\mathrm{v}^{a_\mathrm{v}}}. 
\end{eqnarray}
Finally, the chemical potential $\mu$ is derived in the following equation:
\begin{eqnarray}
 \mu&=&\frac{1}{\beta(N_\mathrm{c}+N_\mathrm{v})}\Biggl\{N_\mathrm{c}\left[\ln \left( \frac{2 \pi \beta}{b_\mathrm{c} N_\mathrm{c}^{a_\mathrm{c}-1}} \right)-a_\mathrm{c}\right]  \nonumber \\
 &+&N_\mathrm{v}\left[\ln \left( \frac{2 \pi \beta}{b_\mathrm{v} N_\mathrm{v}^{a_\mathrm{v}-1}} \right) -a_\mathrm{v} \right]-\left[2a_\mathrm{c}B_\mathrm{cc}\bm{I}_\mathrm{c}\bm{A}_\mathrm{cc}\bm{I}_\mathrm{c}^T \right. \nonumber \\
 &+&2a_\mathrm{v}B_\mathrm{vv}\bm{I}_\mathrm{v}\bm{A}_\mathrm{vv}\bm{I}_\mathrm{v}^T+\left. (a_\mathrm{c}+a_\mathrm{v})B_\mathrm{cv}\bm{I}_\mathrm{c}\bm{A}_\mathrm{cv}\bm{I}_\mathrm{v}^T \right] \Biggl\}. \nonumber \\
\end{eqnarray}
This is associated with eq. (\ref{mu_a}), which is a formulation using spatial coordinates.


%
\bibliographystyle{jpsj}
\bibliography{jpsj_template.bib}


\end{document}